\documentclass[a4paper,11pt]{article}
\pdfoutput=1 

\usepackage{jheppub} 
\usepackage[T1]{fontenc} 
\usepackage{bm}
\usepackage{multirow}
\usepackage{nicefrac}
\usepackage[normalem]{ulem}

\newcommand{\cgr}[6]{
\biggl\langle
\begin{matrix} #1 & #3 \\ #2 & #4 \end{matrix}
\bigg\vert
\begin{matrix} #5 \\ #6 \end{matrix}
\biggr\rangle}

\newcommand{\vket}[4]{
\biggl\vert #1
\begin{matrix} #2 \\ #3 \end{matrix}
#4 \biggr\rangle}

\title{$\bm{\chi_{c1}(3872)}$ and its Partners in the Diabatic Born-Oppenheimer Approximation for QCD}

\author{Fareed Alasiri,}
\author{Eric Braaten,}
\author{Roberto Bruschini}
\affiliation{Department of Physics,
         The Ohio State University, Columbus, OH\ 43210, USA}

\emailAdd{alasiri.6@osu.edu}
\emailAdd{braaten.1@osu.edu}
\emailAdd{bruschini.1@osu.edu}

\abstract{
In the  Born-Oppenheimer approximation for QCD, the exotic hidden-charm tetraquark meson $\chi_{c1}(3872)$ is a near-threshold bound state in Born-Oppenheimer potentials associated with an isospin-0 adjoint meson.
The $\chi_{c1}(3872)$ is the $1^{++}$ member of a heavy-quark spin-symmetry multiplet whose other members have $J^{PC}$ quantum numbers $0^{++}$, $1^{+-}$, and $2^{++}$.
We introduce a simple model for the Born-Oppenheimer potentials that interpolates between the adjoint-meson potential at short distances and the triplet-meson-pair potential at large distances.
We take into account the spin splittings of charm mesons and adjoint mesons nonperturbatively for the first time by solving the diabatic Schr\"odinger equation.
We also take into account the narrow avoided crossing with the quarkonium potential. 
We tune the energy of $\chi_{c1}(3872)$ to the $D^* \bar{D}$ threshold and then calculate the spin splittings of the other members of the multiplet and their decay widths into charm-meson pairs.
We also calculate the energies and decay widths of the corresponding multiplet of hidden-bottom tetraquarks.
These calculations provide a template for the quantitative analysis of all hidden-heavy hadrons using the Born-Oppenheimer approximation for QCD.
}


\begin{document} 
\maketitle
\flushbottom

\section{Introduction}
\label{sec:intro}

The $\chi_{c1}(3872)$ discovered by the Belle Collaboration in 2003 \cite{Belle03} turned out to be just the first of  dozens of {\it exotic  heavy  hadrons} that have been discovered. 
Most of them are {\it hidden-heavy hadrons} whose constituents  include a charm-quark pair ($c \bar c$) 
or a bottom-quark pair ($b \bar b$).
The list of exotic hidden-heavy hadrons includes dozens of $c \bar c$ tetraquark mesons, 5 $b \bar b$ tetraquark mesons, and 5 $c \bar c$ pentaquark baryons \cite{Leb23b}.
These discoveries have revealed an unexpected corner of the QCD spectrum that challenges our understanding of  strongly interacting gauge theories.

Thousands of theory papers have been written on the exotic heavy hadrons.
(See, for example, Refs.~\cite{Ali17,Esp17,Leb17,Guo18,Kar18,Olsen18,Bram20,Chen23,Husk24} and references within.)
Most of these papers are based on constituent models that correspond to assumptions about how the heavy quark, the heavy antiquark, and the other colored constituents are clustered.
These models have had little success in explaining the  pattern of those exotic hidden-heavy hadrons that had already been observed or in predicting new exotic hidden-heavy hadrons before they were discovered.
They also make very little contact with the fundamental field theory QCD.

The Born-Oppenheimer (B\nobreakdash-O) approximation for QCD provides a theoretical framework for multi-heavy hadrons that is based firmly on QCD.
It was introduced by Juge, Kuti, and Morningstar in 1999, who applied it to quarkonium hybrid mesons  \cite{Jug99}. 
The B\nobreakdash-O approximation for QCD exploits the large mass of a heavy quark compared to the energies of gluons and light quarks.
The B\nobreakdash-O approximation applied to a hidden-heavy hadron separates the problem into two steps:
(1) the calculation of {\it Born-Oppenheimer potentials} defined by the energy of QCD in the presence of static color sources separated by a variable distance $r$, 
(2) the solution of the Schr\"odinger equation for a heavy quark and antiquark interacting through the B\nobreakdash-O potentials.
The B\nobreakdash-O approximation for QCD  applied to hidden-heavy and double-heavy hadrons has been formulated as an effective field theory called {\it Born-Oppenheimer Effective Field Theory} (BOEFT) \cite{Ber15,Onc17,Bram17,Soto20b,Ber24}.
BOEFT has been applied to the spectrum of quarkonium hybrids \cite{Ber15,Onc17,Soto23} and to the semi-inclusive decay rates of quarkonium hybrids into quarkonium \cite{Onc17,Bram22}.
It has also been applied to double-heavy baryons \cite{Soto20a}.
However, the development of BOEFT has not immediately revealed the pattern of the exotic heavy hadrons.

A pattern for the exotic hidden-heavy hadrons based on the B\nobreakdash-O approximation for QCD was proposed in Ref.~\cite{Braa24b}.
The B\nobreakdash-O potentials for hidden-heavy hadrons include adjoint-hadron potentials that are repulsive color-Coulomb potentials at small $r$ offset by the energy of an adjoint hadron, which is a hadron consisting of QCD fields bound to a static color-octet source.
The B\nobreakdash-O potentials also include heavy-hadron-pair potentials that approach the threshold for a pair of heavy hadrons at large $r$.
The spectrum of QCD in the presence of two static color sources must be a smooth function of the separation $r$ of the sources.
It has been recognized only recently that this requires each adjoint-hadron potential to connect smoothly to a heavy-hadron-pair potential in an intermediate region of $r$ \cite{Braa24b,Bram24}.
Such a potential could be a monotonically decreasing function of $r$, approaching the heavy-hadron-pair threshold from above.
In this case, the potential would not support any bound states.
However, the potential could instead cross below the heavy-hadron-pair threshold before approaching it.
In this case, the potential would have a minimum in the intermediate region of $r$ whose depth is determined by an adjoint-hadron energy.
The potential could support bound states or resonances if the minimum is deep enough.
In Ref.~\cite{Braa24b}, the proposal was applied to exotic hidden-heavy tetraquark mesons.
It explains a remarkable property of isospin-1 hidden-bottom tetraquarks.
In Ref.~\cite{Alas25}, the proposal was applied to hidden-heavy pentaquark baryons.
It gives remarkable predictions for the quantum numbers of nonstrange hidden-charm pentaquarks.

A particularly remarkable feature of $\chi_{c1}(3872)$, which we refer to more concisely as $X_c$, is that its mass is extremely close to the  threshold for the pair of charm mesons $D^{\ast0} \bar{D}^0$.
The difference between its mass and the threshold is $\varepsilon_X = -50 \pm 93$~keV \cite{PDG24}.
The $J^{PC}$ quantum numbers of  $X_c$ are $1^{++}$  \cite{LHCb13}.
It therefore has an $S$-wave coupling to the charm-meson pairs $D^{\ast0} \bar{D}^0$ and $D^0 \bar{D}^{\ast0}$.
The small binding energy $|\varepsilon_X|$ implies that $X_c$ is a loosely bound charm-meson molecule with the flavor structure $D^{\ast0} \bar{D}^0+D^0 \bar{D}^{\ast0}$ and with universal properties determined by $\varepsilon_X$ \cite{Braa03}.
One of the universal properties is a mean radius $(8\mu |\varepsilon_X|)^{-1/2}$,  where $\mu$ is the reduced mass of $D^{\ast0}$ and $D^0$.
The binding energy of $X_c$ implies an upper bound  $|\varepsilon_X| < 200$~keV at the 90\% confidence level.
Its mean radius is therefore greater than 5~fm, which is more than an order of magnitude larger than that of ordinary hadrons.

In Refs.~\cite{Braa24b,Bram24}, $X_c$ was identified as a bound state near threshold in the two B\nobreakdash-O potentials associated with the lowest-energy isospin-0 adjoint meson. 
One of the two potentials has a narrow avoided crossing with the quarkonium potential, which was taken into account in Ref.~\cite{Bram24}.
The $X_c$ is the $1^{++}$ member of a heavy-quark spin-symmetry (HQSS) multiplet whose other members have $J^{PC}$ quantum numbers $1^{+-}$, $0^{++}$, and $2^{++}$.
In Ref.~\cite{Braa24b}, the spin splittings of $X_c$ and its HQSS partners were estimated by treating the spin splittings of charm mesons as a first-order perturbation.
In Ref.~\cite{Bram24}, the spin splittings of $X_c$ and its HQSS partners were approximated by the spin splittings  of charmonium hybrid mesons.
A quantitative treatment of $X_c$ and its HQSS partners must take into account  their couplings to charm-meson pairs nonperturbatively.
In this paper, we treat them nonperturbatively for the first time by solving Schr\"odinger equations that include the terms that provide charm-meson spin splittings.

The traditional formulation of the B\nobreakdash-O approximation uses the {\it  adiabatic representation} in which the potential energy is diagonalized.
Its diagonal entries have simple physical interpretations as adiabatic energies.
The adiabatic B\nobreakdash-O approximation is most natural for deeply bound molecules.
There is a diametrically opposite representation of the B\nobreakdash-O approximation called the {\it  diabatic representation} in which the kinetic energy is diagonalized \cite{Smit69}. 
The diabatic B\nobreakdash-O approximation is most natural for the scattering of atoms and for loosely bound molecules.
The diabatic B\nobreakdash-O approximation for QCD was introduced by Bruschini and Gonz\'alez \cite{Brus20}.
Many of the exotic hidden-heavy hadrons that have been discovered including $X_c$ have mass close to the threshold for a pair of heavy hadrons.
These hidden-heavy hadrons are most naturally described using the diabatic B\nobreakdash-O approximation.
Their spin splittings are sensitive to the spin splittings of the pair of heavy hadrons.
The problem of taking into account heavy-meson spin splittings systematically was solved by Bruschini in Ref.~\cite{Brus23a} using the diabatic B\nobreakdash-O approximation.
This opened the way to a quantitative treatment of $X_c$ and other exotic hidden-heavy hadrons whose masses are close to a heavy-hadron-pair threshold.

In this paper, we use the diabatic B\nobreakdash-O approximation for QCD to present the first quantitative treatment of $X_c$ and its HQSS partners. 
We introduce a simple model for the B\nobreakdash-O potentials that interpolates between the adjoint-meson potential at small $r$ and the heavy-meson-pair threshold at large $r$.
We include in the potential energy the terms that produce the spin splittings of charm-meson pairs at large $r$ and the spin splittings of the adjoint meson at small $r$.
We also include the transition potential that produces the narrow avoided crossing with the quarkonium potential.
We improve the accuracy of the kinetic energy in heavy-meson-pair channels at large $r$ by replacing the heavy-quark mass by the appropriate heavy-meson mass.
We solve the coupled-channel diabatic Schr\"odinger equations to calculate the spin splittings of $X_c$ and its HQSS partners and to estimate their widths from decays into heavy-meson pairs.

The Schr\"odinger equations in the B\nobreakdash-O approximation for  $X_c$ and its HQSS partners have also been presented recently in Ref.~\cite{Bram26}.
The quarkonium transition potential and the charm-meson-pair spin-splitting terms were both taken into account.
The Schr\"odinger equations were solved numerically for the energies of quarkonium bound states but not for the energies of the HQSS partners of $X_c$, which are resonances or virtual states.

The remainder of this paper is organized as follows.
In Section~\ref{sec:QCDStaticSources}, we describe some relevant aspects of QCD with static color sources.
In Section~\ref{sec:BO-QCD}, we discuss the adiabatic and diabatic representations of the B\nobreakdash-O approximation for QCD.
In Section~\ref{sec:Adiabatic}, we present the adiabatic Hamiltonian that describes $X_c$ and its HQSS partners in the HQSS limit.
In Sections~\ref{sec:Diabatic} and \ref{sec:RadTetraCh}, we present the corresponding diabatic Hamiltonian and we take into account the spins of the heavy quark and antiquark. 
In Sections~\ref{sec:RadDiChannels} and  \ref{sec:SchrEqSpinSplit}, we take into account the  spin splittings of heavy mesons and the spin splittings of the adjoint meson. 
In Section~\ref{sec:AvoidedCrossing}, we take into account the narrow avoided crossing with the quarkonium potential.
In Sections~\ref{sec:RadTetraCh}, \ref{sec:RadDiChannels}, \ref{sec:SchrEqSpinSplit}, and \ref{sec:AvoidedCrossing}, the $1^{++}$ block of the radial diabatic Hamiltonian is presented explicitly.
The $0^{++}$, $1^{+-}$, and  $2^{++}$ blocks of the radial diabatic Hamiltonian are presented in Appendix~\ref{app:DiabaticH}.
In Section~\ref{sec:Models}, we present a simple model for the B\nobreakdash-O potentials and we give all the other parameters that appear in the coupled-channel Schr\"odinger equations.
In Section~\ref{sec:DeltaQDependence}, we compare the predictions of first-order perturbation theory in the spin splittings with nonperturbative results from solving the Schr\"odinger equation.
In Section~\ref{sec:CharmoniumTetraquarks}, we tune the energy of $X_c$ to the $D^* \bar{D}$ threshold and we then calculate the energies of the HQSS partners of $X_c$ and their decay widths into charm-meson pairs. 
In Section~\ref{sec:BottomoniumTetraquarks}, we calculate the energies of the analogous HQSS multiplet of $b \bar b$ tetraquarks and their decay widths into bottom-meson pairs.
We summarize our results in Section~\ref{sec:SummaryProspects} and we describe many of the improvements that would be required for quantitative predictions that can be compared with experiment.
In Appendix~\ref{sec:SPARSE}, we describe briefly the $K$-matrix formalism for scattering in the multichannel Schr\"odinger equation.
The real poles of the $K$-matrix for multiple coupled channels can be calculated with high precision efficiently using the SPARSE algorithm \cite{Brus25} and they provide good first approximations to the energies and widths of narrow resonances.


\section{QCD with Static Color Sources}
\label{sec:QCDStaticSources}

In this section, we discuss QCD in the presence of static color sources.
They include single color-triplet ($\bm{3}$), color-antitriplet ($\bm{3^\ast}$), and color-octet ($\bm{8}$) sources and also a pair of $\bm{3}$ and $\bm{3^\ast}$ color sources separated by a distance $r$.

\subsection{Light QCD}
\label{sec:LightQCD}

We refer to QCD  with only the gluon fields and the  fields for the light quarks $u$, $d$, and $s$ as {\it light QCD}.
Light QCD has translational symmetries in space and time that guarantee conservation of momentum and energy.
Light QCD has a rotational symmetry generated by the total angular momentum $\bm{j}$ of the light-QCD fields.
It also has two discrete symmetries:  parity $P_\mathrm{light}$ and charge conjugation $C_\mathrm{light}$, which satisfy $P_\mathrm{light}^2=1$ and $C_\mathrm{light}^2=1$.
Because of these symmetries, the energy eigenstates of light QCD can be chosen to have definite angular-momentum $(j,\lambda)$, parity $\pi$, and charge conjugation $\gamma$ quantum numbers.
We will specify the light-QCD quantum numbers as $j^{\pi\gamma}$.

Light QCD has an approximate $SU(3)$ flavor symmetry and a more accurate $SU(2)$ isospin symmetry.
The energy eigenstates of light QCD can be organized into $SU(3)$ multiplets and into approximately degenerate $SU(2)$ multiplets. 

\subsection{Triplet and antitriplet mesons}
\label{sec:Static}

The presence of a single static color source at the origin breaks the translational symmetry of QCD, but not the rotational symmetry.
A static $\bm{3}$ ($\bm{3^\ast}$) color source also breaks the charge conjugation symmetry of QCD, but not the parity symmetry. 
The energy eigenstates of light QCD in the presence of the $\bm{3}$ ($\bm{3^\ast}$) color source can be labeled by quantum numbers $j^\pi$.
They can also be organized into multiplets with light-quark flavor quantum numbers, such as the isospin.

The discrete states in the spectrum of light QCD with a static color source are called {\it static hadrons}.
The discrete states bound to a $\bm{3}$ color source and with the flavor of a light antiquark $\bar q$ are called {\it triplet mesons}.
The ground states of the triplet meson are an isospin doublet with quantum numbers $j^\pi=\tfrac12^-$.
The discrete states bound to a $\bm{3^\ast}$ color source and with the flavor of a light quark $q$ are called {\it antitriplet mesons}.
They are related to the triplet mesons by charge conjugation. 
The ground states of the antitriplet meson are an isospin doublet with quantum numbers $j^\pi=\tfrac12^+$.

\subsection{Adjoint mesons and gluelumps}
\label{sec:Adjoint}

The presence of a static {\bf 8} color source at the origin breaks the translational symmetry of QCD, but not the rotational,  charge conjugation, or  parity symmetries.
The energy eigenstates of light QCD with an $\bm{8}$ color source can therefore be chosen to have definite angular momentum, parity, and charge conjugation quantum numbers $j^{\pi\gamma}$.
They can also be organized into multiplets with light-quark flavor quantum numbers, such as isospin. 

The discrete states in the spectrum of light QCD with an $\bm{8}$ color source are called {\it adjoint hadrons}, because the {\bf 8} representation of the gauge group $SU(3)$ is also called the adjoint representation.
An adjoint hadron whose light-QCD fields have baryon number 0 is called an {\it adjoint meson}.
In pure $SU(3)$ gauge theory, the adjoint hadrons are called {\it gluelumps}.
In QCD, there are additional adjoint mesons with the flavor $q \bar{q}$ of a light quark and a light antiquark.
Gluelumps and $SU(3)$-singlet adjoint mesons can mix.
An adjoint hadron with the flavor $qqq$ of three light quarks is called an {\it adjoint baryon}.

The first lattice gauge theory calculations of gluelump energies were by Michael and collaborators using pure $SU(3)$ gauge theory \cite{Camp85,Fos98}.
The ground-state gluelump has $j^{\pi\gamma}= 1^{+-}$.
The first two excited gluelumps have quantum numbers $1^{--}$ and $2^{--}$.
The gluelump energies have been calculated more accurately by Herr, Schlosser, and Wagner \cite{Herr23}.
The energies of the $1^{--}$ and $2^{--}$ gluelumps relative to the $1^{+-}$ ground-state gluelump are 0.34(4) and 0.52(1)~GeV.
The gluelumps with much higher energy include the $0^{-+}$ gluelump with energy 2.35(12)~GeV relative to the $1^{+-}$ gluelump.
Marsh and Lewis have calculated the gluelump energies using lattice QCD with 2+1 flavors of dynamical light quarks and a pion mass of about $3.5\, m_\pi$ but without any extrapolation to zero lattice spacing \cite{Mar13}. 
The ordering in energy of the first few gluelumps is the same as in pure $SU(3)$ gauge theory.

Foster and Michael calculated the energies of adjoint mesons with light-quark flavors using $SU(3)$ gauge theory with massless valence quarks $q \bar q$ \cite{Fos98}.
The lowest-energy adjoint mesons have quantum numbers $J^P = 0^-$ and $1^-$.
Both have energies near that of the $1^{+-}$ ground-state gluelump.
Foster and Michael were unable to determine  the ordering in energy of the $1^{+-}$ ground-state gluelump, the $0^{-+}$ adjoint meson, and the $1^{--}$ adjoint meson. 
The definitive determination of their energies require lattice QCD calculations with dynamical light quarks. 

 As pointed out in Ref.~\cite{Braa24b}, the discoveries of exotic hidden-heavy hadrons imply that there are adjoint hadrons whose energies are hundreds of MeV lower than the $1^{+-}$ ground-state gluelump. 
The discovery of $X(3872)$ near the $D^*\bar{D}$ threshold implies that the isospin-0 $1^{--}$ adjoint meson has much lower energy than the ground-state gluelump.
The discoveries of the hidden-bottom tetraquark mesons $Z_b^+(10610)$ and $Z_b^+(10650)$ near the $B^*\bar{B}$ and $B^*\bar{B}^*$ thresholds by the Belle Collaboration \cite{Belle11} imply that the isospin-1 $0^{-+}$ and $1^{--}$  adjoint mesons have much lower energies than the ground-state gluelump. 
The discovery of the hidden-charm pentaquark baryon $P_{c\bar c}(4312)$ near the $\Sigma_c \bar{D}$ threshold by the LHCb collaboration \cite{LHCb19} implies that the isospin-1 $\tfrac12^+$ adjoint baryon has much lower energy than the ground-state gluelump \cite{Alas25}.
Thus the discoveries of exotic hidden-heavy hadrons provide strong motivation for definitive calculations of the spectrum of adjoint hadrons using lattice QCD.

\subsection{Adiabatic Born-Oppenheimer potentials}
\label{sec:BO-quantum}

The energies of discrete states of light QCD  in the presence of static $\bm{3}$ and $\bm{3^\ast}$ color sources as functions of their separation $r$ are called {\it adiabatic Born-Oppenheimer potentials}.
The presence of a $\bm{3}$ color source and a $\bm{3^\ast}$ color source at the positions $+\tfrac12\bm{r}$ and $-\tfrac12\bm{r}$ breaks the translational symmetry of light QCD.
The sources break the rotational symmetry group $SO(3)$ down to the subgroup $SO(2)$ generated by $\bm{j} \cdot \hat{\bm{r}}$.
The sources also break the discrete symmetries $P_\mathrm{light}$ and $C_\mathrm{light}$ down to the single discrete symmetry $(CP)_\mathrm{light}$.
The eigenstates of the light-QCD Hamiltonian with the $\bm{3}$ and $\bm{3^\ast}$ color sources can be chosen to also be eigenstates of $\bm{j} \cdot \hat{\bm{r}}$ and $(CP)_\mathrm{light}$.
There is also an additional discrete symmetry:  a reflection $R_\mathrm{light}$ through any specific plane containing $\bm{\hat{r}}$.
The rotations generated by $\bm{j} \cdot \hat{\bm{r}}$ and the reflections $R_\mathrm{light}$ together form the \emph{cylindrical symmetry group}, often denoted by $D_{\infty h}$.
 We refer to the cylindrical symmetry and the discrete symmetry $(CP)_\mathrm{light}$ as  {\it Born-Oppenheimer symmetries}.
The eigenvalues $\lambda$ of $\bm{j} \cdot \hat{\bm{r}}$ must be integer or half-integer.
The eigenvalues of $(CP)_\mathrm{light}$ are $+1$ or $-1$.
The eigenvalues of $R_\mathrm{light}$ are $+1$ or $-1$.
We refer to these eigenvalues as {\it Born-Oppenheimer quantum numbers}.
The operator $R_\mathrm{light}$ commutes with $\lvert \bm{j} \cdot \hat{\bm{r}} \rvert$ but not with $\bm{j} \cdot \hat{\bm{r}}$.
The energy eigenstates of light QCD with $\bm{3}$ and $\bm{3^\ast}$ color sources can therefore be chosen to also be eigenstates of $\lvert \bm{j} \cdot \hat{\bm{r}} \rvert$,
$(CP)_\mathrm{light}$, and $R_\mathrm{light}$.
Thus $|\lambda|$, $\eta$, and $\epsilon$ are independent B\nobreakdash-O quantum numbers.

The traditional labels for B\nobreakdash-O quantum numbers have the form $\Lambda_\eta^\epsilon$, where $\Lambda=|\lambda|$, the subscript $\eta$ is $g$ or $u$ if the eigenvalue of $(CP)_\mathrm{light}$ is $+1$ or $-1$, and the superscript $\epsilon$ is $+$ or $-$ if the eigenvalue of $R$  is  $+1$ or $-1$.
If $\lvert\lambda\rvert$ is an integer, $\Lambda$ is usually denoted by $\Sigma,\Pi,\Delta, \ldots$ if $|\lambda|$ is 0,1,2, \ldots.
If $\Lambda$ is not $\Sigma$, the superscript $\epsilon$ is usually omitted because cylindrical symmetry requires the states with $\Lambda_\eta^+$ and $\Lambda_\eta^-$ to be degenerate in energy.
If the symmetries also require the states with $\Lambda_g^\epsilon$ and $\Lambda_u^\epsilon$ to be degenerate in energy, the subscript $\eta$ can be omitted.
The lowest energy adiabatic potential with quantum numbers $\Lambda_\eta^\epsilon$ can be labeled $1\Lambda_\eta^\epsilon$.
The successively higher energy potentials with those quantum numbers are labeled $n\Lambda_\eta^\epsilon$ with successively higher integers $n$.
There are no crossings between different adiabatic potentials with the same quantum numbers $\Lambda_\eta^\epsilon$.

In the limit $r \to 0$, the rotational symmetry and the discrete symmetries of QCD are restored.
Thus the discrete states in this limit can be labeled by quantum numbers $j^{\pi\gamma}$.
As the separation $r$ of the $\bm{3}$ and $\bm{3^\ast}$ color sources decreases to 0, their effect on the light-QCD fields approaches that of a linear combination of  a local color-singlet ({\bf 1}) source and a local color-octet ({\bf 8}) source.
Since a local color-singlet operator is no color source at all, the only discrete state is the light-QCD vacuum with $j^{\pi\gamma} = 0^{++}$.
The associated adiabatic potential is the $1\Sigma_g^+$ potential.
At small $r$, it can be approximated by the attractive color-Coulomb potential proportional to $1/r$.
For the $\bm{8}$ color source, the discrete states are adjoint hadrons.
At small $r$, the associated adiabatic potentials can be approximated by the repulsive color-Coulomb potential proportional to $1/r$ offset by the energy of the adjoint hadron.
We refer to the continuation of such a potential to  larger $r$ as an {\it adjoint-hadron potential}.

The adjoint-meson potentials associated with an adjoint meson with quantum numbers $j^{\pi\gamma}$ have B\nobreakdash-O quantum numbers $\Lambda_\eta^\epsilon$ with $\Lambda=0,\ldots, j$ and subscript $\eta = \gamma\pi$.
The superscript is $\epsilon = \pi (-1)^{j}$ if $\Lambda=0$ and it can be $+$ or $-$ (but is usually omitted) if $\Lambda > 0$.
In  pure $SU(3)$ gauge theory, the  lowest-energy adjoint meson is the ground-state $1^{+-}$ gluelump.
The associated adjoint-meson potentials  are $\Pi_u$ and $\Sigma_u^-$ potentials and they become degenerate as $r \to 0$.
In QCD, the  lowest-energy adjoint mesons have not been established.
The calculations of adjoint-meson energies using $SU(3)$ gauge theory with massless valence quarks suggest that the lowest-energy adjoint mesons 
are the $1^{+-}$ gluelump and the adjoint mesons with flavors $q \bar q$ and quantum numbers $0^{-+}$ and $1^{--}$ \cite{Fos98}.
The adjoint-meson potential associated with a $0^{-+}$ adjoint meson is a $\Sigma_u^-$ potential.
The adjoint-meson potentials associated with a $1^{--}$ adjoint meson are a $\Sigma_g^+$ potential and a $\Pi_g$ potential, and they become degenerate as $r \to 0$.

In the limit $r \to \infty$, the confinement property of QCD implies that the finite-energy discrete states of light QCD with $\bm{3}$ and $\bm{3^\ast}$ color sources consist of a triplet hadron bound to the $\bm{3}$ color source and an antitriplet hadron bound to the $\bm{3^\ast}$ color source.
We will refer to such a state as a {\it triplet-hadron-pair} state.
Its energy is the sum of the energies of the triplet and antitriplet hadrons.
The states with definite B\nobreakdash-O quantum numbers $\Lambda_\eta^\epsilon$ are superpositions of triplet-hadron-pair states.
We refer to the continuation of such an adiabatic potential to smaller $r$ as a {\it triplet-hadron-pair potential}.
The B\nobreakdash-O quantum numbers $\Lambda_\eta^\epsilon$ for the triplet-hadron-pair potentials associated with a pair of triplet hadrons can be determined from their  quantum numbers $j_1^{\pi_1}$ and $j_2^{\pi_2}$ \cite{Braa24a}.
The total angular momentum $j$ ranges from $|j_1-j_2|$ to $j_1+j_2$.
The parity $P_\mathrm{light}$ is $\pi = \pi_1\pi_2$. 
The charge conjugation $C_\mathrm{light}$ can be +1 or $-1$.
For each $j$, the values of $\lambda$ range from $-j$ to $+j$ and $\Lambda = |\lambda|$.
If the triplet and antitriplet hadrons are charge-conjugate static mesons with $j_2^{\pi_2} = j_1^{-\pi_1}$, the triplet-meson pair  has $\pi = -1$ and $\eta = (-1)^{j+1}$. 

In the limit $r \to \infty$, the lowest-energy discrete states of light QCD with $\bm{3}$ and $\bm{3^\ast}$ color sources are a pair of ground-state triplet and antitriplet mesons with  $j_1^{\pi_1} = \tfrac12^-$ and $j_2^{\pi_2}=\tfrac12^+$.
Their total angular momentum $j$ can  be 0 or 1.
The reflection quantum number for $\Lambda = 0$ is $\epsilon = (-1)^{j+1}$.
The $j=0$ state has $(CP)_\mathrm{light}= -1$ and its B\nobreakdash-O quantum numbers are $\Sigma_u^-$.
The $j=1$ state has $(CP)_\mathrm{light}= +1$ and its B\nobreakdash-O quantum numbers are $\Pi_g$ and $\Sigma_g^+$.
Thus the lowest-energy triplet-meson-pair potentials at large $r$ are $1\Sigma_u^-$, $1\Sigma_g^+$, and $1\Pi_g$ \cite{Tar22}.
Note that their B\nobreakdash-O quantum numbers coincide with those for the adjoint-meson potentials associated with the $0^{-+}$ and $1^{--}$ adjoint mesons.
The $1\Sigma_u^-$, $1\Sigma_g^+$, and $1\Pi_g$ potentials at large $r$ approach a constant equal to twice the energy of the ground-state triplet meson.

The continuity of the spectrum of light QCD in the presence of two static color sources as a function of $r$ implies that all the adjoint-hadron potentials at small $r$ 
must connect smoothly to triplet-hadron-pair potentials  at large $r$ \cite{Braa24b,Bram24}.
The simplest possibility for the qualitative behavior of these potentials is that as $r$ increases, the potential approaches the triplet-hadron-pair threshold monotonically from above, in which case it cannot support any bound states or resonances.
The next simplest possibility is that the potential crosses below the triplet-hadron-pair threshold and then approaches it from below, in which case it may support bound states below the threshold if the potential is deep enough.
The next-to-next simplest possibility is that the potential crosses below the triplet-hadron-pair threshold, then crosses back above it, and finally approaches it from above, in which case it may also support resonances above the threshold if the barrier provided by the potential is high enough.
Whether these potentials are deep enough to support bound states or have barriers high enough to support resonances can only be determined by lattice QCD calculations with dynamical light quarks.

\subsection{Born-Oppenheimer potentials from Lattice QCD}
\label{sec:BOLattice}

The adiabatic potentials for QCD can be calculated using lattice QCD.
The first lattice calculations of the lowest $1\Sigma_g^+$ potential in pure $SU(3)$ gauge theory as a function of the distance $r$  between the $\bm{3}$ and $\bm{3^\ast}$ color sources showed that it increased approximately linearly with $r$,
consistent with the energy of a gluon flux tube connecting the sources.
Later calculations with smaller lattice spacing showed that its behavior at small $r$ was consistent with the attractive color-Coulomb potential proportional to $1/r$.

In their pioneering work on the B\nobreakdash-O approximation for QCD in 1999, Juge, Kuti, and Morningstar calculated the lowest-energy $\Pi_u$, $\Sigma_u^-$, and other excited potentials for pure $SU(3)$ gauge theory as functions of the distance $r$  between the sources  \cite{Jug99}. 
At large $r$, these excited potentials increase approximately linearly.
Their behavior at small $r$ could be deduced using the effective field theory pNRQCD \cite{Bram99}. 
The excited potentials all approach a common repulsive color-Coulomb potential proportional to $1/r$.
They form multiplets that approach the repulsive color-Coulomb potential offset by the energy of a gluelump. 
The energy splittings between the multiplets therefore approach the energy splittings of the gluelumps.
If the gluelump has quantum numbers $j^{\pi \gamma}$, the multiplet consists of $j+1$ excited potentials $\Lambda_\eta^\epsilon$ with $\eta = \gamma \pi$ and 
$\Lambda$ ranging from 0 to $j$.
The lowest-energy $\Pi_u$ and $\Sigma_u^-$ potentials form a multiplet associated with the $1^{+-}$ ground-state gluelump.

Juge, Kuti, and Morningstar calculated many more excited B\nobreakdash-O potentials in $SU(3)$ gauge theory in Ref.~\cite{Jug03}.
The next higher multiplet of potentials after $\Pi_u$ and $\Sigma_u^-$ have B\nobreakdash-O quantum numbers $\Pi_g$ and $\Sigma_g^+$ and are associated with a $1^{--}$ adjoint meson.
There have been several more recent calculations of the excited B\nobreakdash-O potentials in $SU(3)$ gauge theory \cite{Cap18,Sch21,Bicu21b,Sha23}.
Parametrizations of many of the excited potentials have been given in Ref.~\cite{Alas24}.

The first calculation of B\nobreakdash-O potentials using lattice QCD with dynamical light quarks was carried out by Bali et al.\  in 2000 \cite{Bal00}.
They calculated the quarkonium $\Sigma_g^+$ potential and the lowest hybrid potential $\Pi_u$ using lattice QCD with two flavors of light quarks.
The most important effect of the dynamical light quarks was a small increase in the $\Sigma_g^+$ potential at small $r$.
The difference can be attributed to the slower running of the QCD coupling constant in the attractive color-Coulomb potential.
The dynamical light quarks had no significant effect on the $\Pi_u$ potential.

The behavior of the lowest $\Sigma_g^+$ potential at large $r$ is very different in pure $SU(3)$ gauge theory and in QCD with dynamical light quarks.
In $SU(3)$ gauge theory, the $1\Sigma_g^+$ potential increases linearly as $r \to \infty$.
In QCD, the isospin-0 $1\Sigma_g^+$ potential at large $r$ must be the lowest triplet-meson-pair $\Sigma_g^+$ potential, which approaches a constant.
The crossover from increasing linearly to approaching a constant occurs through a narrow avoided crossing with the isospin-0 $2\Sigma_g^+$ potential.
The two lowest $\Sigma_g^+$ potentials were calculated by Bali {\it et al.}\ using lattice QCD with 2 flavors of light quarks and a pion mass of about $5\,m_\pi$ \cite{Bal05}. 
They calculated the lowest two adiabatic potentials $1\Sigma_g^+$ and $2\Sigma_g^+$ as functions of the separation $r$ of the static sources.
There is a narrow avoided crossing between the two potentials at a radius near 1.25~fm \cite{Bal05}. 

The calculations of the lowest-energy isospin-0  $\Sigma_g^+$ potentials have been extended to lattice QCD with $2+1$ flavors of dynamical light quarks by Bulava et al.\  \cite{Bul19,Bul24}. 
In Ref.~\cite{Bul24}, the potentials were extrapolated to the physical light-quark masses.
The three lowest $\Sigma_g^+$ adiabatic potentials were calculated as functions of $r$ in the avoided-crossing region.
There is a narrow avoided crossing between the $1\Sigma_g^+$ and $2\Sigma_g^+$ potentials at a radius near 1.21~fm and a narrower avoided crossing between the $2\Sigma_g^+$ and $3\Sigma_g^+$ potentials at a radius near 1.34~fm \cite{Bul24}.


\section{Born-Oppenheimer approximation for QCD}
\label{sec:BO-QCD}

In this section, we describe the Born-Oppenheimer (B\nobreakdash-O) approximation for QCD in the adiabatic representation and in the diabatic representation.

\subsection{Born-Oppenheimer approximation}
\label{sec:BOapprox}

The  B\nobreakdash-O approximation is used in atomic and molecular physics to understand the binding of atoms into molecules and the scattering of atoms \cite{Born27}. 
It exploits the large mass of an atomic nucleus compared to the mass  of the electron. 
The {\it Born-Oppenheimer approximation for QCD}  exploits the large mass $m_Q$  of the  heavy quark compared to the momentum scales associated with the interactions of gluons and light quarks.
In the B\nobreakdash-O approximation for a hidden-heavy hadron, the light-QCD fields are taken into account by treating their total energy in the presence of $\bm{3}$ and $\bm{3^\ast}$ color sources as a potential energy in a Schr\"odinger equation for the heavy quark $Q$ and the heavy antiquark $\bar Q$.
The discrete energies of light QCD with $\bm{3}$ and $\bm{3^\ast}$ color sources separated by $r$ define {\it adiabatic Born-Oppenheimer potentials}. 
They are labeled by B\nobreakdash-O quantum numbers $\Lambda_\eta^\epsilon$ and by an integer excitation number $n$.
The lowest adiabatic potential $1\Lambda_\eta^\epsilon$ can be calculated the most easily using lattice QCD, because it is the lowest-energy state with the quantum numbers $\Lambda_\eta^\epsilon$.
The spectrum of light QCD with $\bm{3}$ and $\bm{3^\ast}$ color sources also includes continuum states that correspond to one or more light hadrons scattering from the $\bm{3}$ and $\bm{3^\ast}$  sources.
These continuum states begin at thresholds that are equal to the sum of a lower B\nobreakdash-O potential and the masses of the light hadrons.

The positions of the $\bm{3}$ and $\bm{3^\ast}$ color sources can be identified with the positions of $Q$ and $\bar Q$. 
Truncating the states of light QCD to include only the discrete states reduces the problem to solving the Schr\"odinger equation with infinitely many adiabatic potentials.
 A further truncation of these adiabatic potentials to a finite number is necessary to get a Schr\"odinger equation for hidden-heavy hadrons that can be solved numerically.
 
\subsection{Adiabatic Born-Oppenheimer Hamiltonian}
\label{sec:AdiabaticH}

The B\nobreakdash-O Hamiltonian  $\mathbf{H}$ in the center-of-momentum frame can be expressed in terms of the separation vector $\bm{r}$ of the two color sources, the corresponding conjugate momentum $\bm{p} = -i\bm{\nabla}$, and the spins $\bm{S}_1$ and $\bm{S}_2$ of $Q$ and $\bar Q$.
It can be organized into an expansion in powers of $1/m_Q$, where $m_Q$ is the mass of the heavy quark.
At leading order in $1/m_Q$, the Hamiltonian  reduces to the {\it static potential} $\mathbf{V}_0(\bm{r})$, which depends on $\bm{r}$ but not on $\bm{p}$ or $\bm{S}_1$ or $\bm{S}_2$.
At first order in $1/m_Q$, the B\nobreakdash-O Hamiltonian $\mathbf{H}_1$ includes also a kinetic term that depends on $\bm{p}$ and $\bm{r}$ and the $1/m_Q$ potential  $\mathbf{V}_1$ that depends on $\bm{r}$ and also on $\bm{S}_1$ and $\bm{S}_2$. 
The eigenvalues of the static potential $\mathbf{V}_0(\bm{r})$ depend only on the radial variable $r =|\bm{r}|$.
They are discrete energies of light QCD with $\bm{3}$ and $\bm{3^\ast}$ color sources separated by a distance $r$.
The eigenvalues of $\mathbf{V}_0(\bm{r})$ are called {\it adiabatic potentials} and we denote them by $V_{n \Lambda_\eta^\epsilon}(r)$.
The adiabatic potentials for $n\Lambda_\eta^\epsilon$ are strictly ordered in energy and have no crossings, i.e.\  $V_{(n+1)\Lambda_\eta^\epsilon}(r) > V_{n\Lambda_\eta^\epsilon}(r) $ at every value of $r$ for all $n$. 
The adiabatic potentials can also be labeled by additional light-quark flavor quantum numbers, such as isospin, but these will usually not be indicated explicitly.

In the {\it adiabatic representation}, a unitary transformation $\mathbf{U}(\bm{r})$ that depends on the vector $\bm{r}$ is used to diagonalize the static potential $\mathbf{V}_0(\bm{r})$.
We denote the diagonal static-potential matrix by $\mathbf{V}_0^{\text{(ad)}}(r)$. 
Its diagonal entries are the adiabatic potentials $V_{n\Lambda_\eta^\epsilon}(r)$.
In the adiabatic representation,  the B\nobreakdash-O Hamiltonian through order $1/m_Q$ has the form
\begin{equation}
\mathbf{H}_1^\mathrm{ad} =  - \frac{1}{m_Q} \big(  \bm{\nabla}+ \bm{\Pi}(\bm{r}) \big)^2 
 +  \mathbf{V}^\mathrm{ad}_0(r) +  \mathbf{V}_1^\mathrm{ad}(\bm{r},\bm{S}_1,\bm{S}_2).
\label{H-adiabatic}
\end{equation}
In the kinetic term, the {\it nonadiabatic coupling} $\bm{\Pi}(\bm{r})$ is a vector whose three components are real antisymmetric matrices that depend on $\bm{r}$.  
It introduces couplings of order $1/m_Q$ between some of the adiabatic potentials $n \Lambda_\eta^\epsilon$.
The effects of these couplings can be important when the differences between the potentials are small.
There are at least two situations in which the nonadiabatic couplings are important.
The adjoint-meson potentials associated with an adjoint meson with quantum numbers $j^{\pi \gamma}$ have B\nobreakdash-O quantum numbers $n \Lambda_\eta^\epsilon$ with $\Lambda = 0,1,\ldots,j$.
These potentials become degenerate as $r \to 0$, so the nonadiabatic couplings between them are important in the small-$r$ region.
Although subsequent adiabatic potentials labeled by $n\Lambda_\eta^\epsilon$ and $(n+1)\Lambda_\eta^\epsilon$ cannot cross, they can have a {\it narrow avoided crossing}.
The nonadiabatic couplings between two such potentials are important in the avoided-crossing region of $r$.
The triplet-hadron-pair potentials associated with any specific pair of triplet hadrons become degenerate in the limit $r \to \infty$.
If the spin splittings of heavy hadrons of order $1/m_Q$ are taken into account, there are transitions between the triplet-hadron-pair potentials from the $1/m_Q$ adiabatic potential $\mathbf{V}_1^\mathrm{ad}$ that are important in the large-$r$ region.

We will refer to the Hamiltonian that includes only the kinetic term and the adiabatic static potential $\mathbf{V}^\mathrm{ad}_0(r)$ in Eq.~\eqref{H-adiabatic} as the {\it adiabatic Hamiltonian in the HQSS limit}, because it respects heavy-quark spin symmetry. 
The $1/m_Q$ adiabatic potential $\mathbf{V}_1^\mathrm{ad}(\bm{r},\bm{S}_1,\bm{S}_2)$ breaks HQSS through its dependence on $\bm{S}_1$ and $\bm{S}_2$.
In the large-$r$ limit, the spin-dependent term in $\mathbf{V}_1^\mathrm{ad}$ depends only on $\bm{S}_1$ and $\bm{S}_2$ and it gives spin splittings to well-separated heavy hadrons.
In the small-$r$ limit, the spin-dependent term in $\mathbf{V}_1^\mathrm{ad}$ depends only on $\bm{S}_1+\bm{S}_2$ and it gives spin splittings to adjoint hadrons with $j>0$.
In the intermediate region of $r$, the spin-dependent terms in $\mathbf{V}_1^\mathrm{ad}$ are more complicated, because they depend also on $\bm{r}$.

\begin{figure}[t]
\centerline{ \includegraphics[width=12cm,clip=true]{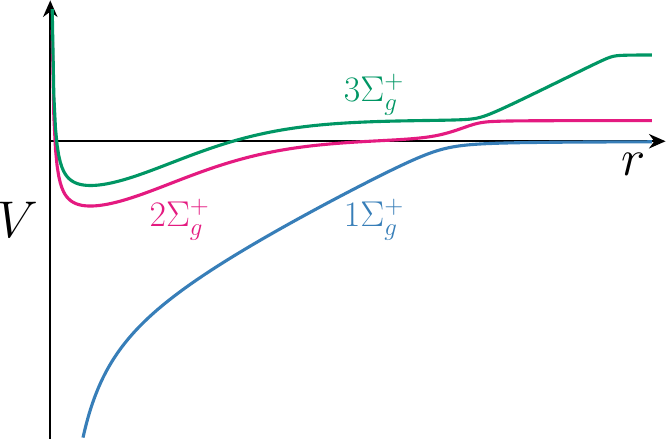} }
\caption{
Qualitative behaviors of the adiabatic isospin-0 $1\Sigma_g^+$, $2\Sigma_g^+$, and $3\Sigma_g^+$ potentials.
The horizontal axis is twice the energy of the ground-state triplet meson.
There are narrow avoided crossings between the $1\Sigma_g^+$ and $2\Sigma_g^+$ potentials and between the $2\Sigma_g^+$ and $3\Sigma_g^+$ potentials. 
}
\label{fig:AdiabaticPot}
\end{figure}

The adiabatic potentials most relevant to $X_c$ and its HQSS partners are the isospin-0 $1\Pi_g$ and $2\Sigma_g^+$ potentials. 
The qualitative behavior of the $2\Sigma_g^+$ potential is illustrated in Fig.~\ref{fig:AdiabaticPot}.
At small $r$,  the $2\Sigma_g^+$ potential and the $1\Pi_g$ potential (which is not shown in Fig.~\ref{fig:AdiabaticPot}) both approach the repulsive color-Coulomb potential offset by the energy of the lowest-energy isospin-0 $1^{--}$ adjoint meson.
In the intermediate region of $r$, they both approach the ground-state triplet-meson-pair threshold. 
The isospin-0 $1\Sigma_g^+$ potential is also somewhat relevant to $X_c$ and its HQSS partners, because it has a narrow avoided crossing with the $2\Sigma_g^+$ potential, as illustrated in Fig.~\ref{fig:AdiabaticPot}.
The narrow avoided crossing was first calculated using lattice QCD by Bali et al.\  \cite{Bal00}.
At small $r$, the $1\Sigma_g^+$ potential is the attractive color-Coulomb potential.  
In the intermediate region of $r$, it crosses over to increasing linearly with $r$. 
In the avoided-crossing region, the $1\Sigma_g^+$ potential crosses over from increasing linearly to approaching a constant equal to the ground-state triplet-meson-pair threshold.
The $2\Sigma_g^+$ potential has the opposite behavior,  crossing over from approximately constant  to increasing linearly.
The $1\Pi_g$ potential remains near the triplet-meson-pair threshold through the avoided-crossing region and beyond.
The avoided crossing between the $1\Sigma_g^+$ and $2\Sigma_g^+$ potentials is narrow in the sense that the minimum energy gap between the two potentials is much smaller than the nonperturbative energy scale $\Lambda_\text{QCD}$.
It is narrow because the crossover of the two potentials requires the creation or annihilation of a light-quark pair ($u \bar u$ or $d \bar d$).
Such a process is observed empirically to be suppressed in low-energy hadron physics presumably from nonperturbative effects in QCD.

The isospin-0 $2\Pi_g$ and $3\Sigma_g^+$ potential may be relevant to flavor partners of $X_c$ in which the light-quark pair is replaced by a strange-quark pair ($s \bar s$).
The qualitative behavior of the $3\Sigma_g^+$ adiabatic potential is illustrated in Fig.~\ref{fig:AdiabaticPot}.
The $3\Sigma_g^+$ potential has a narrow avoided crossing with the $2\Sigma_g^+$ potential.
That narrow avoided crossing was first calculated using lattice QCD by Bulava et al.\ \cite{Bul19}.
At small $r$, the $3\Sigma_g^+$ potential and the $2\Pi_g$ potential (which is not shown in Fig.~\ref{fig:AdiabaticPot}) both approach the repulsive color-Coulomb potential offset by the energy of the lowest-energy isospin-0 $1^{--}$ adjoint meson with an $s \bar s$ pair.
In the intermediate region of $r$, they are both approximately equal to the strange-triplet-meson-pair threshold.
In the avoided-crossing region, the $2\Sigma_g^+$ potential  crosses over from increasing linearly to approaching a constant equal to the strange-triplet-meson-pair threshold.
The $3\Sigma_g^+$ potential has the opposite behavior, crossing over from approximately constant to increasing linearly.
The $2\Pi_g$ potential remains near the strange-triplet-meson-pair threshold through the avoided-crossing region and beyond.
At still larger $r$, the $3\Sigma_g^+$ potential has another avoided crossing where it crosses over to approaching a constant that is approximately equal to the threshold for a ground-state $S$-wave triplet meson and a $P$-wave triplet meson.

The isospin-0 $1\Sigma_u^-$ potential is also relevant to the HQSS partners of $X_c$, because it approaches the ground-state triplet-meson-pair threshold at large $r$ and therefore becomes degenerate with the $1\Pi_g$ and $1\Sigma_g^+$ potentials as $r \to \infty$.
At small $r$, it approaches the repulsive color-Coulomb potential offset by the energy of the lowest-energy isospin-0 $0^{-+}$ adjoint meson.

\subsection{Diabatic Born-Oppenheimer Hamiltonian}
\label{sec:DiabaticH}

The form of the adiabatic Hamiltonian in Eq.~\eqref{H-adiabatic} can be changed by a unitary transformation $\mathbf{A}(\bm{r})$ that depends on the vector $\bm{r}$.
In the {\it diabatic representation}, the kinetic term in Eq.~\eqref{H-adiabatic} is simplified by a unitary transformation that satisfies $\mathbf{A}(\bm{r}) \big(  \bm{\nabla}+ \bm{\Pi}(\bm{r}) \big)\mathbf{A}(\bm{r}) =\bm{\nabla}$.
In the diabatic representation, the B\nobreakdash-O Hamiltonian through order $1/m_Q$ has the form
\begin{equation}
\mathbf{H}_1^\mathrm{di} = - \frac{1}{m_Q} \, \bm{\nabla}^2 
+ \mathbf{V}^\mathrm{di}_0(\bm{r}) + \mathbf{V}_1^\mathrm{di}(\bm{r},\bm{S}_1,\bm{S}_2 ) .
\label{H-diabatic}
\end{equation}
The diabatic static potential $\mathbf{V}^\mathrm{di}_0(\bm{r}) = \mathbf{A}(\bm{r}) \mathbf{V}^\mathrm{ad}_0(r) \mathbf{A}(\bm{r})^\dagger$ depends on the angles of $\bm{r}$.
Since the adiabatic and diabatic representations are related by a unitary transformation, the diabatic Hamiltonian in Eq.~\eqref{H-diabatic} and the adiabatic Hamiltonian in Eq.~\eqref{H-adiabatic} are completely equivalent as long as there is no truncation of the infinitely many coupled  B\nobreakdash-O potentials.
However, in practical applications of the B\nobreakdash-O approximation, it is necessary to truncate the infinite number of potentials to a finite number.
The truncated adiabatic Hamiltonian is no longer exactly equivalent to the truncated diabatic Hamiltonian.
At best, the truncated Hamiltonians may be related by a transformation that is approximately unitary in a limited region of $r$ and in a limited range of the energy.

We will refer to the Hamiltonian that includes only the kinetic term and the diabatic static potential $\mathbf{V}^\mathrm{di}_0(\bm{r})$ in Eq.~\eqref{H-diabatic} as the {\it diabatic Hamiltonian in the HQSS limit}, because it has exact heavy-quark spin symmetry.
The $1/m_Q$ diabatic potential $\mathbf{V}_1^\mathrm{di}(\bm{r},\bm{S}_1,\bm{S}_2 )$ breaks HQSS through its dependence on $\bm{S}_1$ and $\bm{S}_2$.
An angular decomposition can be used to make the entries of the diabatic static potential functions of the radial variable $r$ only \cite{Brus23a}, in which case we denote it by $\mathbf{V}^\mathrm{di}_0(r)$.
The diagonal entries of $\mathbf{V}^\mathrm{di}_0(r)$ are called {\it diabatic  potentials}.
Unlike adiabatic potentials, diabatic potentials with the same B\nobreakdash-O quantum numbers $\Lambda_\eta^\epsilon$ can cross.
The off-diagonal entries of $\mathbf{V}^\mathrm{di}_0(r)$ are called {\it transition potentials}.
The vanishing of the nonadiabatic coupling terms implies that the nature of the states in the diabatic basis change slowly as functions of $r$. 
If a pair of adiabatic potentials has a narrow avoided crossing, the corresponding pair of diabatic potentials cross.

The transformations of the diabatic light-QCD states with energies $V_{n\,\Lambda_\eta^\epsilon}(r)$ under the symmetries of light QCD are very complicated except in the limits $r \to 0$ and $r \to \infty$.
The symmetries of light QCD include rotations generated by $\bm{j}$ and the discrete symmetries $P_\text{light}$ and $C_\text{light}$.
In the limit $r \to 0$, the adjoint-meson potentials associated with an adjoint meson with quantum numbers $j^{\pi\gamma}$ are labeled by $n\Lambda_\eta^\epsilon$ with $\eta = \gamma\pi$ and $\Lambda = 0,1,\ldots,j$.
In the limit $r \to \infty$, the triplet-meson-pair potentials associated with triplet and anti-triplet mesons with quantum numbers $j_1^{\pi_1}$ and $j_2^{\pi_2}$ have parity $\pi = \pi_1 \pi_2$ and total angular momentum $j$ ranging from $|j_1-j_2|$ to $j_1+j_2$.
In both limits, the light-QCD states have simple transformations under the symmetries of light QCD.
In the intermediate region of $r$, the expansion of a state labeled by $n\Lambda_\eta^\epsilon$ in terms of states with simple rotational transformations includes all $j \ge \Lambda$.
The expansion in terms of states with simple parity transformations includes $- \pi$ as well as $\pi$.

The diabatic representation of the B\nobreakdash-O approximation provides the flexibility to truncate the diabatic potentials so that the corresponding light-QCD states all have simple transformation properties under the symmetries of light QCD \cite{Brus23a,braa25x}. 
Specifically, the potentials can be chosen to have definite light-QCD quantum numbers $j^{\pi \gamma}$ not only as $r \to 0$ and $r \to \infty$ but for all $r$.
A diabatic potential with light-QCD quantum numbers $j^{\pi \gamma}$ and with B\nobreakdash-O quantum numbers $\Lambda_\eta^\epsilon$ with $\eta = \gamma \pi$ can be labeled by $j^{\pi \gamma}\Lambda_\eta^\epsilon$.
We refer to these labels as {\it diabatic quantum numbers}.
Note that there is some redundancy in the diabatic quantum numbers because $\gamma\pi = \eta$.
If the truncation includes more than one channel with the same diabatic quantum numbers $j^{\pi \gamma}\Lambda_\eta^\epsilon$, the higher-energy channels can be distinguished by inserting primes in the exponent of $j^{\pi \gamma}$, as in $j^{\pi \gamma \prime}$.
The truncated diabatic potentials $j^{\pi \gamma}\Lambda_\eta^\epsilon$ can be chosen so that the rotational and discrete symmetries of QCD remain exact symmetries of the Schr\"odinger equation \cite{Brus23a,braa25x}. 
In particular, the quantum numbers $J^{PC}$ remain exactly conserved. 

\begin{figure}[t]
\centerline{ \includegraphics[width=12cm,clip=true]{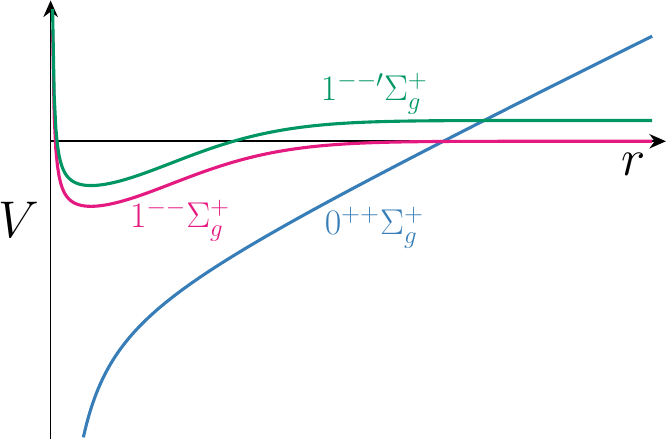}}
\caption{
Qualitative behaviors of the diabatic isospin-0 $0^{++}\Sigma_g^+$, $1^{--}\Sigma_g^+$, and $1^{--\prime}\Sigma_g^+$ potentials.
The horizontal axis is twice the energy of the ground-state triplet meson.
}
\label{fig:Diabatic}
\end{figure}

The diabatic potentials most relevant to $X_c$ and its HQSS partners are the $1^{--}\Pi_g$ and $1^{--}\Sigma_g^+$ potentials.
The qualitative behavior of the $1^{--}\Sigma_g^+$ diabatic potential is illustrated in Fig.~\ref{fig:Diabatic}.
At small $r$, the $1^{--}\Sigma_g^+$ potential and the $1^{--}\Pi_g$ potential (which is not shown in Fig.~\ref{fig:Diabatic}) both approach the repulsive color-Coulomb potential offset by the energy of the isospin-0 $1^{--}$ adjoint meson.
In the intermediate region of $r$, they both approach the ground-state triplet-meson-pair threshold and they continue to approach that threshold as $r \to \infty$.
In this paper, we assume the energy of the isospin-0 $1^{--}$ adjoint meson is low enough that  the $1^{--}\Pi_g$ and $1^{--}\Sigma_g^+$ potentials support a bound state near the ground-state triplet-meson-pair threshold.
The isospin-0 $0^{++}\Sigma_g^+$ potential is also relevant to $X_c$ and its HQSS partners, because it crosses the $1^{--}\Sigma_g^+$ potential and it has the same B\nobreakdash-O quantum numbers.
The qualitative behavior of the $0^{++}\Sigma_g^+$ diabatic potential is illustrated in Fig.~\ref{fig:Diabatic}.
The $0^{++}\Sigma_g^+$ potential is the quarkonium potential.
At small $r$, it is the attractive color-Coulomb potential.
In the intermediate region of $r$, it crosses over to increasing linearly as $r \to \infty$.
The narrow avoided crossing between the $1\Sigma_g^+$ and $2\Sigma_g^+$ adiabatic potentials implies that the transition potential between the $0^{++}\Sigma_g^+$ and $1^{--}\Sigma_g^+$ diabatic potentials is small.

The isospin-0 $1^{--\prime}\Pi_g$ and $1^{--\prime}\Sigma_g^+$ potentials are relevant to flavor partners of $X_c$ in which the light-quark pair is replaced by a strange-quark pair ($s \bar s$).
The qualitative behavior of the $1^{--\prime}\Sigma_g^+$ diabatic potential is illustrated in Fig.~\ref{fig:Diabatic}.
At small $r$, the $1^{--\prime}\Sigma_g^+$ potential and the $1^{--\prime}\Pi_g$ potential (which is not shown in Fig.~\ref{fig:Diabatic}) both approach the repulsive color-Coulomb potential offset by the energy of the lowest-energy isospin-0 $1^{--}$ adjoint meson with an $s \bar s$ pair.
In the intermediate region of $r$, they both approach a constant that is equal to the strange-triplet-meson-pair threshold and they continue to approach that threshold as $r \to \infty$.
The narrow avoided crossing between the $2\Sigma_g^+$ and $3\Sigma_g^+$ adiabatic potentials in Fig.~\ref{fig:AdiabaticPot} implies that the transition potential between the $0^{++}\Sigma_g^+$ and $1^{--\prime}\Sigma_g^+$ diabatic potentials is small.

The isospin-0 $0^{-+}\Sigma_u^-$ potential is relevant to the HQSS partners of $X_c$ because it approaches the triplet-meson-pair threshold at large $r$ and  therefore becomes degenerate with the $1^{--}\Pi_g$ and $1^{--}\Sigma_g^+$ potentials as $r \to \infty$.
At small $r$, the  $0^{-+}\Sigma_u^-$ potential approaches the repulsive color-Coulomb potential offset by the energy of the isospin-0 $0^{-+}$ adjoint meson.
In this paper, we assume the $0^{-+}\Sigma_u^-$  potential does not support any bound states.

\subsection{Angular Momenta}
\label{sec:AngularMomenta}

The total angular momentum $\bm{J}$ for a hidden-heavy tetraquark system is the sum of the angular momentum $\bm{j}$ of light QCD, the relative orbital angular momentum $\bm{L}_{Q\bar Q}$ of the $Q \bar Q$ pair, and the $Q \bar Q$ spin $\bm{S}_{Q\bar Q} = \bm{S}_1 + \bm{S}_2$. 
It can be expressed as 
\begin{equation}
\bm{J} = (\bm{j} + \bm{L}_{Q\bar Q}) + \bm{S}_{Q\bar Q} =  \bm{L} + \bm{S}_{Q\bar Q},
\label{J=L+S}
\end{equation}
where $\bm{L} = \bm{j} + \bm{L}_{Q\bar Q}$ is the {\it Born-Oppenheimer  angular momentum}. 
In the HQSS limit, $\bm{S}_{Q\bar Q}$ is conserved.
Since $\bm{J}$ is always conserved and $\bm{S}_{Q\bar Q}$  is conserved in the HQSS limit, $\bm{L}$  is also conserved  in the HQSS limit.
In the B\nobreakdash-O  Schr\"odinger equation, $\bm{L}$ plays a role similar to an orbital angular momentum in a conventional Schr\"odinger equation.
We therefore denote the quantum number for $\bm{L}^2$ by $L$.
In the case of hidden-heavy tetraquarks, the possible values of $L$ are nonnegative integers: $L=0,1,2,\ldots$.
Note that $L$ is not the quantum number for the square $ \bm{L}_{Q\bar Q}^2$ of the orbital angular momentum, which we denote instead by $L_{Q\bar Q}$.
While $L$ is conserved in the HQSS limit, $L_{Q\bar Q}$ is only conserved in the HQSS limit if $j=0$.

The angular-momentum quantum numbers associated with the various angular momenta in Eq.~\eqref{J=L+S} are constrained by triangle relations.  
The {\it triangle symbol} $\{ j_1,j_2,j_3\}$ is defined to be 1 if the three angular-momentum quantum numbers satisfy the {\it triangle relation} $|j_1 - j_3| \le j_2 \le j_1 + j_3$ and 0 otherwise.
The triangle symbol  $\{ j_1,j_2,j_3\}$ is invariant under permutations of $j_1$, $j_2$, and $j_3$.
The B\nobreakdash-O angular momentum $L$ is constrained by the triangle relations $\{ J, L, S_{Q\bar Q}  \}=1$ and $\{ j, L_{Q\bar Q,} L \}=1$.

The total angular momentum $\bm{J}$ in Eq.~\eqref{J=L+S} can also be expressed as 
\begin{equation}
\bm{J} = (\bm{j} + \bm{S}_{Q\bar Q}) + \bm{L}_{Q\bar Q}  = \bm{S} + \bm{L}_{Q\bar Q} ,
\label{J=J+L}
\end{equation}
where $\bm{S} = \bm{j} + \bm{S}_{Q\bar Q}$ is the {\it static angular momentum}. 
The static angular momentum was used by Bruschini in Ref.~\cite{Brus23a} to take into account the spin splittings of heavy mesons within the diabatic  B\nobreakdash-O  framework.
In the B\nobreakdash-O Schr\"odinger equation, $\bm{S}$ plays a role similar to a spin in a conventional Schr\"odinger equation.
We therefore denote the quantum number for $\bm{S}^2$ by $S$. 
In the case of hidden-heavy tetraquark mesons,  the possible values of $S$ are nonnegative integers: $S=0,1,2,\ldots$.
Note that $S$ is not the quantum number for the square $\bm{S}_{Q\bar Q}^2$ of the $Q \bar Q$ spin, which we denote instead by $S_{Q\bar Q}$. 
The possible values of $S_{Q\bar Q}$ are 0 and 1.
While $S_{Q\bar Q}$ is conserved in the HQSS limit, $S$ is only conserved if $L_{Q\bar Q} = 0$.
The static angular momentum $S$ is constrained by the triangle relations $\{ J, S, L_{Q\bar Q}  \}=1$ and $\{ j, S_{Q\bar Q,} S \}=1$.

\subsection{Conserved Quantum Numbers}
\label{sec:Conserved}

The B\nobreakdash-O Hamiltonian with all of the infinitely many potentials conserves $J^{PC}$.
In the HQSS limit, the B\nobreakdash-O Hamiltonian without any truncation of the infinitely many potentials conserves $L^P$ as well as $S_{Q \bar Q}$. 
If the adiabatic B\nobreakdash-O Hamiltonian is truncated to a finite number of potentials, the quantum numbers $J^{PC}$ (or $L^P$ in the HQSS limit) are not exactly conserved.
They will only be approximately conserved in the energy region where the truncated B\nobreakdash-O Hamiltonian gives a good approximation to the complete Hamiltonian.
In this energy region, the truncated adiabatic Hamiltonian can be organized into a block-diagonal form with blocks labeled by $J^{PC}$ (or by $L^P$ in the HQSS limit).
These quantum numbers will be conserved only if the off-diagonal blocks that break the corresponding symmetries are neglected.

In the diabatic representation, the conservation of $L^P$ in the HQSS limit and the exact conservation of $J^{PC}$ can be achieved by appropriate truncations of the B\nobreakdash-O Hamiltonian. 
In the HQSS limit, conservation of $L^P$ requires including all the diabatic potentials associated with an adjoint meson at small $r$. 
If the truncation includes a specific $j^{\pi \gamma}\Lambda_\eta^\epsilon$ potential, it must include all such potentials with $\Lambda \le j$.
Beyond the HQSS limit, conservation of $J^{PC}$ requires including all the diabatic potentials with the same value of the static angular momentum $S$ \cite{Brus23a}.
If the truncation includes diabatic potentials that approach the threshold for a pair of static hadrons with quantum numbers $j_1^{\pi_1}$ and $j_2^{\pi_2}$ at large $r$, conservation of $J^{PC}$ requires including all the diabatic potentials $j^{\pi\gamma}\Lambda_\eta^\epsilon$ with $\pi = \pi_1 \pi_2$ and $\pi \gamma = \eta$ that approach that threshold with $j$ ranging from $|j_1 - j_2|$ to $j_1 + j_2$.

We will use truncations of the diabatic Hamiltonian for which $L^P$ is conserved in the HQSS limit and $J^{PC}$ is exactly conserved.
In the HQSS limit, truncating to the $1^{--}\Pi_g$ and $1^{--}\Sigma_g^+$ potentials ensures the conservation of $L^P$. 
The $1^{--}\Pi_g$ and $1^{--}\Sigma_g^+$ potentials approach the ground-state triplet-meson-pair threshold at large $r$. 
The $0^{-+}\Sigma_u^-$ potential also  approaches the triplet-meson-pair threshold at large $r$.
Beyond the HQSS limit, the $0^{-+}\Sigma_u^-$ potential must be included in the truncation to ensure the conservation of $J^{PC}$.
Since the $0^{++}\Sigma_g^+$ potential crosses the $1^{--}\Sigma_g^+$ potential, the $0^{++}\Sigma_g^+$ potential must also be included in the truncation for it to be a good approximation in the energy region near the triplet-meson-pair threshold.

\subsection{Kinetic Improvement}
\label{sec:KineticImprovement}

In the diabatic representation of the B\nobreakdash-O approximation, the kinetic term in the Hamiltonian is diagonal.
At first order in $1/m_Q$, the kinetic term has the simple form in Eq.~\eqref{H-diabatic} with a multiplicative factor of $1/m_Q$.
The  kinetic term remains hermitian and diagonal if the operator $\bm{\nabla}^2/m_Q$ is replaced by $\bm{\nabla} \cdot (\mathbf{M}(r)^{-1} \bm{\nabla})$, where $\mathbf{M}(r)$ is a diagonal mass matrix that depends on the radial variable $r$.
We refer to this replacement as a {\it kinetic improvement}.
The diabatic Hamiltonian has additional kinetic terms that are higher order in $1/m_Q$.
There are  important effects of higher order kinetic terms that can be absorbed into the diagonal entries of $\mathbf{M}(r)$.
A heavy-hadron-pair potential at large $r$ approaches the threshold for a pair of heavy hadrons.
If the two heavy hadrons have relative momentum $p$, the diabatic Hamiltonian in Eq.~\eqref{H-diabatic} implies that their kinetic energy is $p^2/m_Q$.
However, the physical kinetic energy of two well-separated heavy hadrons with relative momentum $p$  is $p^2/(2\mu)$, where $\mu$ is the reduced mass of the two heavy hadrons.
The difference between the denominators $1/m_Q$ and $1/(2\mu)$ comes from kinetic terms of order $1/m_Q^2$ and higher.
The accuracy of the Hamiltonian at large $r$ can be improved by a resummation that replaces $m_Q$ by an interpolation between $m_Q$ at small $r$ and $2\mu$ at large $r$.

A particularly simple form of kinetic improvement can be obtained by using a constant mass matrix $\mathbf{M}$ that does not depend on $r$.
The $0^{++}\Sigma_g^+$ diabatic quarkonium potential is the attractive color-Coulomb potential at small $r$ and it is a linear repulsive potential at large $r$.
A natural choice for the diagonal entry of $\mathbf{M}$ for such a channel is the heavy-quark mass $m_Q$.
The other hidden-heavy diabatic potentials are repulsive color-Coulomb potentials at small $r$ and they approach the threshold for a pair of heavy hadrons at large $r$.
A natural choice for the diagonal entry of $\mathbf{M}$ for such a channel  is twice the reduced mass $\mu$ of the pair of heavy hadrons.
Replacing the diagonal entry of $\mathbf{M}$ by $2 \mu$ can give a dramatic  improvement in the accuracy of the Hamiltonian at large $r$.
Since the wavefunction at small $r$  is suppressed by the repulsive color-Coulomb potential, the decrease in the accuracy of the Schr\"odinger equation at small $r$ is a small price to pay for the dramatic  improvement in accuracy at large $r$.
In the case of kinetic improvement with constant mass matrix $\mathbf{M}$, the diabatic Hamiltonian in Eq.~\eqref{H-diabatic} becomes
\begin{equation}
\mathbf{H}_1^\mathrm{di} = - \mathbf{M}^{-1}\, \bm{\nabla}^2
+ \mathbf{V}^\mathrm{di}_0(\bm{r}) + \mathbf{V}_1^\mathrm{di}(\bm{r},\bm{S}_1,\bm{S}_2 ) .
\label{Hdiabatic-kiM}
\end{equation}

Kinetic improvement with a constant mass matrix $\mathbf{M}$ is simple in the diabatic representation.
Kinetic improvement is much more complicated in the adiabatic representation.
The $\bm{r}$-dependent unitary transformation $\mathbf{A}(\bm{r})^\dagger$ that transforms the diabatic Hamiltonian into the adiabatic Hamiltonian does not commute with $\mathbf{M}$.
To ensure that the adiabatic Hamiltonian is hermitian, the kinetic term in the diabatic Hamiltonian in Eq.~\eqref{Hdiabatic-kiM} has to be expressed as $-\bm{\nabla} \cdot (\mathbf{M}^{-1} \bm{\nabla})$ before carrying out the unitary transformation.
After the transformation, the mass matrix $\mathbf{A}(\bm{r})^\dagger \mathbf{M}\, \mathbf{A}(\bm{r})$ in the adiabatic Hamiltonian depends on the vector $\bm{r}$.


\section{Adiabatic Hamiltonian in the HQSS Limit}
\label{sec:Adiabatic}

In this section, we write down the radial adiabatic Hamiltonians in the HQSS limit for the truncations of the adiabatic potentials to the isospin-0 $1\Pi_g$,  $2\Sigma_g^+$, and $1\Sigma_u^-$ potentials.

\subsection{\texorpdfstring{$1\Pi_g$}{1 Pi g} and \texorpdfstring{$2\Sigma_g^+$}{2 Sigma g plus} Potentials}
\label{sec:Sigmag+Pig}

The adiabatic Hamiltonian in the HQSS limit includes only the kinetic term and the static potential in Eq.~\eqref{H-adiabatic}.
We first consider the truncation of the adiabatic potentials to the $1 \Pi_g$ and $2 \Sigma_g^+$ potentials associated with the $j^{\pi \gamma}=1^{--}$ adjoint meson.
The adiabatic Schr\"odinger equation for the $1 \Pi_g$ and $2 \Sigma_g^+$ potentials is given in Appendix~I of Ref.~\cite{Ber24}.
The only quantum number that is exactly conserved is $CP$.
However, the B\nobreakdash-O angular momentum $L$ and the parity $P$ are approximately conserved.
The radial adiabatic Hamiltonian can be arranged into blocks with rows and columns labeled by $L^P$.
The quantum numbers $L^P$ will be exactly conserved if the off-diagonal blocks are ignored.

For any block label $L^P$, the parity can satisfy either $P = (-1)^L$ or $P = (-1)^{L+1}$.
The $L^P$ block of the radial adiabatic Hamiltonian has a  different form in the two cases.
It is therefore convenient to introduce an alternative to the block label $L^P$: $(L,\pm)$, where the second label is the sign $\pm$ in $P = \pm (-1)^L$.
The $L^P = 0^-$ block of $\mathbf{H}_0$ is a $1 \times 1$ matrix:
\begin{equation}
\mathbf{H}_0^{(0,-)} =  
  -\frac{1}{m_Q\, r}\bigg(\frac{\mathrm{d}\hphantom{r}}{\mathrm{d}r}\bigg)^2r  
+ \frac{2}{m_Q\, r^2}  + V_{2\Sigma_g^+}(r),
\label{H0-}
\end{equation}
where $V_{2\Sigma_g^+}(r)$ is the $2 \Sigma_g^+$ potential.
The centrifugal term indicates that the orbital angular momentum is $L_{Q \bar Q} = 1$.
If $L\ge 1$, the $L^P$ block of $\mathbf{H}_0$ with parity $P=(-1)^L$ is a $1\times 1$ matrix:
\begin{equation}
\mathbf{H}_0^{(L,+)}  =  
  -\frac{1}{m_Q\, r}\bigg(\frac{\mathrm{d}\hphantom{r}}{\mathrm{d}r}\bigg)^2r  
+ \frac{L(L+1)}{m_Q\, r^2} +  V_{1\Pi_g}(r),
\label{HL+}
\end{equation}
where $ V_{1\Pi_g}(r)$ is the $1 \Pi_g$ potential.
The centrifugal term indicates that $L_{Q \bar Q} = L$.
If $L\ge 1$, the $L^P$ block of $\mathbf{H}_0$ with $P= (-1)^{L+1}$ is a $2 \times 2$ matrix:
\begin{equation}
\mathbf{H}_0^{(L,-)} = 
  -\frac{1}{m_Q\, r}\bigg(\frac{\mathrm{d}\hphantom{r}}{\mathrm{d}r}\bigg)^2r  
+ \frac{1}{m_Q\, r^2} \begin{pmatrix}
L(L+1)+2         & -2 \sqrt{L(L+1)}	\\
-2 \sqrt{L(L+1)} & L(L+1)	        \\
\end{pmatrix}
+ \begin{pmatrix}
V_{2\Sigma_g^+}(r)		& 0	\\
0	& V_{1\Pi_g}(r)	\\
\end{pmatrix}.
\label{HL-}
\end{equation}
The static potential is a diagonal matrix whose diagonal entries are the $2 \Sigma_g^+$ and $1 \Pi_g$ potentials.
Similar $2\times2$ adiabatic Hamiltonian matrices were first obtained in Ref.~\cite{Ber15} for the case of heavy-quarkonium hybrids.
In the  centrifugal term in Eq.~\eqref{HL-}, the off-diagonal entries of the matrix come from nonadiabatic couplings.
The eigenvalues of that matrix are $(L-1)L$ and $(L+1)(L+2)$, which correspond to $L_{Q \bar Q} = L-1$ and $L+1$.

\begin{table}[t]
\centering
\begin{tabular}{|c|c|c|c|c|}
\hline
~$j^{\pi\gamma}$~ &   ~$L^P$~ &  $J^{PC}$ &  ~B-O potentials~ &  ~$L_{Q \bar Q}$~ \\
\hline
\multirow{4}{*}{$1^{--}$} & 
 $1^+$ & ~$1^{+-}$, $(0,1,2)^{++}$~ & $\Pi_g$, $\Sigma_g^+$ & 0, 2\\
\cline{2-4}
& $0^-$ & $0^{-+}$,  $1^{--}$ & $\Sigma_g^+$ & 1\\
\cline{2-4}
& $1^-$ & $1^{-+}$,  $(0,1,2)^{--}$ & $\Pi_g$ & 1  \\
\cline{2-4}
& $2^-$ & $2^{-+}$, $(1,2, 3)^{--}$ & $\Pi_g$, $\Sigma_g^+$ & 1, 3 \\
\cline{2-4}
& $2^+$ & $2^{+-}$, $(1,2, 3)^{++}$ & $\Pi_g$ & 2\\
\cline{2-4}
& $3^+$ & $3^{+-}$, $(2, 3,4)^{++}$ & $\Pi_g$, $\Sigma_g^+$  & 2, 4\\
\hline
\multirow{4}{*}{$0^{-+}$} 
& $0^+$ & $0^{++}$, $1^{+-}$ & \multirow{4}{*}{$\Sigma_u^-$}  & 0\\
\cline{2-3}
& $1^-$  & $1^{--}$, $(0, 1, 2)^{-+}$ & & 1\\
\cline{2-3}
& $2^+$  & $2^{++}$, $(1, 2, 3)^{+-}$ & & 2\\
\cline{2-3}
& $3^-$  & $3^{--}$, $(2, 3,4)^{-+}$ & & 3\\
\hline
\end{tabular}
\caption{Heavy-quark spin-symmetry multiplets in adiabatic B-O potentials associated with 
adjoint mesons with $j^{\pi\gamma} =1^{--}$ and $0^{-+}$.
The HQSS multiplets are labeled by $L^P$, where $L$ is the B-O angular momentum and $P$ is the parity.
The first $J^{PC}$ is for a spin-singlet state ($S_{Q \bar Q} = 0$) and the others are for spin-triplet states ($S_{Q \bar Q} = 1$).
The column labeled $\Lambda_\eta^\epsilon$ lists the B-O quantum numbers of the potentials that appear in the Schr\"odinger equation.
The column labeled $L_{Q \bar Q}$ lists the orbital angular momenta.
The $L^P$ multiplets for $0^{-+}$ are listed in order of increasing $L_{Q\bar{Q}}=L$.
The $L^P$ multiplets for $1^{--}$ are listed in order of increasing minimum $L_{Q\bar{Q}}$, then increasing $L$.
}
\label{tab:QQbarqqbar}
\end{table}

The eigenvalues of the Hamiltonians in Eqs.~\eqref{H0-}, \eqref{HL+}, and \eqref{HL-} labeled by $(L,\pm)$ or by $L^P$ are the energies of HQSS multiplets of hidden-heavy tetraquark mesons.
The $J^{PC}$ quantum numbers in the multiplets can be deduced from the triangle relation $\{ J, L, S_{Q\bar Q}  \}=1$ and the eigenvalue $CP = (-1)^{S_{Q\bar Q}+1}$, which is the product of $\eta = +1$ for $2 \Sigma_g^+$ or $1 \Pi_g$ and the $Q \bar Q$ spin factor.
We refer to $S_{Q \bar Q} = 0$ as {\it spin singlet} and $S_{Q \bar Q} = 1$ as {\it spin triplet}. 
The $L^P = 0^-$ multiplet consists of one spin-singlet state $J^{PC} = 0^{-+}$ and one spin-triplet state $1^{--}$ (so in this case spin-triplet is a misnomer).
The $L^+$ multiplet with $L\ge1$ consists of one spin-singlet state $L^{+-}$ and three spin-triplet states $(L-1,L,L+1)^{++}$.
The $L^-$ multiplet with $L\ge1$ consists of one spin-singlet state $L^{-+}$ and three spin-triplet states $(L-1,L,L+1)^{--}$.
The $J^{PC}$ quantum numbers for several $L^P$ multiplets are listed in Table~\ref{tab:QQbarqqbar}. 
The last three $L^P$ multiplets are listed in a possible order of increasing energies.
Their actual order depends on the $1\Pi_g$ and $2\Sigma_g^+$ potentials.

The quantum numbers $1^{++}$ of $X_c$ appear in the multiplets for the ground state with $L^P=1^+$ and the excited state with $L^P=2^+$  in the $1 \Pi_g$ and $2 \Sigma_g^+$ potentials.
We identify $X_c$ with the $1^{++}$ spin-triplet state in the ground-state multiplet.
The other states in the ground-state multiplet are a spin-singlet state and spin-triplet $0^{++}$ and $2^{++}$ states.
The $L^P=1^+$ block of $\mathbf{H}_0$ is obtained by setting $L=1$ in Eq.~\eqref{HL-}.
The $L^P=2^+$ block of $\mathbf{H}_0$ is obtained by setting $L=2$ in Eq.~\eqref{HL+}.

\subsection{\texorpdfstring{$1\Sigma_u^-$}{1 Sigma u minus} Potential}
\label{sec:Sigmau-}

We next consider the truncation of the adiabatic potentials to the $1 \Sigma_u^-$ potential associated with the $0^{-+}$ adjoint meson.
In this case, the B\nobreakdash-O angular momentum $L$ is equal to the $Q\bar{Q}$ orbital angular momentum $L_{Q \bar Q}$ and the parity is $P = (-1)^{L_{Q \bar Q}}$. 
The $L$ block of the radial adiabatic Hamiltonian is a $1\times 1$ matrix:
\begin{equation}
\mathbf{H}_0^L =  
  -\frac{1}{m_Q\, r}\bigg(\frac{\mathrm{d}\hphantom{r}}{\mathrm{d}r}\bigg)^2r  
+ \frac{L(L+1)}{m_Q\, r^2} +  V_{1\Sigma_u^-}(r),
\label{HL}
\end{equation}
where $V_{1\Sigma_u^-}(r)$ is the $1 \Sigma_u^-$ potential.
The centrifugal term indicates that the orbital angular momentum is $L_{Q \bar Q} = L$.

The eigenvalues of the Hamiltonians in Eq.~\eqref{HL} labeled by $L$ are the energies of HQSS multiplets of hidden-heavy tetraquark mesons.
The $J^{PC}$ quantum numbers in the HQSS multiplets can be deduced from the triangle relation $\{ J, L, S_{Q\bar Q}  \}=1$, the parity $P=(-1)^L$, and the  eigenvalue $CP = (-1)^{S_{Q\bar Q}}$, which is the product of $\eta = -1$ for $1 \Sigma_u^-$ and the $Q \bar Q$ spin factor.
The $L^P=0^+$ multiplet consists of one spin-singlet state $0^{++}$ and one spin-triplet state $1^{+-}$.
The $L^-$  multiplet with $L$ odd consists of one spin-singlet state $L^{--}$ and three spin-triplet states  $(L-1,L,L+1)^{-+}$.
The $L^+$ multiplet with $L$ even and $L\ge 2$ consists of one spin-singlet state $L^{++}$ and three spin-triplet states $(L-1,L,L+1)^{+-}$.

The $J^{PC}$ quantum numbers in the lowest-energy $L^P$ multiplets for the $1 \Sigma_u^-$ potential are listed in Table~\ref{tab:QQbarqqbar}.
The $L^P$ multiplets are listed in order of increasing energy.
There is no state with the quantum numbers $1^{++}$ of $X_c$, but there are states with the same $J^{PC}$ as other states in the $X_c$ multiplet.
The $0^+$ multiplet consists of a spin-singlet $0^{++}$ state and a spin-triplet $1^{+-}$ state.  
The $2^+$ multiplet includes a spin-singlet $2^{++}$ state and a spin-triplet $1^{+-}$ state.


\section{Diabatic Hamiltonian in the HQSS Limit}
\label{sec:Diabatic}

In this section, we write down the radial diabatic Hamiltonians in the HQSS limit for the truncation of the diabatic potentials to the isospin-0 $1^{--}\Sigma_g^+/\Pi_g$ and $0^{-+}\Sigma_u^-$ potentials.
We determine the blocks of the radial diabatic Hamiltonian with definite  B\nobreakdash-O angular momentum $L$ and parity $P$ and we present the blocks for $L^P= 1^+$ and $2^+$ explicitly.

\subsection{Angular decomposition}
\label{sec:AngularDecomp}

The diabatic Hamiltonian in the HQSS limit includes only the kinetic term and the static potential in Eq.~\eqref{H-diabatic}:
\begin{equation}
\mathbf{H}_0 = - \frac{1}{m_Q} \, \bm{\nabla}^2  + \mathbf{V}_0(\bm{r}) .
\label{H0-diabatic}
\end{equation}
We have suppressed the superscript (di) on the static potential $\mathbf{V}_0(\bm{r})$.  
It depends on the vector $\bm{r}$ but not on the $Q$ and $\bar Q$ spins $\bm{S}_1$ and $\bm{S}_2$, so the Hamiltonian respects HQSS.

If a Hamiltonian and a conserved angular momentum are simultaneously diagonalized, the projections of that  angular momentum are related by rotational symmetry and therefore do not need to be described explicitly in the Schr\"odinger equation.
We refer to the states that are described explicitly in the Schr\"odinger equation as a {\it Schr\"odinger channel}.
The states acted on by the Hamiltonian $\mathbf{H}_0$ in Eq.~\eqref{H0-diabatic} can be expressed as  direct products of light-QCD states and $Q \bar Q$ spin states.
The light-QCD states can be chosen to have definite B\nobreakdash-O  quantum numbers $\Lambda_\eta^\epsilon$. 
In the diabatic representation, the truncation of the diabatic potentials can be chosen so that the light-QCD states have definite light-QCD quantum numbers $j^{\pi \gamma}$ at all $\bm{r}$ \cite{Brus24,braa25x}. 
We denote these diabatic light-QCD states by $| j^{\pi \gamma}\Lambda_\eta^\epsilon;\,\bm{r} \rangle$.
In the HQSS limit, the $Q \bar Q$ spin states do not need to be described explicitly. 
The Schr\"odinger channels can therefore be specified by the diabatic quantum numbers $j^{\pi \gamma}\Lambda_\eta^\epsilon$. 

In the light-QCD states $| j^{\pi \gamma}\Lambda_\eta^\epsilon;\,\bm{r} \rangle$, the labels $\Lambda$ and $\epsilon$ specify the eigenvalues of $|\bm{j} \cdot \hat{\bm{r}}|$ and $R_\mathrm{light}$.
If $\Lambda > 0$, an alternative to the labels $\Lambda$ and $\epsilon$ is the eigenvalue $\lambda$ of $\bm{j} \cdot \hat{\bm{r}}$, which can be $+\Lambda$ or $-\Lambda$.
Another alternative is the eigenvalue $m$ of $\bm{j} \!\cdot\! \hat{\bm{z}}$.
The  light-QCD states $| j^{\pi \gamma}, m  ; \bm{r}\rangle$ depend on the direction $\hat{\bm{r}}$ only through the label $\bm{r}$.
The  light-QCD states $| j^{\pi \gamma}, \lambda  ; \bm{r}\rangle$ depend on $\hat{\bm{r}}$ also through the eigenvalue $\lambda$ of $\bm{j} \cdot \hat{\bm{r}}$.
The $2j+1$ states $| j^{\pi \gamma}, \lambda;\, \bm{r} \big\rangle$ are related to the states $| j^{\pi \gamma}, m;\, \bm{r} \rangle$ by a rotation that can be expressed as
\begin{equation}
\bigl\lvert j^{\pi \gamma}, \lambda  ; \bm{r} \big\rangle 
= \sum_m D^j_{m,\lambda}(\hat{\bm{r}}) \, \bigl\lvert j^{\pi \gamma}, m  ; \bm{r} \big\rangle ,
\label{m-lambda}
\end{equation}
where $D^j_{m,\lambda}(\hat{\bm{r}}) =  D^j_{m,\lambda}(\phi,\theta,\psi=0)$ is a Wigner $D$-function. 
Its arguments are Euler angles $\theta$ and $\phi$ that specify the direction $\hat{\bm{r}}$ of the vector $\bm{r}$.
If the third Euler angle $\psi$ was not set to 0, it would introduce an additional phase factor $e^{-i\lambda\psi}$.

An advantage of the  light-QCD states $| j^{\pi \gamma}, \lambda  ; \bm{r}\rangle$ is that they simplify the  constraints from the cylindrical and $(CP)_\mathrm{light}$ symmetries.
The cylindrical symmetry implies that the matrix elements of $\mathbf{V}_0(\bm{r})$ are diagonal in $\lambda$ and that they depend only on $r= |\bm{r}|$. 
The $(CP)_\mathrm{light}$ symmetry implies that the matrix elements are diagonal in $\gamma\pi$.
The matrix elements therefore reduce to
\begin{equation}
\left\langle j^{\pi \gamma}, \lambda ; \bm{r} \Big| \mathbf{V}_0(\bm{r}) \Big| (j^{\pi \gamma})^\prime,  \lambda^\prime ; \bm{r}\right\rangle 
= \delta(\lambda,\lambda^\prime)\, \delta\big( \gamma\pi, \gamma^\prime\pi^\prime \big)\, 
\big( \mathbf{V}_0^\lambda(r) \big)_{j^{\pi \gamma},(j^{\pi \gamma})^\prime} ,
\label{V0-lambda}
\end{equation}
where $\delta (i,j) = \delta_{ij}$ is the Kronecker delta symbol. 
We refer to the diagonal entries of the matrix $ \mathbf{V}_0^\lambda(r)$ as {\it diabatic potentials} and we denote them more concisely by $\big( \mathbf{V}_0^\lambda(r) \big)_{j^{\pi \gamma},j^{\pi \gamma}}  = V_{j^{\pi \gamma}\Lambda^\epsilon_\eta}(r)$, where $\Lambda =|\lambda|$ and $\eta = \gamma\pi$.
We refer to the off-diagonal entries of $ \mathbf{V}_0^\lambda(r)$ as {\it transition potentials} and we denote them by $\big( \mathbf{V}_0^\lambda(r) \big)_{j^{\pi \gamma},(j^{\pi \gamma})^\prime}  = G_{(j^{\pi \gamma},(j^{\pi \gamma})^\prime)\Lambda^\epsilon_\eta}(r)$, where $\Lambda =|\lambda|$ and $\eta = \gamma\pi$.

The radial variable $r$ and the angular variables $\theta$ and $\phi$ can be separated in matrix elements of $\mathbf{V}_0(\bm{r})$ by first using the rotational symmetry in Eq.~\eqref{m-lambda} to express the states $| j^{\pi \gamma}, m  ; \bm{r}\rangle$ in terms of  the states $| j^{\pi \gamma}, \lambda  ; \bm{r}\rangle$ and then using the constraints from the cylindrical and $(CP)_\mathrm{light}$ symmetries in Eq.~\eqref{V0-lambda}:
\begin{equation}
\left\langle j^{\pi \gamma} ,m ; \bm{r} \Big\rvert \mathbf{V}_0(\bm{r}) \Bigl\lvert (j^{\pi \gamma})^\prime, m^\prime  ; \bm{r} \right\rangle 
=\sum_\lambda D^j_{m,\lambda}(\hat{\bm{r}})\, D^{j^\prime}_{m^\prime,\lambda}(\hat{\bm{r}})^\ast \,
\big( \mathbf{V}_0^\lambda(r) \big)_{j^{\pi \gamma},(j^{\pi \gamma})^\prime} ,
\label{V0-m}
\end{equation}
where $m$ and $m^\prime$ are eigenvalues of $\bm{j}\!\cdot\! \hat{\bm{z}}$ and the sum is over  eigenvalues $\lambda$ of $\bm{j} \!\cdot\! \hat{\bm{r}}$ with $|\lambda| \le \mathrm{min}(j,j^\prime)$.
If the third Euler angle $\psi$ in the Wigner $D$-functions was not set to 0, it would cancel between the two Wigner $D$-functions.

The angular decomposition of the  diabatic potential in Eq.~\eqref{V0-m} can be exploited to reduce the diabatic Schr\"odinger equation to a coupled set of radial differential equations \cite{Brus23a,braa25x}. 
The radial diabatic Hamiltonian in the HQSS limit has the form
\begin{equation}
\mathbf{H}_0 =  
\frac{1}{m_Q} \left[ - \frac{1}{r} \bigg(\frac{\mathrm{d}\hphantom{r}}{\mathrm{d}r}\bigg)^2  r + \frac{1}{r^2} \mathbf{L}_{Q \bar Q}^2 \right]
+ \mathbf{V}_0(r) ,
\label{H0-r}
\end{equation}
where $\mathbf{L}_{Q \bar Q}^2$ is the diagonal orbital-angular-momentum matrix whose diagonal entry for a channel with orbital angular momentum $L_{Q \bar Q}$ is $L_{Q \bar Q}(L_{Q \bar Q}+1)$.
We refer to $\mathbf{V}_0(r)$ as the {\it radial diabatic static potential}.
Its diagonal entries are diabatic potentials and its off-diagonal entries are transition potentials.

In Ref.~\cite{Brus23a}, Fierz transformations were used to determine the angular-momentum coefficients entering the radial diabatic static potential in the case of quarkonium coupling to pairs of $S$-wave heavy mesons.
The path to the results of Ref.~\cite{Brus23a} can be simplified by avoiding Fierz transformations and just using angular-momentum algebra.
The angular-momentum coefficients can be expressed in general as sums of products of Clebsch-Gordan coefficients and Wigner $6j$ and $9j$ symbols, as shown in Refs.~\cite{Braa24a,Brus24}. 
This also allows the results of Ref.~\cite{Brus23a} to be generalized to any set of diabatic potentials and to any pair of heavy hadrons.

\subsection{Projection onto B-O Angular Momentum}
\label{sec:LProject}

The rows and columns of the radial diabatic Hamiltonian $\mathbf{H}_0$ in Eq.~\eqref{H0-r} can be labeled by the light-QCD quantum numbers $(j^{\pi \gamma},m)$ with $|m| \le j$ and the orbital-angular-momentum quantum numbers $(L_{Q \bar Q},m_L)$ with $|m_L| \le L_{Q \bar Q}$.
The B\nobreakdash-O angular momentum $\bm{L} = \bm{j} + \bm{L}_{Q \bar Q}$ was introduced in Eq.~\eqref{J=L+S}.
We denote its quantum numbers by $(L,M_L)$.
The rows and columns of $\mathbf{H}_0$ can alternatively be labeled by $\big( (j^{\pi \gamma}, L_{Q \bar Q}) L,M_L \big)$, with $|M_L| \le L$ and with the angular momenta satisfying the triangle relation $\{ j, L_{Q \bar Q}, L \} = 1$.

As shown in Refs.~\cite{Brus23a,braa25x}, the B\nobreakdash-O angular momentum $\bm{L}$ is conserved if the truncated diabatic potentials include all the  B\nobreakdash-O quantum numbers $\Lambda_\eta^\epsilon$ associated with each  $j^{\pi \gamma}$ or equivalently all the light-QCD states labeled by $(j^{\pi \gamma},\lambda)$ with $|\lambda| \le j$.
The Hamiltonian $\mathbf{H}_0$  is then diagonal in $(L,M_L)$ and its matrix elements do not depend on $M_L$.
The radial Schr\"odinger channels can therefore be labeled by $(j^{\pi \gamma},L_{Q \bar Q})L$.
If $L$ is specified, the radial Schr\"odinger channels can be labeled simply by $(j^{\pi \gamma},L_{Q \bar Q})$.
The radial diabatic static potential $\mathbf{V}_0(r)$ can  be decomposed into commuting blocks $\mathbf{V}_0^L(r)$  labeled by the conserved quantum number $L$ and with rows and columns labeled by $(j^{\pi \gamma}, L_{Q \bar Q})$.
The methods in Ref.~\cite{Brus23a} can be used to show that the entries of the $L$ block of $\mathbf{V}_0(r)$ are 
\begin{equation}
\left( \mathbf{V}_0^L(r) \right)_{(j^{\pi \gamma},L_{Q\bar Q}),((j^{\pi \gamma})^\prime, L_{Q\bar Q}^\prime)} 
= (-1)^{j - j^\prime}   \sum_\lambda
\cgr{j}{\lambda}{L}{-\lambda}{L_{Q\bar Q}}{0} \cgr{j^\prime}{\lambda}{L}{-\lambda}{L_{Q\bar Q}^\prime}{0} 
\big( \mathbf{V}_0^\lambda(r) \big)_{j^{\pi \gamma},(j^{\pi \gamma})^\prime} ,
\label{V0-L}
\end{equation}
where $\mathbf{V}_0^\lambda(r)$ are the matrices of diabatic potentials and transition potentials defined by Eq.~\eqref{V0-lambda}.
The Clebsch-Gordan coefficients imply the triangle relations $\{ j, L_{Q\bar Q}, L \} = 1$ and $\{ j^\prime, L_{Q\bar Q}^\prime, L \} = 1$.
The static potential matrices $\mathbf{V}_0^L(r)$ need not be diagonal in $j$ or $\pi$ or $L_{Q\bar Q}$, because  these quantum numbers are not in general conserved.
They are however diagonal in $\gamma\pi$, because $(CP)_\mathrm{light}$ is conserved in the HQSS limit.

\subsection{Projection onto Parity}
\label{sec:PProject}

As shown in Refs.~\cite{Brus23a,braa25x}, the total parity $P$  is conserved if the truncated adiabatic potentials $(j^{\pi \gamma},\lambda)$ include both  values $\lambda=\pm \Lambda$ for every $\Lambda \le j$.
The radial Hamiltonian $\mathbf{H}_0$  can then be arranged into two commuting blocks with $P=+$ and $-$.
After the angular decomposition, the total parity is determined by $P_\mathrm{light} = \pi$ and $L_{Q \bar Q}$:
\begin{equation}
P = \pi \, (-1)^{L_{Q \bar Q}+1}.
\label{P-total}
\end{equation}
This includes a factor of $-1$ from the relative parities of $Q$ and $\bar Q$.
The B\nobreakdash-O angular momentum $\bm{L}$ is conserved in the HQSS limit and it commutes with the total parity $P$. 
Thus the $L$ block of $\mathbf{H}_0$ can be arranged into two commuting blocks labeled by $L^P$, where $P$ is + and $-$.
The $L^P$ block of $\mathbf{V}_0(r)$ can be obtained  from the $L$ block in Eq.~\eqref{V0-L} simply by restricting the channels $(j^{\pi \gamma},L_{Q \bar Q})$ to those for which $L_{Q \bar Q}$ is either even or odd in accord with the parity constraint in Eq.~\eqref{P-total}.
The possibility to choose the truncation of the diabatic potentials so that $L^P$ is exactly conserved in the HQSS limit provides an advantage over the adiabatic representation.

\subsection{Truncation of Diabatic Potentials}
\label{sec:Truncation}

We proceed to truncate the diabatic potentials to the $1^{--}\Sigma_g^+/\Pi_g$ and $0^{-+}\Sigma_u^-$ potentials.
The diabatic potentials in Eq.~\eqref{V0-L} with $\lambda=0$ are $\big( \mathbf{V}_0^0(r) \big)_{1^{--},1^{--}}= V_{1^{--}\Sigma_g^+}(r)$ and $\big( \mathbf{V}_0^0(r) \big)_{0^{-+},0^{-+}} = V_{0^{-+}\Sigma_u^-}(r)$. 
The diabatic potentials with $\lambda=\pm1$ are $\big( \mathbf{V}_0^{\pm1}(r) \big)_{1^{--},1^{--}} = V_{1^{--}\Pi_g}(r)$. 
There are no transition potentials.
The diabatic potentials $ V_{1^{--}\Pi_g}(r)$ and $V_{0^{-+}\Sigma_u^-}(r)$ approach the adiabatic potentials $V_{1 \Pi_g}(r)$ and $V_{1\Sigma_u^-}(r)$ both at large $r$ and at small $r$.
The diabatic potential $V_{1^{--}\Sigma_g^+}(r)$ approaches $V_{2 \Sigma_g^+}(r)$ at small $r$ and it approaches $V_{1 \Sigma_g^+}(r)$ at large $r$ beyond the narrow avoided crossing between the $1\Sigma_g^+$ and $2 \Sigma_g^+$ adiabatic potentials.
The diabatic potentials interpolate smoothly between the adiabatic potentials at intermediate $r$.

The general expression for the $L$ block of $\mathbf{V}_0(r)$ is given in Eq.~\eqref{V0-L}. 
Its rows and columns are labeled by $(0^{-+},L_{Q \bar Q})$ with orbital angular momentum $L_{Q \bar Q}=L$ and by $(1^{--},L_{Q \bar Q})$ with $L_{Q \bar Q}$ constrained to at most 3 values by $\{ 1, L_{Q \bar Q},L \} =1$.
The entries of $\mathbf{V}_0^L(r)$ are diagonal in $j^{\pi \gamma}$, but they need not be diagonal in $L_{Q\bar Q}$.
The entries diagonal in $j^{\pi \gamma}$ are
\begin{subequations}
\begin{align}
\left( \mathbf{V}_0^{L} (r) \right)_{(1^{--},L_{Q \bar Q}), (1^{--},L_{Q \bar Q}^\prime)} =&
 \{ L, 1, L_{Q \bar Q} \} \,\delta(L_{Q \bar Q},L_{Q \bar Q}^\prime)   \,   V_{1^{--}\Pi_g}(r) 
\nonumber\\
& \hspace{-1cm}
+\, \cgr{1}{0}{L}{0}{L_{Q\bar Q}}{0} \cgr{1}{0}{L}{0}{L_{Q\bar Q}^\prime}{0} \,
\big( V_{1^{--}\Sigma_g^+}(r) - V_{1^{--}\Pi_g}(r) \big) ,
\label{VLQQbar1--}
\\\left( \mathbf{V}_0^{L} (r) \right)_{(0^{-+},L_{Q \bar Q}), (0^{-+},L_{Q \bar Q}^\prime)} =& 
 \delta(L, L_{Q \bar Q}) \, \delta(L,L_{Q \bar Q}^\prime)  \, V_{0^{-+}\Sigma_u^-}(r) .
\label{VLQQbar0-+}
\end{align}
\label{VLQQbar}
\end{subequations}
The Clebsch-Gordan coefficients with integer $L$ in Eq.~\eqref{VLQQbar1--} are 
\begin{subequations}
\begin{align}
\cgr{1}{0}{L}{0}{L-1}{0} &= - \sqrt{L/(2L+1)},
\label{CG:L-1}
\\
\cgr{1}{0}{L}{0}{L}{0} &= 0,
\label{CG:L}
\\
\cgr{1}{0}{L}{0}{L+1}{0} &= +\sqrt{(L+1)/(2L+1)}.
\label{CG:L+1}
\end{align}
\label{CG}
\end{subequations}

\subsection{\texorpdfstring{$L^P$}{L P} Blocks of the Radial Diabatic Static Potential}
\label{sec:V0LP:Di}

Given the truncation to diabatic potentials with $j^{\pi \gamma}=1^{--}$ or $0^{-+}$, the $L^P$ blocks of $\mathbf{V}_0(r)$ can be obtained  from the $L$ blocks in Eq.~\eqref{VLQQbar} simply by restricting the channels $(j^{\pi \gamma},L_{Q \bar Q})$ to those for which $L_{Q \bar Q}$ is either even or odd in accord with $P = (-1)^{L_{Q \bar Q}}$.
The $L^P$ block of the radial diabatic Hamiltonian has the form
\begin{equation}
\mathbf{H}_0^{L^P} =  
\frac{1}{m_Q} \left[ - \frac{1}{r} \bigg(\frac{\mathrm{d}\hphantom{r}}{\mathrm{d}r}\bigg)^2  r + \frac{1}{r^2} \mathbf{L}_{Q \bar Q}^2 \right]
+ \mathbf{V}_0^{L^P}(r) ,
\label{H0-LP}
\end{equation}
where $\mathbf{V}_0^{L^P}(r)$ is the $L^P$ block of  $\mathbf{V}_0(r)$ and $\mathbf{L}_{Q \bar Q}^2$ is the $L^P$ block of the diagonal orbital-angular-momentum matrix, whose diagonal entries are $L_{Q \bar Q}(L_{Q \bar Q}+1)$.
An alternative label for the $L^P$ block is $(L,\pm)$, where the second label is the sign $\pm$ in $P = \pm (-1)^L$.

We first consider the $1^{--}\Sigma_g^+/\Pi_g$ potentials for which the $L$ block is given in Eq.~\eqref{VLQQbar1--}.
The radial diabatic channels are labeled $(1^{--}, L_{Q \bar Q})$, where $L_{Q \bar Q}$ is constrained by the triangle relation $\{ 1, L_{Q \bar Q}, L \} = 1$.
If $L=0$, the only channel is $(1^{--}, 1)$, which has parity $P=-$.
The $0^-$ or $(0,-)$ block of $\mathbf{H}_0$ is a $1\times1$ matrix.
The $(0,-)$ block of $\mathbf{L}_{Q \bar Q}^2$ is 2.
The diabatic potential is
\begin{equation}
\mathbf{V}_0^{(0,-)} (r) = V_{1^{--}\Sigma_g^+}(r).
\label{V0-0-}
\end{equation}
The diabatic Hamiltonian $\mathbf{H}_0^{(0,-)}$ differs from the adiabatic Hamiltonian in Eq.~\eqref{H0-} only by replacing $V_{2\Sigma_g^+}(r)$ by the diabatic potential $V_{1^{--}\Sigma_g^+}(r)$.
If $L\ge 1$, there are three channels $(1^{--}, L_{Q \bar Q})$ with $L_{Q \bar Q} = L-1$, $L$, and $L+1$.
The channel $(1^{--}, L)$ has parity $P=(-1)^L$.
The $(L,+)$ block of $\mathbf{H}_0$ is a $1\times1$ matrix.
The $(L,+)$ block of $\mathbf{L}_{Q \bar Q}^2$ is $L(L+1)$.
The diabatic potential is
\begin{equation}
\mathbf{V}_0^{(L,+)} (r) = V_{1^{--}\Pi_g}(r).
\label{V0-LL}
\end{equation}
The diabatic Hamiltonian $\mathbf{H}_0^{(L,+)}$ differs from the adiabatic Hamiltonian in Eq.~\eqref{HL+}  only by replacing $V_{1\Pi_g}(r)$ by the diabatic potential $V_{1^{--}\Pi_g}(r)$.
The channels $(1^{--}, L-1)$ and $(1^{--}, L+1)$ have parity $P=(-1)^{L+1}$.
The $(L,-)$ block of $\mathbf{H}_0$ is a $2 \times 2$ matrix.
The $(L,-)$ block of the orbital-angular-momentum matrix is a diagonal matrix:
\begin{equation}
\mathbf{L}_{Q \bar Q}^2 = \mathrm{diag}\big( (L-1)L,(L+1)(L+2) \big).
\label{V0-L+1}
\end{equation}
The radial diabatic static-potential matrix is
\begin{align}
\mathbf{V}_0^{(L,-)}(r) =& 
V_{1^{--}\Pi_g}(r) \bm{1}_{2 \times 2} 
\nonumber\\
&+ \frac{1}{2L+1} \big[V_{1^{--}\Sigma_g^+}(r) - V_{1^{--}\Pi_g}(r) \big]
\begin{pmatrix}
L                    & -\sqrt{L(L+1)} \\
-\sqrt{L(L+1)} & L+1  \\
\end{pmatrix},
\label{V0-LL+1}
\end{align}
where $\bm{1}_{2 \times 2}$ is the $2 \times 2$ unit matrix.
The diabatic Hamiltonian $\mathbf{H}_0^{(L,-)}$ can be obtained from the adiabatic Hamiltonian in Eq.~\eqref{HL+} by the unitary transformation that diagonalizes the orbital-angular-momentum matrix in  Eq.~\eqref{HL+} and by replacing $V_{1\Pi_g}(r)$ and $V_{2\Sigma_g^+}(r)$ by the diabatic potentials $V_{1^{--}\Pi_g}(r)$ and $V_{1^{--}\Sigma_g^+}(r)$.
A similar $2\times2$ diabatic Hamiltonian matrix was first obtained in Ref.~\cite{Onc17} for the case of heavy-quarkonium hybrids.

We next consider the $0^{-+}\Sigma_u^-$ potential for which the $L$ block is given in Eq.~\eqref{VLQQbar0-+}.
The radial channels are labeled $(0^{-+}, L_{Q \bar Q})$. 
The only channel with B\nobreakdash-O angular momentum $L$ is $(0^{-+}, L)$, which has parity $P=(-1)^L$.
The $L$ block of $\mathbf{H}_0$ is a $1\times1$ matrix.
The $L$ block of $\mathbf{L}_{Q \bar Q}^2$ is $L(L+1)$.
The diabatic potential is
\begin{equation}
\mathbf{V}_0^L (r) = V_{0^{-+}\Sigma_u^-}(r).
\label{V0-LP}
\end{equation}
The diabatic Hamiltonian $\mathbf{H}_0^L$ differs from the adiabatic Hamiltonian  in Eq.~\eqref{HL} only by replacing $V_{1\Sigma_u^-}(r)$ by the diabatic potential $V_{0^{-+}\Sigma_u^-}(r)$.

We identify $X_c$ with the $1^{++}$ spin-triplet state in the $j^{\pi\gamma} = 1^{--}$ ground-state multiplet with $L^P = 1^+$ in the $1^{--}\Sigma_g^+/\Pi_g$ potentials.
The $1^+$ block of $\mathbf{H}_0$ is a $2 \times 2$ matrix whose radial diabatic static potential is obtained by setting $L=1$ in Eq.~\eqref{V0-LL+1}:
\begin{align}
\mathbf{H}_0^{1^+} =& 
  -\frac{1}{m_Q\, r}\bigg(\frac{\mathrm{d}\hphantom{r}}{\mathrm{d}r}\bigg)^2r  
+ \frac{1}{m_Q\, r^2} \begin{pmatrix}
~0~ &~0~ \\
0 & 6 \\
\end{pmatrix}
\nonumber\\
& 
+ \frac{1}{3} 
\begin{pmatrix}
2\, V_{1^{--}\Pi_g}(r)+ V_{1^{--}\Sigma_g^+}(r)& - \sqrt{2} \big[  V_{1^{--}\Sigma_g^+}(r) - V_{1^{--}\Pi_g}(r) \big]	\\
-\sqrt{2} \big[ V_{1^{--}\Sigma_g^+}(r) - V_{1^{--}\Pi_g}(r) \big] & V_{1^{--}\Pi_g}(r)+ 2\, V_{1^{--}\Sigma_g^+}(r)	\\
\end{pmatrix}.
\label{H1+:diabatic}
\end{align}

This Hamiltonian can be obtained from the $(L,-)=(1,-)$ block of the adiabatic Hamiltonian in Eq.~\eqref{HL-}  by a unitary transformation and by replacing $V_{1\Pi_g}(r)$ and $V_{2\Sigma_g^+}(r)$ by the diabatic potentials $V_{1^{--}\Pi_g}(r)$ and $V_{1^{--}\Sigma_g^+}(r)$.
There is also a $1^{++}$ state in the excited-state $2^+$ multiplet.
The $2^+$ block of $\mathbf{H}_0$ is the $1\times 1$ matrix whose diabatic potential is given in Eq.~\eqref{V0-LL}:
\begin{equation}
\mathbf{H}_0^{2^+} = 
  -\frac{1}{m_Q\, r}\bigg(\frac{\mathrm{d}\hphantom{r}}{\mathrm{d}r}\bigg)^2r  + \frac{6}{m_Q\, r^2} + V_{1^{--}\Pi_g}(r).
\label{H2+:diabatic}
\end{equation}
This Hamiltonian differs from the $(L,+)=(2,+)$ block of the adiabatic Hamiltonian in Eq.~\eqref{HL+} only by replacing $V_{1\Pi_g}(r)$ by the diabatic potential $V_{1^{--}\Pi_g}(r)$.


\section{Radial Tetraquark Channels}
\label{sec:RadTetraCh}

In this section, we introduce radial tetraquark channels whose labels include the B\nobreakdash-O angular momentum $L$ and the $Q \bar Q$ spin $S_{Q \bar Q}$.
We determine the $J^{PC}$ block of the diabatic Hamiltonian in the HQSS limit in the radial tetraquark basis and we present the $1^{++}$ block explicitly.

\subsection{Radial Tetraquark Channels}
\label{sec:RadTetraChannels}

If our truncation of the diabatic potentials conserves the total angular momentum $\bm{J}$, the total parity $P$, and the total charge conjugation $C$, the radial diabatic Hamiltonian $\mathbf{H}_0$ in the HQSS limit in Eq.~\eqref{H0-r} can be arranged into commuting  blocks labeled by $L$ and $J^{PC}$.
To determine  the $J^{PC}$ blocks of $\mathbf{H}_0$, we must take into account the spin states of the heavy quark and antiquark.
The heavy-quark-pair spin states  $|m_1,m_2 \rangle$ can be labeled by the projections $m_1$ and $m_2$ of $\bm{S}_1$ and $\bm{S}_2$ onto the $\hat{\bm{z}}$ axis.
The conservation of $\bm{S}_1$ and $\bm{S}_2$ implies that $\mathbf{H}_0$ is diagonal in $m_1$ and $m_2$ and its matrix elements do not depend on $m_1$ and $m_2$.
A convenient basis for the $Q \bar Q$ spin states are the states $|S_{Q \bar Q},m_S\rangle$ labeled by the angular-momentum quantum numbers for $\bm{S}_{Q \bar Q} = \bm{S}_1 + \bm{S}_2$.
When the $Q \bar Q$ spin states are taken into account explicitly, the rows and columns of the radial diabatic Hamiltonian in Eq.~\eqref{H0-r} can be labeled by light-QCD, orbital-angular-momentum, and B\nobreakdash-O angular-momentum quantum numbers $\big( (j^{\pi \gamma},L_{Q \bar Q})L,M_L\big)$ and by $Q \bar Q$ spin quantum numbers $(S_{Q \bar Q},m_S)$.

In the HQSS limit, the conditions for a truncation of the diabatic potentials to conserve the total angular momentum $\bm{J} = \bm{L}+ \bm{S}_{Q \bar Q}$ are the same as those for conserving the B\nobreakdash-O angular momentum $\bm{L}$  \cite{Brus23a,braa25x}.
We denote the angular-momentum quantum numbers for $\bm{J}$ by $(J,M_J)$.
Since $\bm{J}$ is conserved, states with different  projections $M_J$ do not need to be described explicitly in the Schr\"odinger equation because they are related by the rotational symmetry.
The radial diabatic Schr\"odinger channels with $Q \bar Q$ spin states taken into account can therefore be labeled by $\big( (j^{\pi \gamma},L_{Q \bar Q})L,S_{Q \bar Q} \big)J$, where the angular momenta are constrained by the triangle relations $\{ j, L_{Q \bar Q}, L \}=1$ and $\{ L, S_{Q \bar Q}, J \} =1$.
We refer to these radial diabatic Schr\"odinger channels as {\it radial tetraquark channels}.
If $J$ is specified, the radial tetraquark channels can be labeled simply by $\big( (j^{\pi \gamma},L_{Q \bar Q})L,S_{Q \bar Q} \big)$.
We refer to the set of  channels $\bigl( (j^{\pi\gamma},L_{Q\bar Q})L,S_{Q \bar Q} \bigr)$ as the {\it radial tetraquark basis}.

Conservation of charge conjugation $C$ follows automatically from conservation of $CP$ and parity.
The total charge conjugation $C$ is determined by $P_\mathrm{light}$, $C_\mathrm{light}$, $L_{Q \bar Q}$, and $S_{Q \bar Q}$.
It is most conveniently expressed as an equation for $CP$, which is determined by $(C P)_\mathrm{light} = \eta$ and $S_{Q \bar Q}$:
\begin{equation}
CP = \eta\, (-1)^{S_{Q \bar Q}+1}.
\label{CP-total}
\end{equation}
Since $CP$ is exactly conserved and $S_{Q \bar Q}$ is conserved in the HQSS limit, this $CP$ constraint implies that $\eta$ is conserved in the heavy-quark limit.

\subsection{\texorpdfstring{$J^{PC}$}{J P C} Block of Hamiltonian}
\label{sec:JPCBlock}

Since $J$ is exactly conserved, the radial diabatic Hamiltonian $\mathbf{H}_0$ can be arranged into commuting blocks labeled by $J$.
Their rows and columns can be labeled by the radial tetraquark channels $\big( (j^{\pi \gamma},L_{Q \bar Q})L,S_{Q \bar Q} \big)$.
The $J^P$ block of $\mathbf{H}_0$ can be obtained  from the $J$ block by restricting the channels to those for which $L_{Q \bar Q}$ is either even or odd in accord with the parity constraint in Eq.~\eqref{P-total}. 
The $J^{PC}$ block can then be obtained by further restricting the channels to those for which $S_{Q \bar Q}$ is either 0 or 1 in accord with the $CP$ constraint in Eq.~\eqref{CP-total}.
Since $L$ and $S_{Q \bar Q}$ are conserved in the HQSS limit, $\mathbf{H}_0$ is diagonal in $L$ and $S_{Q \bar Q}$ and its entries do not depend on  $S_{Q \bar Q}$.
The $J^{PC}$ block of $\mathbf{H}_0$ can therefore be arranged into commuting blocks labeled by $(L,S_{Q \bar Q})J^{PC}$, where the angular momenta satisfy the triangle relation $\{ L, S_{Q \bar Q} , J\} = 1$.
Their rows and columns can be labeled by $(j^{\pi \gamma},L_{Q\bar Q})$.
The  entries of the $(L,S_{Q \bar Q})J^{PC}$ block of the radial diabatic static potential $\mathbf{V}_0(r)$ are
\begin{align}
\biggl( \mathbf{V}_0^{(L,S_{Q \bar Q})J^{PC}}(r)&  \biggr)_{(j^{\pi \gamma},L_{Q\bar Q}),((j^{\pi \gamma})^\prime, L_{Q\bar Q}^\prime)} 
\nonumber\\
&= \{ L, S_{Q \bar Q} , J\} \, \delta(P ,\pi\, (-1)^{L_{Q \bar Q}+1})\, \delta(P ,\pi^\prime\, (-1)^{L_{Q \bar Q}^\prime+1})\,
\nonumber\\
&
\times  \delta\big(C P,\gamma \pi\, (-1)^{S_{Q \bar Q}+1}\big)\, \delta\big(\gamma \pi,\gamma^\prime \pi^\prime \big)
\left( \mathbf{V}_0^L(r) \right)_{(j^{\pi \gamma},L_{Q\bar Q}),((j^{\pi \gamma})^\prime, L_{Q\bar Q}^\prime)} .
\label{V0JPC}
\end{align}
There are explicit constraints on the quantum numbers through Kronecker delta symbols and triangle symbols.
There are also additional triangle relations $\{ j, L_{Q\bar Q}, L \} =1$ and $\{ j^\prime, L_{Q\bar Q}^\prime, L \}=1$ implicit in the Clebsch-Gordan coefficients in the entries of $\mathbf{V}_0^L(r)$ in Eq.~\eqref{V0-L}.
The static potential matrices $\mathbf{V}_0^{(L,S_{Q \bar Q})J^{PC}}(r)$ need not be diagonal in $j$ or $\pi$ or $L_{Q \bar Q}$ because these quantum numbers are not in general conserved, but they are diagonal in $\gamma \pi$.
The $J^{PC}$ block of $\mathbf{V}_0(r)$ can be obtained by combining the $(L,S_{Q \bar Q})J^{PC}$ blocks for $L$ and $S_{Q \bar Q}$ that satisfy $ \{ L, S_{Q \bar Q} , J\} = 1$.

The $J^{PC}$ block of the radial diabatic Hamiltonian $\mathbf{H}_0$ consists of commuting blocks labeled by quantum numbers $(L,S_{Q \bar Q})$ satisfying $\{ L, S_{Q \bar Q} , J\} = 1$.
Our final form for the $J^{PC}$ block  is
\begin{equation}
\mathbf{H}_0^{J^{PC}} = 
\frac{1}{m_Q} \left[- \frac{1}{r} \bigg(\frac{\mathrm{d}\hphantom{r}}{\mathrm{d}r}\bigg)^2  r
+ \frac{1}{r^2} \mathbf{L}_{Q \bar Q}^2 \right]
 + \mathbf{V}_0^{J^{PC}}(r) ,
\label{H0-JPC}
\end{equation}
where $\mathbf{V}_0^{J^{PC}}(r)$ is the $J^{PC}$ block of the radial diabatic static potential and $\mathbf{L}_{Q \bar Q}^2$ is the $J^{PC}$ block of the orbital-angular-momentum matrix, whose diagonal entries are $L_{Q \bar Q}(L_{Q \bar Q}+1)$.
The rows and columns are labeled by radial tetraquark channels $\big( (j^{\pi \gamma},L_{Q \bar Q})L,S_{Q \bar Q} \big)$, with the quantum numbers constrained by the triangle relation $ \{ L, S_{Q \bar Q} , J\} = 1$, the parity constraint in Eq.~\eqref{P-total}, and the $CP$ constraint in Eq.~\eqref{CP-total}.

With our truncation of the diabatic potentials, the light-QCD quantum numbers in the radial tetraquark channels $\big( (j^{\pi \gamma},L_{Q \bar Q}) L, S_{Q \bar Q} \big)$ are $j^{\pi \gamma} = 1^{--}$ and $0^{-+}$.
The parity constraint, $\pi =-$, requires $L_{Q \bar Q}$ to be even if $P=+$ and odd if $P=-$.
The $CP$ constraint requires $S_{Q \bar Q}$ to satisfy $CP =(-1)^{S_{Q \bar Q}}$ if $j^{\pi \gamma} = 1^{--}$ and $CP =(-1)^{S_{Q \bar Q}+1}$ if  $j^{\pi \gamma} = 0^{-+}$.

\subsection{\texorpdfstring{$J^{PC} = 1^{++}$}{JPC = 1++} Block}
\label{sec:1++Channel}

Here we present the $1^{++}$ block of the diabatic Hamiltonian in the HQSS limit in the radial tetraquark basis.
The radial tetraquark channels can be labeled by $\big( (j^{\pi \gamma},L_{Q \bar Q}) L, S_{Q \bar Q} \big)$.
The two triangle relations are $\{ L, S_{Q \bar Q}, 1 \} =1$ and $\{ j, L_{Q \bar Q}, L \} =1$.
Given that $P=+$ and $\pi = -$, the parity constraint in Eq.~\eqref{P-total} implies that $L_{Q \bar Q}$ is even.
Given that $CP=+$, the constraint from Eq.~\eqref{CP-total} is that $j^{\pi \gamma} = 0^{-+}$ if  $S_{Q \bar Q}=0$ and that $j^{\pi \gamma} = 1^{--}$ if $S_{Q \bar Q}=1$.

We enumerate the radial tetraquark channels in order of increasing $L_{Q \bar Q}$, increasing $L$, and then increasing $S_{Q \bar Q}$.
We first consider $L_{Q\bar Q}=0$.
The two triangle relations reduce to $\{ L, j, 1 \} =1$ and $j = L$.
Given that $j$ is 0 or 1, the only possibility is $j = L=1$.
We next consider $L_{Q\bar Q}=2$.
The two triangle relations reduce to $\{ L, j, 1 \} =1$ and $\{ j, 2, L \} =1$.
Given that $j$ is 0 or 1, the only possibilities are $j = L=1$ or $j = 1$ and $L=2$.
Thus there are 3 radial tetraquark channels $\big( (j^{\pi \gamma},L_{Q \bar Q}) L, S_{Q \bar Q} \big)$ for  $J^{PC} = 1^{++}$:\\
\indent \indent \indent 1: $\big( (1^{--},0) 1, 1 \big)$,\\
\indent \indent \indent 2: $\big( (1^{--},2) 1, 1 \big)$,\\
\indent \indent \indent 3: $\big( (1^{--},2) 2, 1 \big)$.\\
Channel 1 is $S$-wave and the other two channels are $D$-wave.
All three channels are spin-triplet ($S_{Q \bar Q} =1$).

The general  form of the $1^{++}$ block of the radial diabatic Hamiltonian is given by Eq.~\eqref{H0-JPC}. 
The orbital-angular-momentum matrix is a diagonal $3\times 3$ matrix:
\begin{equation}
\mathbf{L}_{Q \bar Q}^2 =  \mathrm{diag}\big( 0,6,6 \big) .
\label{L2:1++3}
\end{equation}
The $1^{++}$ block of the radial diabatic static potential in the radial tetraquark basis is
\begin{equation}
 \mathbf{V}^{1^{++}}_0(r) =
V_{1^{--}\Pi_g}(r) \bm{1}_{3 \times 3} + \frac{1}{3}\big[V_{1^{--}\Sigma_g^+}(r) - V_{1^{--}\Pi_g}(r) \big]
\begin{pmatrix}
1          & -\sqrt{2} & ~0~	          \\
-\sqrt{2} & 2          &0            \\
0	    & 0          & 0	          \\
\end{pmatrix},
\label{V0:1++}
\end{equation}
where $\bm{1}_{3 \times 3}$ is the $3 \times 3$ unit matrix.
The $1^{++}$ block does not depend on the $0^{-+}\Sigma_u^-$ potential.
The $1^{++}$ block of the radial Hamiltonian consists of two commuting blocks:  a $2 \times 2$ block with $(L,S_{Q \bar Q})= (1,1)$ and a $1 \times 1$ block with $(L,S_{Q \bar Q})= (2,1)$.
The $(1,1)$ block is the $L^P=1^+$ Hamiltonian in Eq.~\eqref{H1+:diabatic}.
The $(2,1)$ block is the $L^P=2^+$ Hamiltonian in Eq.~\eqref{H2+:diabatic}.
If $V_{1^{--}\Sigma_g^+}(r) = V_{1^{--}\Pi_g}(r)$, the three radial tetraquark channels are all decoupled.

We have identified $X_c$ as the $1^{++}$ member of a HQSS multiplet whose other members are $0^{++}$, $1^{+-}$, and $2^{++}$.
The radial tetraquark channels for $0^{++}$, $1^{+-}$, and $2^{++}$ are enumerated and the corresponding blocks of the diabatic Hamiltonian in the radial tetraquark basis are given explicitly in Sections~\ref{sec:0++}, \ref{sec:1+-}, and \ref{sec:2++} of Appendix~\ref{app:DiabaticH}.


\section{Radial Dimeson Channels}
\label{sec:RadDiChannels}

In this section, we introduce radial dimeson channels whose labels include the spins $J_1$ and $J_2$ of a pair of heavy mesons. 
We determine the $J^{PC}$ block of the diabatic Hamiltonian in the HQSS limit in the radial dimeson basis and we present the $1^{++}$ block explicitly. 

\subsection{Heavy-Hadron Spin Splittings}
\label{sec:SpinSplittings}

The energy of $X_c$ is close to the threshold for the $S$-wave charm-meson pair $D^{\ast0} \bar{D}^0$.
It is therefore necessary to take into account its couplings to pairs of  $S$-wave charm mesons.
The ground-state $S$-wave charm mesons are $D^0$ and $D^+$ with $J^P = 0^-$ and  $D^{\ast0}$ and $D^{\ast+}$ with $J^P = 1^-$.
We ignore isospin splittings, so we denote the $S$-wave charm mesons by $D$ and $D^*$.
We choose the charge-conjugation phase conventions $C D = \bar D$ and  $C D^* = \bar D^*$.
We denote the isospin-averaged $D^*$-$D$ mass splitting by $\Delta_c$ and the isospin-averaged $B^*$-$B$ mass splitting by $\Delta_b$.

A heavy hadron containing a heavy quark $Q$ of mass $m_Q$ reduces at leading order in $1/m_Q$  to a static hadron and the decoupled spin $\bm{S}_Q$ of the heavy quark. 
The static hadron has light-QCD angular-momentum and parity quantum numbers $j^\pi$.
The heavy quark has spin and parity quantum numbers $\tfrac12^+$. 
The spin-splitting term in the Hamiltonian for the heavy hadron is
\begin{equation}
H_{\text{SS},j^\pi} = \frac{C_{3,j^\pi}}{m_Q} \, \bm{j} \cdot \bm{S}_Q,
\label{HSS-meson}
\end{equation}
where $ \bm{j}$ is the angular momentum of light QCD.
The numerator $C_{3,j^\pi}$ of the coefficient has a finite limit as $m_Q \to \infty$.
A heavy hadron has parity $P = +\pi$. 
If the heavy hadron has spin $J$, the spin splitting from the Hamiltonian in Eq.~\eqref{HSS-meson} is
\begin{equation}
E_\text{SS}^{J^P} = \frac{C_{3,j^\pi}}{2\, m_Q} \bigl\{j, \tfrac12, J \bigr\} \left[ J(J+1) - j(j+1) - \tfrac34 \right].
\label{ESS-JP}
\end{equation}

An $S$-wave heavy meson reduces at leading order in $1/m_Q$ to a ground-state triplet meson with $j^\pi = \tfrac12^-$ and the decoupled spin $\bm{S}_Q$ of the heavy quark.
The $J^P$ quantum numbers of the heavy meson are $0^-$ and $1^-$. 
The spin splitting between the $S$-wave heavy mesons with $J^P= 1^-$ and $0^-$ from Eq.~\eqref{ESS-JP} is
\begin{equation}
\Delta_Q  = C_{3,\nicefrac{1}{2}^-}/m_Q.
\label{DeltaQ}
\end{equation}
The spin splitting in Eq.~\eqref{ESS-JP} is $-\tfrac34\Delta_Q$ for $J=0$ and $+\tfrac14\Delta_Q$ for $J=1$.
Their spin-weighted average is zero. 

\subsection{Heavy-Hadron-Pair Channels}
\label{sec:SpinSplittingSchrEq}

The B\nobreakdash-O potentials with finite energy as $r \to \infty$ are static-hadron-pair potentials that approach the threshold for a triplet hadron and an antitriplet hadron.
The adiabatic potentials that approach the threshold for a pair of  ground-state triplet and antitriplet mesons are $1\Pi_g$, $1\Sigma_g^+$, and $1\Sigma_u^-$.
The corresponding diabatic triplet-meson-pair potentials at large $r$ are $1^{--}\Pi_g$, $1^{--}\Sigma_g^+$, and $0^{-+}\Sigma_u^-$.

Through first order in $1/m_Q$,  the diabatic potentials are the diagonal elements of the diabatic potential matrix $\mathbf{V}_0(\bm{r}) +\mathbf{V}_1(\bm{r},\bm{S}_1,\bm{S}_2)$.
The eigenvalues of those diabatic potentials with finite energy as $r \to \infty$ approach the energies of pairs of heavy hadrons, which include spin splittings of order $1/m_Q$.
We refer to the continuation of these diabatic potentials to smaller $r$ as {\it heavy-hadron-pair potentials}.
The angular-momentum couplings of the angular momenta of the triplet and antitriplet hadrons, the spins of the $Q$ and $\bar Q$, and the spins of the heavy hadrons are $(j_1, \tfrac12)J_1$ and $(j_2, \tfrac12)J_2$.
The static-hadron-pair potentials are split into heavy-hadron-pair potentials at large $r$.
The $1\Sigma_u^-$, $1\Sigma_g^+$, and $1\Pi_g$ adiabatic potentials that approach twice the energy of a ground-state static meson are split into heavy-hadron-pair potentials that approach the three thresholds for the $S$-wave heavy-meson pairs $D \bar D$, $D^* \bar D$, and $D^* \bar D^*$, which are $-\tfrac32\Delta_Q$, $-\tfrac12\Delta_Q$, and $+\tfrac12\Delta_Q$, respectively.

The radial tetraquark channels $\big( (j^{\pi\gamma},L_{Q \bar Q}) L, S_{Q \bar Q} \big)J$ are labeled by light-QCD quantum numbers, the orbital angular momentum, the B\nobreakdash-O angular momentum, the $Q \bar Q$ spin, and the total angular momentum.
The expressions for the spin-dependent terms in the $1/m_Q$ diabatic potential at large $r$ in the radial tetraquark basis are very complicated, because the spins of the heavy quark and antiquark must enter through $S_{Q \bar Q}$ instead of through the spins $J_1$ and $J_2 $ of two heavy hadrons.
When the spin splittings of heavy hadrons are taken into account, it is more convenient to include the spins $J_1$ and $J_2$ of the two heavy hadrons among the labels for the radial diabatic channels.
One basis for the radial diabatic states consists of products of the spin and parity states $(J_1^{P_1},m_1)$ and $(J_2^{P_2},m_2)$ of the two heavy hadrons and their orbital-angular-momentum states $(L_{Q\bar Q}, m_L)$.
An alternative basis consists of products of $\big( ( J_1^{P_1},J_2^{P_2})S,M_S\big)$ and $(L_{Q\bar Q}, m_L)$, where $(S,M_S)$ are the quantum numbers for the static angular momentum $\bm{S}$ introduced in Eq.~\eqref{J=J+L}.
Note that $S$ coincides with the total spin of the two heavy hadrons.
The most convenient basis consists of the states $\big( ((J_1^{P_1},J_2^{P_2})S ,L_{Q \bar Q} )J,M_J\big)$, where $(J,M_J)$ are the total angular-momentum quantum numbers.
The labels $\big( (J_1^{P_1},J_2^{P_2})S ,L_{Q \bar Q} \big)J$ define a radial Schr\"odinger channel that we refer to as a {\it radial dihadron channel}.
If both of the heavy hadrons are mesons, we refer to it as a  {\it radial dimeson channel}.
If $J$ is specified, the radial dimeson channels can be labeled simply by $\big( (J_1^{P_1},J_2^{P_2})S ,L_{Q \bar Q} \big)$.
We refer to the set of  channels $\big( (J_1^{P_1},J_2^{P_2})S ,L_{Q \bar Q} \big)$ as the {\it radial dimeson basis}.

The radial dihadron channels $\big( (J_1^{P_1},J_2^{P_2})S, L_{Q\bar Q} \big)$ have definite total parity:  
\begin{equation}
P = P_1\, P_2\, (-1)^{L_{Q \bar Q}}.
\label{P-dimeson}
\end{equation}
If the two heavy hadrons consist of two mesons or of a baryon and an antibaryon, the radial dihadron channels with definite total charge conjugation are even or odd under interchange of the labels $J_1^{P_1}$ and $J_2^{P_2}$ with $CP= (-1)^{S-J_1-J_2}$ or $(-1)^{S-J_1-J_2+1}$, respectively \cite{Braa24a}. 

If the two heavy hadrons are $S$-wave heavy mesons whose quantum numbers $J_1^{P_1}$ and $J_2^{P_2}$ are $0^-$ or $1^-$, the radial dimeson channels are $\big( (J_1^-,J_2^-)S, L_{Q\bar Q} \big)$.
The total parity for the channel is $P = (-1)^{L_{Q \bar Q}}$.
An alternative to the spins $(J_1,J_2)$ of the heavy mesons in the label for the radial dimeson channel are the names $D^{(*)} \bar{D}^{(*)}$ of the corresponding charm mesons.
There are four radial dimeson channels with orbital angular momentum $L_{Q \bar Q}$ and $CP = +1$:
\begin{subequations}
\begin{align}
\big( (0^-,0^-)0,L_{Q \bar Q} \big) &\equiv D \bar D(0,L_{Q \bar Q}),
\\
\tfrac{1}{\sqrt{2}} \big[ \big( (1^-,0^-)1,L_{Q \bar Q} \big) + \big( (0^-,1^-)1,L_{Q \bar Q} \big) \big]
&\equiv \tfrac{1}{\sqrt{2}} \big[ D^* \bar D(1,L_{Q \bar Q}) + D \bar D^*(1,L_{Q \bar Q}) \big],
\\
\big( (1^-,1^-)0,L_{Q \bar Q} \big) &\equiv D^* \bar D^*(0,L_{Q \bar Q}) , 
\\
\big( (1^-,1^-)2,L_{Q \bar Q} \big) &\equiv D^* \bar D^*(2,L_{Q \bar Q}).
\end{align}
\label{dimesonCP+}
\end{subequations}
There are two radial dimeson channels with orbital angular momentum $L_{Q \bar Q}$ and $CP = -1$:
\begin{subequations}
\begin{align}
\tfrac{1}{\sqrt{2}} \big[ \big( (1^-,0^-)1,L_{Q \bar Q} \big) - \big( (0^-,1^-)1,L_{Q \bar Q} \big) \big]
&\equiv \tfrac{1}{\sqrt{2}}\big[ D^* \bar D(1,L_{Q \bar Q}) - D \bar D^*(1,L_{Q \bar Q}) \big],
\\
\big( \big( 1^-,1^-)1,L_{Q \bar Q} \big) &\equiv D^* \bar D^*(1,L_{Q \bar Q}) .
\end{align}
\label{dimesonCP-}
\end{subequations}

\subsection{Transformation between Radial Tetraquark and Dimeson Channels}
\label{sec:Tetra-Dimeson}

The most convenient basis for solving the radial Schr\"odinger equation with heavy-hadron spin splittings taken into account is the radial dihadron basis.
The radial dihadron channels $\big( (J_1^{P_1},J_2^{P_2})S ,L_{Q \bar Q} \big)J$ and the radial tetraquark channels  $\big( ( j^{\pi\gamma},L_{Q \bar Q}) L, S_{Q \bar Q} \big)J$ are related by a change in the basis for the angular-momentum states.
The fundamental angular momenta are the spins $s_1=\tfrac12$ and $s_2=\tfrac12$ of $Q$ and $\bar Q$, the light-QCD angular momenta $j_1$ and $j_2$ of the pair of static hadrons, and the orbital angular momentum $L_{Q \bar Q}$.
The angular-momentum couplings for the radial dihadron channels with angular momentum $J$ are $\big( (J_1,J_2) S, L_{Q \bar Q} \big)J$ with heavy-hadron angular momenta $(j_1,s_1)J_1$ and $(j_2,s_2)J_2$.
The angular-momentum couplings for the radial tetraquark channels with angular momentum $J$ are $\big( (j,L_{Q \bar Q}) L, S_{Q \bar Q} \big)J$ with light-QCD angular momentum $(j_1,j_2)j$ and $Q \bar Q$ spin $(s_1,s_2)S_{Q \bar Q}$.

The change of the angular-momentum basis from the radial dimeson channels to the radial tetraquark channels can be carried out conveniently in two steps.
We will consider general values for all the angular momenta so the change of basis can be applied to general radial dihadron channels and general radial multiquark channels.
The first step is the recoupling of $j_1$,  $j_2$, $s_1$, and $s_2$ using an identity involving a Wigner $9j$ symbol:
\begin{align}
\vket{\big((j_1,s_1)J_1,(j_2,s_2)J_2\big)}{S}{M_S}{}
=& 
(-1)^{4 j_2 s_1} 
\sqrt{ \tilde{J}_1 \tilde{J}_2 }
\sum_{j,S_{Q \bar Q}} \sqrt{\tilde{j} \tilde{S}_{Q \bar Q}}
\nonumber\\
& \times 
\begin{Bmatrix}
j_1                & s_1                      & J_1 \\
j_2                & s_2                      & J_2 \\
j & S_{Q \bar Q} & S \end{Bmatrix}
\vket{  \big((j_1,j_2)j, (s_1,s_2)S_{Q \bar Q} \big)}{S}{M_S}{} .
\label{di-tetra1}
\end{align}
We have  introduced the compact notation $\tilde j = 2j+1$ for the number of projections with angular momentum $j$.
The sign $(-1)^{4 j_2 s_1}$ in Eq.~\eqref{di-tetra1} takes into account the interchange of the order of a heavy-quark operator, which is fermionic, and a light-QCD operator,  which can be bosonic or fermionic.
Given that $s_1 = \tfrac12$, the sign is $-1$ only if $j_2$ is a half integer.

The second step in the change of the angular-momentum basis is the recoupling of $j$, $S_{Q \bar Q}$, and  $L_{Q \bar Q}$ using an identity involving a Wigner $6j$ symbol:
\begin{align}
\vket{\big( (j ,S_{Q \bar Q})S, L_{Q \bar Q} \big)}{J}{M_J}{}
=& 
(-1)^{4 S_{Q \bar Q} L_{Q \bar Q}}  \sqrt{ \tilde{S}} \sum_L (-1)^{S_{Q \bar Q} + S  + L_{Q \bar Q} + L} \sqrt{  \tilde{L}} 
\nonumber\\
& \times 
\begin{Bmatrix}
j  & S_{Q \bar Q} & S \\
J  & L_{Q \bar Q}  & L  \end{Bmatrix}
\vket{\big( (j,L_{Q \bar Q}) L, S_{Q \bar Q} \big)}{J}{M_J}{} .
\label{ketjSL-ketjLS}
\end{align}
The sign $(-1)^{4 S_{Q \bar Q} L_{Q \bar Q}}$ takes into account the ordering of operators that could be fermionic in the general case, but it is +1 if either $S_{Q \bar Q}$ or $L_{Q \bar Q}$ is an integer.
The triangle relations implicit in the $6j$ symbol imply that $S_{Q \bar Q} + S  + L_{Q \bar Q} + L$ is an integer.
Note that $S_{Q \bar Q} + S  + L_{Q \bar Q}$ is not necessarily an integer, so the sign $(-1)^{S_{Q \bar Q} + S  + L_{Q \bar Q}}$ cannot be moved outside the sum in Eq.~\eqref{ketjSL-ketjLS}.
If $S_{Q \bar Q} = 0$ or $L_{Q \bar Q} = 0$, identities for $6j$ symbols with a zero entry imply that the right side of Eq.~\eqref{ketjSL-ketjLS} reduces to a single angular-momentum state whose coefficient is +1.
If $j=0$, the right side of Eq.~\eqref{ketjSL-ketjLS} reduces to a single angular-momentum state multiplied by the sign $(-1)^{4 S_{Q \bar Q} L_{Q \bar Q}} (-1)^{J - L_{Q \bar Q} - S_{Q \bar Q}}$ that comes from the reordering of the $S_{Q \bar Q}$ and $L_{Q \bar Q}$ angular-momentum states.

The heavy quark and antiquark spins are   $s_1 = s_2 = \tfrac12$.
The change of basis from the angular momenta in the radial dihadron channels to those in radial multiquark channels is
\begin{align}
\vket{\big( ( (j_1, \tfrac12) J_1,&  (j_2, \tfrac12) J_2)S, L_{Q \bar Q}\big)}{J}{M_J}{}
= (-1)^{2j_2} \sqrt{\tilde{J}_1 \tilde{J}_2 \tilde{S}}
\sum_{j,S_{Q \bar Q},L} (-1)^{S_{Q \bar Q} + S  + L_{Q \bar Q} + L} \sqrt{\tilde{j} \tilde{S}_{Q \bar Q} \tilde{L} }
\nonumber\\
&\times 
\begin{Bmatrix}
j &  S_{Q \bar Q} & S  \\
J &  L_{Q \bar Q} & L \end{Bmatrix}
\begin{Bmatrix}
j_1          & \tfrac12                     & J_1 \\
j_2          & \tfrac12                     & J_2 \\
j &  S_{Q \bar Q} & S \end{Bmatrix}
\vket{\big( ( (j_1,j_2)j, L_{Q \bar Q}) L, (\tfrac12,\tfrac12)S_{Q \bar Q} \big)}{J}{M_J}{} .
\label{di-tetra12}
\end{align}
There are 10 triangle relations implicit in the Wigner $6j$ and $9j$ symbols.
In the case of an $S$-wave dimeson channel, $j_1 = j_2 = \tfrac12$ and $J_1$ and $J_2$ are 0 or 1.

\subsection{Radial Diabatic Static Potential in the Radial Dimeson Basis}
\label{sec:VStaticDimeson}

The $L$ block of $\mathbf{V}_0 (r)$ is given in Eq.~\eqref{V0-L}.
The block of $\mathbf{V}_0 (r)$ with conserved quantum numbers $(L, S_{Q \bar Q})J^{PC}$ is given in terms of the $L$ block in Eq.~\eqref{V0JPC}.
Its rows and columns are labeled by $(j^{\pi \gamma},L_{Q \bar Q})$.
This equation also gives the $J^{PC}$ block of $\mathbf{V}_0 (r)$, whose rows and columns labeled by the radial tetraquark channels $\big( (j^{\pi \gamma},L_{Q \bar Q}) L, S_{Q \bar Q} \big)$.
The $J$ block of $\mathbf{V}_0 (r)$ in the radial tetraquark basis is obtained from the $J^{PC}$ block by combining the blocks for $P$ and $C$ that satisfy the parity constraint in Eq.~\eqref{P-total} and the $CP$ constraint in Eq.~\eqref{CP-total}.
The inverse of the formula for the recoupling of angular momenta in Eq.~\eqref{di-tetra12} can be used to transform the $J$ block of $\mathbf{V}_0 (r)$  to one whose rows and columns are labeled by the radial dimeson channels $\big( (J_1^{P_1},J_2^{P_2})S ,L_{Q \bar Q} \big)$.
The $J$ block of $\mathbf{V}_0 (r)$ in the radial dimeson angular-momentum basis is
\begin{align}
\big( \mathbf{V}_0^J(r)& \big)_{\big( (J_1,J_2)S,L_{Q\bar Q} \big), \big( (J_1,J_2)S^\prime,L_{Q\bar Q}^\prime \big)} 
= (-1)^{S - S^\prime + L_{Q \bar Q} - L_{Q \bar Q}^\prime} \sqrt{\tilde{J}_1 \tilde{J}_2 \tilde{S}}\sqrt{\tilde{J}_1^\prime \tilde{J}_2^\prime \tilde{S}^\prime}
\nonumber\\
&
\times \sum_{j^{\pi\gamma}}   \sum_{(j^{\pi\gamma})^\prime} \sqrt{\tilde{j}\tilde{j}^\prime} \sum_{S_{Q \bar Q},L}  \tilde{S}_{Q \bar Q} \tilde{L}  
\begin{Bmatrix}
j &  S_{Q \bar Q} & S  \\
J &  L_{Q \bar Q} & L \end{Bmatrix}
\begin{Bmatrix}
j^\prime &  S_{Q \bar Q} & S^\prime  \\
J &  L_{Q \bar Q}^\prime & L \end{Bmatrix}
\nonumber\\
&\times
\begin{Bmatrix}
j_1          & \tfrac12                     & J_1 \\
j_2          & \tfrac12                     & J_2 \\
j &  S_{Q \bar Q} & S \end{Bmatrix}
\begin{Bmatrix}
j_1^\prime          & \tfrac12                     & J_1^\prime \\
j_2^\prime         & \tfrac12                     & J_2^\prime \\
j^\prime &  S_{Q \bar Q} & S^\prime \end{Bmatrix}
\left( \mathbf{V}_0^L(r) \right)_{(j^{\pi \gamma},L_{Q\bar Q}),((j^{\pi \gamma})^\prime, L_{Q\bar Q}^\prime)} .
\label{Vstatic-J}
\end{align}

The $J^P$ block of $\mathbf{V}_0 (r)$ in the radial dimeson basis is obtained from the $J$ block in Eq.~\eqref{Vstatic-J} by restricting $L_{Q\bar Q}$ and $L_{Q\bar Q}^\prime$ to be either even or odd in accord with the parity constraint in Eq.~\eqref{P-dimeson}.
The $J^{PC}$ block is obtained by further restricting the channels to those of the form in Eqs.~\eqref{dimesonCP+} for $CP = +1$ or Eqs.~\eqref{dimesonCP-} for $CP = -1$.

\subsection{\texorpdfstring{$J^{PC} = 1^{++}$}{JPC = 1++} Block}
\label{sec:1++Channeldimeson}

Here we present the $1^{++}$ block of the diabatic Hamiltonian in the HQSS limit in the radial dimeson basis explicitly.
The radial dimeson channels have the form $\big( ( J_1^-,J_2^-)S,L_{Q \bar Q} \big)$, where $J_1$ and $J_2$ can each be 0 or 1.
The angular momenta satisfy the triangle relations $\{ J_1,J_2, S \}=1$ and $\{ S,L_{Q \bar Q} ,1 \}=1$. 
We first impose the condition $P = +1$.
The parity constraint in Eq.~\eqref{P-dimeson} requires $L_{Q \bar Q}$  to be even.
If $L_{Q \bar Q} = 0$, the triangle relations require $S=1$, so the only allowed channel is $\big( ( J_1^-,J_2^-)1,0 \big)$. 
If $L_{Q \bar Q} = 2$, the triangle relations require $S$ to be 1 or 2, so the allowed channels are $\big( ( J_1^-,J_2^-)1,2 \big)$ and $\big( ( J_1^-,J_2^-)2,2 \big)$. 
We next impose the condition $CP=+$, which restricts the channels to those of the form listed in Eqs.~\eqref{dimesonCP+}.
If $(J_1,J_2)=(0,0)$, then $S=0$ so there are no channels with $CP=+$.
If $(J_1,J_2)$ is (0,1) or (1,0), then $S=1$ so the channels with $CP=+$ are the even superpositions of $\big( (1^-,0^-)1, L_{Q \bar Q} \big)$ and $\big( (0^-,1^-)1, L_{Q \bar Q} \big)$ with $L_{Q \bar Q}  = 0$ or 2.
If $(J_1,J_2)=(1,1)$,  the channel with $CP=+$ is $\big( (1^-,1^-)2,2 \big)$.

We enumerate the radial dimeson channels in order of increasing $L_{Q \bar Q}$, increasing $J_1+J_2$, and then increasing $S$.
We take channels 1 and 2 to be the even superpositions of $\big( (1^-,0^-)1, L_{Q \bar Q} \big)$ and $\big( (0^-,1^-)1, L_{Q \bar Q}  \big)$ with $L_{Q \bar Q}  = 0$ and 2.
We take channel 3 to be $\big( (1^-,1^-)2, 2 \big)$. 
Our 3 radial dimeson channels labeled by charm-meson pairs and by $(S,L_{Q \bar Q})$ are\\
\indent \indent \indent 1:  $[D^* \bar D(1,0) + D \bar D^*(1,0)]/\sqrt2$,\\
\indent \indent \indent 2:  $[D^* \bar D(1,2) + D \bar D^*(1,2)]/\sqrt2$, \\
\indent \indent \indent 3:   $D^* \bar D^*(2,2)$.\\
The thresholds for channels 1 and 2 are the $D^* \bar{D}$ threshold.
The threshold for channel 3 is the $D^* \bar{D}^*$ threshold.

The general form of the $J^{PC}$  block of the radial diabatic Hamiltonian in the HQSS limit is given in Eq.~\eqref{H0-JPC}.
The $1^{++}$ block is a $3\times 3$ matrix.
The orbital-angular-momentum matrix $\mathbf{L}_{Q \bar Q}^2$ is the same as that for the radial tetraquark channels in Eq.~\eqref{L2:1++3}.
The $J$ block of $\mathbf{V}_0(r)$ in the radial dimeson basis can be obtained from the $L$ block in the radial tetraquark basis using Eq.~\eqref{Vstatic-J}.
The  $J^{PC}$ block of $\mathbf{V}_0(r)$ can then be obtained from the $J$ block by restricting the channels to those enumerated above.
The $1^{++}$ block of $\mathbf{V}_0(r)$ in the radial dimeson basis is
\begin{equation}
 \mathbf{V}^{1^{++}}_0(r) =
V_{1^{--}\Pi_g}(r) \bm{1}_{3 \times 3} + \frac{1}{6}\big[ V_{1^{--}\Sigma_g^+}(r) - V_{1^{--}\Pi_g}(r) \big]
\begin{pmatrix}
2		& \sqrt{2}	& \sqrt{6}	\\
\sqrt{2}	& 1		& \sqrt{3}	\\
\sqrt{6}	& \sqrt{3}	& 3		\\
\end{pmatrix} .
\label{V0:1++3}
\end{equation}
The $1^{++}$ block does not depend on the $0^{-+}\Sigma_u^-$ potential.
The $1^{++}$ block of $\mathbf{V}_0(r)$ in Eq.~\eqref{V0:1++3} can be obtained from the $1^{++}$ block in the radial tetraquark basis in Eq.~\eqref{V0:1++} by a unitary transformation.
That the $1^{++}$ block can be arranged into two commuting blocks with $L^P = 1^+$ and $2^+$ is not obvious in the radial dimeson basis.
The advantage of the radial dimeson basis becomes clear only when the spin splittings of heavy mesons are taken into account.
If $V_{1^{--}\Sigma_g^+}(r) = V_{1^{--}\Pi_g}(r)$, the three radial dimeson channels are all decoupled.

We have identified $X_c$ as the $1^{++}$ member of a HQSS multiplet whose other members are $0^{++}$, $1^{+-}$, and $2^{++}$.
The radial dimeson channels for $0^{++}$, $1^{+-}$, and $2^{++}$ are enumerated and the corresponding blocks of the radial diabatic Hamiltonian are presented explicitly in Sections~\ref{sec:0++}, \ref{sec:1+-}, and \ref{sec:2++} of Appendix~\ref{app:DiabaticH}.


\section{Diabatic Hamiltonian with Spin Splittings}
\label{sec:SchrEqSpinSplit}

In this section, we write down the radial diabatic Hamiltonians with spin splittings for the truncation of the diabatic potentials to the isospin-0 $1^{--}\Sigma_g^+/\Pi_g$ and $0^{-+}\Sigma_u^-$ potentials.
We take into account the spin splittings of heavy-meson pairs at large $r$ and the spin splittings of adjoint mesons at small $r$. 
We also use kinetic improvement to take into account the differences between the heavy-meson masses and the heavy-quark mass $m_Q$.

\subsection{Long-Distance Spin-Splitting Term}
\label{sec:SpinSplitting-Long}

The confinement property of QCD implies that the only states of light QCD with $\bm{3}$ and $\bm{3^\ast}$ color sources separated by a distance $r$ with finite energy in the limit  $r \to \infty$ are well-separated triplet and antitriplet hadrons.
The action of the $1/m_Q$ diabatic potential $\mathbf{V}_1(\bm{r}, \bm{S}_1, \bm{S}_2)$ on states with finite energy must reduce in the limit $r \to \infty$ to the sum of two terms corresponding to the triplet and antitriplet hadrons.
The spin-dependent terms in $\mathbf{V}_1(\bm{r}, \bm{S}_1, \bm{S}_2)$ reduce in the limit $r\to\infty$ to the sum of the spin-splitting term in Eq.~\eqref{HSS-meson} for the triplet hadron and a corresponding term for the antitriplet hadron.
We denote this limiting spin-dependent potential by $\mathbf{V}_{\text{SS},\text{long}} (\bm{S}_1,\bm{S}_2)$ and we refer to it as the {\it long-distance spin-splitting term}.
For a $Q\bar{Q}$ pair with energy near the threshold for triplet and antitriplet hadrons with quantum numbers $j_1^{\pi_1}$ and $j_2^{\pi_2}$, the long-distance  spin-splitting term can be expressed as
\begin{equation}
\mathbf{V}_{\text{SS},\text{long}} (\bm{S}_1,\bm{S}_2) 
= \frac{C_{3,j_1^{\pi_1}}}{m_Q} \,\bm{j}_1 \cdot \bm{S}_1 +  \frac{C_{3^\ast,j_2^{\pi_2}}}{m_Q} \,\bm{j}_2 \cdot \bm{S}_2,
\label{HSS-long}
\end{equation}
where $\bm{j}_1$ and $\bm{j}_2$ are the light-QCD angular momenta of the triplet and antitriplet hadrons.
The numerators $C_{3,j_1^{\pi_1}}$ and $C_{3^\ast,j_2^{\pi_2}}$ of the coefficients have finite limits as $m_Q \to \infty$.
The total light-QCD angular-momentum is $\bm{j} = \bm{j}_1 + \bm{j}_2$ and the $Q \bar Q$ spin is $\bm{S}_{Q \bar{Q}}= \bm{S}_1 + \bm{S}_2$.
The static angular momentum $\bm{S} = \bm{j} +\ \bm{S}_{Q \bar{Q}}$ commutes with $\mathbf{V}_{\text{SS},\text{long}}$. 

The long-distance spin-splitting term $\mathbf{V}_{\text{SS},\text{long}}$ in Eq.~\eqref{HSS-long} is a diagonal matrix in the radial dimeson channels $\big( (J_1^{P_1},J_2^{P_2})S, L_{Q\bar Q} \big) J$ with diagonal entries that depend on $J_1$ and $J_2$.
The $J$ block of $\mathbf{V}_{\text{SS},\text{long}}$ in the radial dimeson angular-momentum basis is diagonal and its diagonal entries are
\begin{multline}
\big( \mathbf{V}_{\text{SS},\text{long}}^J(\bm{S}_1,\bm{S}_2) \big)_{\big( (J_1,J_2)S,L_{Q\bar Q}\big),\big( (J_1,J_2)S,L_{Q\bar Q}\big)} 
= \{ J, S, L_{Q\bar Q} \}\,  \{ J_1, J_2, S \}\, 
\\ 
\times \left(\frac{C_{3,j_1^{\pi_1}}}{2\, m_Q} \left[ J_1(J_1+1) - j_1(j_1+1) - \tfrac34   \right] 
+ \frac{C_{3^\ast,j_2^{\pi_2}}}{2\, m_Q} \left[ J_2(J_2+1) - j_2(j_2+1) - \tfrac34   \right]  \right).
\label{VSS-J1J2}
\end{multline}
These diagonal entries are the spin splittings of heavy-hadron-pair thresholds relative to their spin-weighted average.

In the case of ground-state triplet and antitriplet mesons with $j_1^{\pi_1}= \tfrac12^-$ and $j_2^{\pi_2} = \tfrac12^+$, both coefficients in Eq.~\eqref{HSS-long} are determined by the spin splitting $\Delta_Q$ between the $S$-wave heavy mesons with $J^P=1^-$ and $0^-$:  $C_{3,(1/2)^-} = C_{3^\ast,(1/2)^+}  = m_Q \Delta_Q$. 
The diagonal entries of the $J$ block of $\mathbf{V}_{\text{SS},\text{long}}$ reduce to
\begin{align}
\big( \mathbf{V}_{\text{SS},\text{long}}^J(\bm{S}_1,\bm{S}_2) \big)_{\big( (J_1,J_2)S,L_{Q\bar Q}\big),\big( (J_1,J_2)S,L_{Q\bar Q}\big)} 
&=  \{ J, S, L_{Q\bar Q} \}\,  \{ J_1, J_2, S \} 
 \nonumber\\
& \hspace{0cm}
\times \tfrac12 \,  \big[ J_1(J_1+1) +  J_2(J_2+1)  - 3 \big]\, \Delta_Q.
\label{VSSlong-J}
\end{align}
The spin splittings are $-\tfrac{3}{2}\Delta_Q$ if $(J_1,J_2)=(0,0)$, $-\tfrac{1}{2}\Delta_Q$ if $(J_1,J_2)$ is (0,1) or (1,0), and $+\tfrac{1}{2}\Delta_Q$ if $(J_1,J_2)=(1,1)$.
Their spin-weighted average is 0.
With this long-distance spin-splitting term, the $1^{--}\Pi_g$, $1^{--}\Sigma_g^+$, and $0^{-+}\Sigma_u^-$ potentials no longer become degenerate as $r \to \infty$,
but they are  split into potentials that approach the three thresholds for the charm-meson pairs $D \bar D$, $D^* \bar D$, and $D^* \bar D^*$. 

\subsection{Short-Distance Spin-Splitting Term}
\label{sec:SpinSplitting-Short}

In the limit $|\bm{r}| \to 0$, the spin dependence of the $1/m_Q$ diabatic potential $\mathbf{V}_1(\bm{r}, \bm{S}_1, \bm{S}_2)$ can be simplified.
The heavy quark and antiquark can be approximated by $\bm{3}$ and $\bm{3^\ast}$ color sources with spins $\bm{S}_1$ and  $\bm{S}_2$ and with the same mass $m_Q$ . 
When $r$ is sufficiently small, the effect of the heavy quark and antiquark on light-QCD fields can be approximated by a single $\bm{8}$ color source with $Q \bar Q$ spin $\bm{S}_{Q \bar Q} = \bm{S}_1 + \bm{S}_2$. 
The action of the spin-dependent terms in  $\mathbf{V}_1(\bm{r}, \bm{S}_1, \bm{S}_2)$ on light-QCD states and $Q \bar Q$ spin states with energy near the repulsive color-Coulomb potential becomes independent of $\bm{r}$ in the limit $r \to 0$.
We denote this limiting spin-dependent potential by $\mathbf{V}_{\text{SS},\text{short}} (\bm{S}_1+\bm{S}_2)$ and we refer to it as the {\it short-distance spin-splitting term}.
This potential takes into account the spin splittings of adjoint hadrons.

For a $Q\bar{Q}$ pair with energy near the B\nobreakdash-O potential associated with an adjoint meson with quantum numbers $j^{\pi\gamma}$, the short-distance spin-splitting term can be expressed as
\begin{equation}
\mathbf{V}_{\text{SS},\text{short}} (\bm{S}_1+\bm{S}_2) 
= \frac{C_{8,j^{\pi\gamma}}}{2\,m_Q} \,\bm{j} \cdot \bm{S}_{Q \bar Q},
\label{HSS-short}
\end{equation}
where $\bm{j}$ is the light-QCD angular-momentum. 
The denominator $2\,m_Q$ of the coefficient is the total mass of the $Q\bar{Q}$ pair.
The numerator $C_{8,j^{\pi\gamma}}$ has a finite limit as $m_Q \to \infty$.

The short-distance spin-splitting term in Eq.~\eqref{HSS-short} for a state with definite static angular momentum $S$ is
\begin{equation}
\mathbf{V}_{\text{SS},\text{short}} (\bm{S}_1+\bm{S}_2) 
= \frac{C_{8,j^{\pi\gamma}}}{4\,m_Q} \,\bigl\{j, S_{Q \bar Q}, S \bigr\} \, \big[ S (S+1) - j (j+1) - S_{Q \bar Q} (S_{Q \bar Q}+1) \big],
\label{HSS-short-S}
\end{equation}
The spin splitting is 0 for any adjoint meson with $j=0$.
For the $j^{\pi \gamma} = 1^{--}$ adjoint meson, the spin splitting is 0 for a spin-singlet state ($S_{Q \bar Q} = 0$) and proportional to $S (S+1) - 4$ for spin-triplet states ($S_{Q \bar Q} = 1$).
The spin splittings of the spin-triplet states are $-\frac23 \Delta_{8,Q}$, $-\frac13 \Delta_{8,Q}$, and $+\frac13 \Delta_{8,Q}$ for $S=0$, 1, and 2.
Their spin-weighted average is 0.
The spin splitting between the $S=2$ and 0 states is
\begin{equation}
\Delta_{8,Q} = 3\, C_{8,1^{--}}/(2\,m_Q).
\label{Delta8Q}
\end{equation}

The matrix elements of the short-distance spin-splitting term $\mathbf{V}_{\text{SS},\text{short}}$ between radial tetraquark channels $\big( ( j^{\pi\gamma},L_{Q \bar Q}) L, S_{Q \bar Q} \big)J$ can be obtained by applying the inverse of the recoupling formula in Eq.~\eqref{ketjSL-ketjLS} to the expression for $\mathbf{V}_{\text{SS},\text{short}}$ in Eq.~\eqref{HSS-short-S}.
The matrix elements are diagonal in $L_{Q \bar Q}$, because $\mathbf{V}_{\text{SS},\text{short}}$ does not depend on $\bm{r}$.
They are also diagonal in $j^{\pi \gamma}$ and $S_{Q \bar Q}$, but they need not be diagonal in $L$.
The nonzero matrix elements in the $J$ block of $\mathbf{V}_{\text{SS},\text{short}}$ are
\begin{multline}
\big( \mathbf{V}_{\text{SS},\text{short}}^J(\bm{S}_1+\bm{S}_2) \big)_{\big( ( j^{\pi\gamma},L_{Q \bar Q}) L, S_{Q \bar Q} \big),\big( ( j^{\pi\gamma},L_{Q \bar Q}) L^\prime, S_{Q \bar Q} \big)} 
= (-1)^{L - L^\prime}\sqrt{\tilde{L} \tilde{L}^\prime}  \sum_S \tilde{S} 
\\
\times
\begin{Bmatrix} j & L_{Q \bar Q} & L             \\ J & S_{Q \bar Q}  & S  \end{Bmatrix}
\begin{Bmatrix} j & L_{Q \bar Q} & L^\prime \\ J & S_{Q \bar Q}  & S  \end{Bmatrix}
\frac{C_{8,j^{\pi\gamma}}}{4\,m_Q} \big[ S (S+1) - j (j+1) - S_{Q \bar Q} (S_{Q \bar Q}+1) \big] .
\label{VSS-tet}
\end{multline}
The matrix elements depend on $J$ and $L_{Q \bar Q}$ only through the Wigner $6j$ symbols.

The matrix elements of $\mathbf{V}_{\text{SS},\text{short}}$ between radial dimeson channels $\big( (J_1, J_2) S, L_{Q \bar Q} \big)J$ can be obtained by applying the recoupling formula in Eq.~\eqref{di-tetra1} to the expression for the $J$ block of $\mathbf{V}_{\text{SS},\text{short}}$ in  Eq.~\eqref{VSS-tet}.
The matrix elements are diagonal in $S$ and $L_{Q \bar Q}$, but they need not be diagonal in $J_1$ and $J_2$.
The nonzero matrix elements in the $J$ block of $\mathbf{V}_{\text{SS},\text{short}}$ are
\begin{eqnarray}
\big( \mathbf{V}_{\text{SS},\text{short}}^J(\bm{S}_1+\bm{S}_2) \big)_{\big( (J_1, J_2) S, L_{Q \bar Q} \big),\big( (J_1^\prime, J_2^\prime) S, L_{Q \bar Q} \big)} 
&=& \{ S, L_{Q \bar Q} , J \}\,   \sqrt{\tilde{J}_1 \tilde{J}_2 \tilde{J}_1^\prime \tilde{J}_2^\prime}\sum_{j,S_{Q \bar Q}} \tilde{j} \tilde{S}_{Q \bar Q} 
\nonumber\\
&& \hspace{-8cm} \times
\begin{Bmatrix}
j_1 & s_1                & J_1 \\
j_2 & s_2                & J_2 \\
j     & S_{Q \bar Q} & S     \end{Bmatrix}
\begin{Bmatrix}
j_1 & s_1                & J_1^\prime \\
j_2 & s_2                & J_2^\prime \\
j     & S_{Q \bar Q} & S                \end{Bmatrix}
\frac{C_{8,j^{\pi\gamma}}}{4\,m_Q} \big[ S (S+1) - j (j+1) - S_{Q \bar Q} (S_{Q \bar Q}+1) \big] .
\label{VSS-di}
\end{eqnarray}
The matrix elements depend on $J$ and $L_{Q \bar Q}$ only through the triangle symbol $ \{ S, L_{Q \bar Q} , J \}$.
The sum over $j$ and $S_{Q \bar Q}$ is constrained by the triangle relation $ \{ S, j, S_{Q \bar Q} \} = 1$ implied by the Wigner 9j symbols.
The last factor in Eq.~\eqref{VSS-di} is 0 if $S_{Q \bar Q}=0$, because $j=S$.
The sum over $S_{Q \bar Q}$ in Eq.~\eqref{VSS-di} therefore reduces to the single term with $S_{Q \bar Q}=1$.

The last factor in Eq.~\eqref{VSS-di} is 0 if $j=0$, because $S_{Q \bar Q} = S$.
The sum over $j$ and $S_{Q \bar Q}$ in Eq.~\eqref{VSS-di} therefore reduces to the single term with $j=1$ and $S_{Q \bar Q}=1$.
The nonzero matrix elements between radial dimeson channels in the $J$ block of $\mathbf{V}_{\text{SS},\text{short}}$ are
\begin{multline}
\big( \mathbf{V}_{\text{SS},\text{short}}^J(\bm{S}_1+\bm{S}_2) \big)_{\big( (J_1^-, J_2^-) S, L_{Q \bar Q} \big),\big( (J_1^{\prime -}, J_2^{\prime -}) S, L_{Q \bar Q} \big)} 
=  \{ S, L_{Q \bar Q} , J \}\,   \sqrt{\tilde{J}_1 \tilde{J}_2 \tilde{J}_1^\prime \tilde{J}_2^\prime}
\\
\times 
\begin{Bmatrix}
~\tfrac{1}{2}~ & ~\tfrac{1}{2}~ & J_1 \\
  \tfrac{1}{2}   & \tfrac{1}{2}     & J_2 \\
1                    & 1                    & S     \end{Bmatrix}
\begin{Bmatrix}
~\tfrac{1}{2}~ & ~\tfrac{1}{2}~ & J_1^\prime \\
  \tfrac{1}{2}   & \tfrac{1}{2}     & J_2^\prime \\
1                    & 1                    & S                \end{Bmatrix}
\frac{3}{2}\, \big[ S (S+1) - 4 \big]  \, \Delta_{8,Q}.
\label{VSSshort-J}
\end{multline}
We have used Eq.~\eqref{Delta8Q} to eliminate the coefficient $C_{8,1^{--}}$ in favor of the adjoint-meson spin splitting $\Delta_{8,Q}$. 
The eigenvalues of the $S$ block of $\mathbf{V}_{\text{SS},\text{short}}^J$ are the spin splittings of a $1^{--}$ adjoint meson with static angular momentum $S$.
For $S=0$, the dimeson spins $(J_1,J_2)$ are (0,0) and (1,1) and the two eigenvalues are $-\tfrac{2}{3}  \Delta_{8,Q}$ and 0. 
For $S=1$, the dimeson spins  are (0,1), (1,0), and (1,1) and the three eigenvalues are $-\tfrac{1}{3}  \Delta_{8,Q}$, 0, and 0.  
For $S=2$, the only dimeson spins are (1,1) and the spin splitting is $+\tfrac{1}{3}  \Delta_{8,Q}$. 
The spin-weighted spin splitting is 0.

\subsection{Spin-Splitting Potential}
\label{sec:SpinSplittingPotential}

The spin-dependent terms in the potential $\mathbf{V}_1$ can in principle be calculated using lattice QCD. 
In the absence of those calculations, a simple approximation that may be sufficiently accurate for many purposes is an interpolation between the short-distance spin-splitting term $\mathbf{V}_{\text{SS},\text{short}}$ in Eq.~\eqref{HSS-short} at small $r$ and the long-distance spin-splitting term $\mathbf{V}_{\text{SS},\text{long}}$ in Eq.~\eqref{HSS-long} at large $r$.
We will refer to this interpolation as the {\it spin-splitting potential}.
A possible interpolating potential using Gaussian relaxation factors is 
\begin{align}
\mathbf{V}_{\text{SS}} (r, \bm{S}_1,\bm{S}_2) =&  
\mathbf{V}_{\text{SS},\text{short}} (\bm{S}_1+\bm{S}_2) \exp\big( -r^2/R_\text{SS}^2 \big) 
\nonumber\\
&+\,  \mathbf{V}_{\text{SS},\text{long}} (\bm{S}_1,\bm{S}_2)\, \left[ 1 - \exp\big( -r^2/R_\text{SS}^2 \big) \right] .
\label{VSS-r}
\end{align}
The relaxation length $R_\text{SS}$ can be treated as an adjustable parameter. 
The entries of the $J$ block of the spin-splitting potential in Eq.~\eqref{VSS-r} in the radial dimeson basis can be obtained by inserting the diagonal entries of the $J$ block of $\mathbf{V}_{\text{SS},\text{long}}$ in Eq.~\eqref{VSSlong-J} and the entries of the $J$ block of $\mathbf{V}_{\text{SS},\text{short}}$ in Eq.~\eqref{VSSshort-J}.
The $J^P$ block of $\mathbf{V}_{\text{SS}}$ is obtained from the $J$ block by restricting $L_{Q \bar Q}$ to be either even or odd in accord with $P=(-1)^{L_{Q \bar Q}}$.
The  $J^{PC}$ block is obtained by further restricting the channels to those of the form in Eqs.~\eqref{dimesonCP+} for $CP = +1$ or Eqs.~\eqref{dimesonCP-} for $CP = -1$.

The effect of the spin-splitting potential in Eq.~\eqref{VSS-r} as $r \to \infty$ is that the $0^{-+}\Sigma_u^-$, $1^{--}\Pi_g$, and $1^{--}\Sigma_g^+$ potentials are split into potentials that approach the thresholds for the charm-meson pairs $D \bar D$, $D^* \bar D$, and $D^* \bar D^*$.
The effect of the spin-splitting potential as $r \to 0$ is that the $1^{--}\Pi_g$ and $1^{--}\Sigma_g^+$ potentials are split into potentials that differ from the repulsive color-Coulomb potential by the four possible spin splittings of the $1^{--}$ adjoint meson.
The $0^{-+}\Sigma_u^-$ potential is not affected by the spin-splitting potential as $r \to 0$.

The spin-splitting potential in Eq.~\eqref{VSS-r} gives spin splittings to the tetraquark states in a HQSS multiplet.
If the spin-splitting potential is treated as a first-order perturbation, the spin splitting of a tetraquark with light-QCD state $j^{\pi \gamma}, L^P$ and quantum numbers $J^{PC}$ is the expectation value of $\mathbf{V}_{\text{SS}}$ in Eq.~\eqref{VSS-r}:
\begin{equation}
\Delta E^{J^{PC}} = \Delta E^{J^{PC}}_{\text{SS},\text{short}}\,  \big( 1 - P_\text{long} \big) + \Delta E^{J^{PC}}_{\text{SS},\text{long}}\, P_\text{long} ,
\label{DeltaE-SS}
\end{equation}
where $\Delta E^{J^{PC}}_{\text{SS},\text{short}}$ and $\Delta E^{J^{PC}}_{\text{SS},\text{long}}$ are the short-distance and long-distance spin splittings and $P_\text{long}$ is the long-distance probability for the light-QCD state:
\begin{equation}
P_\text{long} = \Big\langle j^{\pi \gamma} ,L^P \Big| 1 - \exp\big( -r^2/R_\text{SS}^2 \big) \Big| j^{\pi \gamma} ,L^P \Big\rangle. 
\label{Plong}
\end{equation}
(We have suppressed the dependence of every term in Eq.~\eqref{DeltaE-SS} on $j^{\pi \gamma}, L^P$.)
If a bound state approaches a threshold to which it has an $S$-wave coupling, $P_\text{long}$ approaches 1.
If there are light-QCD states whose energies are nearly degenerate, it is necessary to instead use degenerate perturbation theory.
The  first-order expressions for the energy shifts are then more complicated.

\subsection{Kinetic Improvement}
\label{sec:ProjectionJPC}

The $J^{PC}$ block of the radial diabatic Hamiltonian in the HQSS limit has the general form in Eq.~\eqref{H0-JPC}.
We approximate the $1/m_Q$ diabatic potential by the spin-splitting potential $\mathbf{V}_{\text{SS}} (r, \bm{S}_1,\bm{S}_2)$ in Eq.~\eqref{VSS-r}.
In the radial dimeson basis, kinetic improvement can be carried out by replacing the multiplicative factor of $1/m_Q$ in the kinetic term in Eq.~\eqref{H0-JPC} by the inverse of a diagonal mass matrix $\mathbf{M}_{J^{PC}}$ as in Eq.~\eqref{Hdiabatic-kiM}.
The diagonal entries of $\mathbf{M}_{J^{PC}}$ for radial dimeson channels $\big( (J_1^-,J_2^-)S ,L_{Q \bar Q} \big)$ are $m_D$ for $D \bar D$ channels, $2\mu_{D^* D}=2m_D m_{D^*}/(m_D + m_{D^*})$ for $D^* \bar D$ and $D \bar D^*$ channels, and $m_{D^*}$ for $D^* \bar D^*$ channels.
The $J^{PC}$ block of the radial diabatic Hamiltonian with  spin splittings and kinetic improvement is
\begin{equation}
\mathbf{H}_1^{J^{PC}} =  
 \mathbf{M}_{J^{PC}} ^{-1} \left[-\frac{1}{r} \bigg(\frac{\mathrm{d}\phantom{r}}{\mathrm{d}r}\bigg)^2  r +\frac{1}{r^2} \mathbf{L}_{Q \bar Q}^2 \right]
+ \mathbf{V}^{J^{PC}}_0(r) + \mathbf{V}_{\text{SS}}^{J^{PC}} (r, \bm{S}_1,\bm{S}_2),
\label{H1-JPC}
\end{equation}
where $\mathbf{V}_{\text{SS}}^{J^{PC}}$ is the $J^{PC}$ block of the spin-splitting potential.

\subsection{\texorpdfstring{$J^{PC} = 1^{++}$}{JPC = 1++} Block}
\label{sec:1++ChannelSpinSplitting}

We will present the $1^{++}$ block of the radial diabatic Hamiltonian with spin splittings and with kinetic improvement in the radial dimeson basis explicitly.
The three radial dimeson channels for $1^{++}$ are enumerated in Section~\ref{sec:1++Channeldimeson}.

The general form of the $J^{PC}$  block of the radial diabatic Hamiltonian is given in Eq.~\eqref{H1-JPC}.
The $1^{++}$ block is a $3\times 3$ matrix.
The orbital-angular-momentum matrix $\mathbf{L}_{Q \bar Q}^2$ is the diagonal matrix in Eq.~\eqref{L2:1++3}.
The $1^{++}$ block of $\mathbf{V}_0(r)$ in the radial dimeson basis is given in Eq.~\eqref{V0:1++3}.
The $1^{++}$ block of the spin-splitting potential $\mathbf{V}_\text{SS}$ is a diagonal matrix that interpolates between the short-distance spin-splitting term $\mathbf{V}_{\text{SS},\text{short}}$ in Eq.~\eqref{VSSshort-J} and the long-distance spin-splitting term $\mathbf{V}_{\text{SS},\text{long}}$ in Eq.~\eqref{VSSlong-J}:
\begin{equation}
\mathbf{V}^{1^{++}}_\text{SS}(r) = \mathrm{diag}\big( -1,-1, 1 \big)
\left( \frac{1}{3}\Delta_{8,Q} \exp\big( -r^2/R_\text{SS}^2 \big)
+ \frac{1}{2}\Delta_Q \left[ 1 - \exp\big( -r^2/R_\text{SS}^2 \big) \right] \right),
\label{VSS:1++3}
\end{equation}
where $\Delta_{8,Q}$ is the adjoint-meson spin splitting and $\Delta_Q$ is the heavy-meson spin splitting.
The $1^{++}$ block of the kinetically improved mass matrix is the diagonal matrix
\begin{equation}
\mathbf{M}_{1^{++}} =  \mathrm{diag}\big( 2\mu_{D^* D},2\mu_{D^* D},m_{D^*} \big) .
\label{M:1++3}
\end{equation}

The $0^{++}$, $1^{+-}$, and $2^{++}$ blocks of the spin-splitting potential and the kinetically improved mass matrix are presented explicitly in Sections~\ref{sec:0++}, \ref{sec:1+-}, and \ref{sec:2++} of Appendix~\ref{app:DiabaticH}.


\section{Diabatic Hamiltonian with Avoided Crossing}
\label{sec:AvoidedCrossing}

In this section, we determine the radial diabatic Hamiltonian for the truncation of the diabatic potentials to the isospin-0 $1^{--}\Sigma_g^+/\Pi_g$, $0^{-+} \Sigma_u^-$, and $0^{++}\Sigma_g^+$ potentials.
We take into account the transitions between the $1^{--} \Sigma_g^+$  and $0^{++} \Sigma_g^+$ diabatic potentials that produce the narrow avoided crossing between the $1\Sigma_g^+$ and $2 \Sigma_g^+$ adiabatic  potentials. 

\subsection{Truncation of Diabatic Potentials}
\label{sec:OniumTruncation}

We have truncated our system to the $1^{--}\Sigma_g^+/\Pi_g$ and $0^{-+}\Sigma_u^-$ diabatic potentials.
We now extend it to include the $0^{++}\Sigma_g^+$ potential.
We refer to $1^{--}\Pi_g$, $1^{--}\Sigma_g^+$, and $0^{-+}\Sigma_u^-$ as {\it tetraquark potentials} and $0^{++}\Sigma_g^+$ as the {\it quarkonium potential}.
The matrices $\mathbf{V}_0^\lambda(r)$ of diabatic potentials and transition potentials that follow from the cylindrical and $(CP)_\mathrm{light}$ symmetries are defined by Eq.~\eqref{V0-lambda}.
There are additional potentials only for $\lambda=0$.
The additional diabatic potential is  $\big( \mathbf{V}_0^0(r) \big)_{0^{++},0^{++}} = V_{0^{++}\Sigma_g^+}(r)$ and there is a transition potential $\big( \mathbf{V}_0^0(r) \big)_{0^{++},1^{--}} = G_{(0^{++},1^{--})\Sigma_g^+}(r)$ between $1^{--}\Sigma_g^+$ and $0^{++}\Sigma_g^+$. 

The general expression for the entries of the $L$ block of the radial diabatic static potential $\mathbf{V}_0(r)$ is given in Eq.~\eqref{V0-L}. 
Its rows and columns are labeled by radial diabatic channels $(j^{\pi\gamma},L_{Q\bar Q})$ where $j^{\pi\gamma}$ is the quantum numbers of an adjoint meson and the orbital angular momentum $L_{Q \bar Q}$ is restricted by the triangle relation $\{ j,L_{Q \bar Q},L \}=1$.
The  entries of the $L$ block of $\mathbf{V}_0(r)$ with factors of the $1^{--}\Pi_g$, $1^{--}\Sigma_g^+$, and $0^{-+}\Sigma_u^-$  potentials are given in Eqs.~\eqref{VLQQbar}.
The additional entries of the $L$ block with factors of the $0^{++}\Sigma_g^+$ potential and the $(0^{++},1^{--})\Sigma_g^+$ transition potential are
\begin{subequations}
\begin{align}
\left( \mathbf{V}_0^L  (r) \right)_{(0^{++},L_{Q \bar Q}), (0^{++},L_{Q \bar Q}^\prime)} =&
\delta(L, L_{Q \bar Q}) \, \delta(L,L_{Q \bar Q}^\prime)   \,  V_{0^{++}\Sigma_g^+}(r) ,
\label{VLQQbar0++AC}
\\
\left( \mathbf{V}_0^L (r) \right)_{(0^{++},L_{Q \bar Q}),(1^{--},L_{Q \bar Q}^\prime)} =& 
(-1)\,  \delta(L,L_{Q \bar Q}) \,  \cgr{1}{0}{L}{0}{L_{Q\bar Q}^\prime}{0}  \,
G_{(0^{++},1^{--})\Sigma_g^+}(r),
\label{VLQQbar0++1--AC}
\\
\left( \mathbf{V}_0^L (r) \right)_{(1^{--},L_{Q \bar Q}),(0^{++},L_{Q \bar Q}^\prime)} =& 
(-1)\,   \cgr{1}{0}{L}{0}{L_{Q\bar Q}}{0}\,  \delta(L,L_{Q \bar Q}^\prime)  \,
G_{(0^{++},1^{--})\Sigma_g^+}(r).
\end{align}
\label{VLQQbarAC}
\end{subequations}
The Clebsch-Gordan coefficients are given in Eqs.~\eqref{CG}.

\subsection{Radial Tetraquark Basis}
\label{sec:OniumRadialTetraquark}

The radial  tetraquark channels with total angular momentum $J$ can be labeled by the channels $\bigl( (j^{\pi\gamma},L_{Q\bar Q})L,S_{Q \bar Q} \bigr)$.
The $(L,S_{Q \bar Q})J^{PC}$ block of $\mathbf{V}_0(r)$ is given in terms of the $L$ block in Eq.~\eqref{V0JPC}.
Its rows and columns are labeled by $(j^{\pi \gamma},L_{Q\bar Q})$.
The radial quarkonium channels are those with $j^{\pi\gamma}=0^{++}$, which implies $L = L_{Q\bar Q}$.

The quarkonium block within the $(L,S_{Q \bar Q})J^{PC}$ block of $\mathbf{V}_0(r)$ is obtained by inserting the $L$ block in Eq.~\eqref{VLQQbar0++AC} into Eq.~\eqref{V0JPC}.
The quarkonium block is diagonal in $L_{Q\bar Q}$.
Its diagonal entries in the radial tetraquark basis are
\begin{multline}
\left( \mathbf{V}_0^{(L,S_{Q \bar Q})J^{PC}}(r)  \right)_{(0^{++},L_{Q\bar Q}),(0^{++}, L_{Q\bar Q})} 
= \{ L, S_{Q \bar Q} , J\} \,\delta(L, L_{Q \bar Q}) \,  \delta(P ,\ (-1)^{L_{Q\bar Q}+1})\,
\\
\times  \delta\big(C P,(-1)^{S_{Q \bar Q}+1}\big)\, 
V_{0^{++}\Sigma_g^+}(r) .
\label{V0JPConium}
\end{multline}
Each nonzero diagonal entry is the $0^{++}\Sigma_g^+$ quarkonium potential. 
The quarkonium $J^{PC}$ block of $\mathbf{V}_0$ can be obtained by combining the $(L,S_{Q \bar Q})J$ blocks in Eq.~\eqref{V0JPConium} for $L$ and $S_{Q \bar Q}$ that satisfy $ \{ L, S_{Q \bar Q} , J\} = 1$.
The $0^{++}$ and $1^{++}$ blocks each consist of a single spin-triplet channel with $L_{Q \bar Q} = 1$.
The $1^{+-}$ block consists of a single spin-singlet channel with $L_{Q \bar Q} = 1$.
The $2^{++}$ block consists of two spin-triplet channels with $L_{Q \bar Q} = 1$ and 3.

The block within the $(L,S_{Q \bar Q})J^{PC}$ block of $\mathbf{V}_0(r)$ that corresponds to transitions between $0^{++} \Sigma_g^+$ and $1^{--} \Sigma_g^+$ can be obtained by inserting the $L$ block in Eq.~\eqref{VLQQbar0++1--AC} into Eq.~\eqref{V0JPC}.
The entries of the transition block in the radial tetraquark basis are
\begin{multline}
\left( \mathbf{V}_0^{(L,S_{Q \bar Q})J^{PC}}(r)  \right)_{(0^{++},L_{Q\bar Q}),(1^{--}, L_{Q\bar Q}^\prime)} 
= \{ L, S_{Q \bar Q} , J \} \, (-1)\, \delta(L, L_{Q \bar Q}) \,   \cgr{1}{0}{L}{0}{L_{Q\bar Q}^\prime}{0}  
\\
\times 
\delta \big( P , (-1)^{L_{Q \bar Q}+1} \big)\,  \delta\big(P , (-1)^{L_{Q \bar Q}^\prime} \big)\,
 \delta\big(C P,(-1)^{S_{Q \bar Q}+1}\big)\,
G_{(0^{++},1^{--})\Sigma_g^+}(r).
\label{V0JPConium1--}
\end{multline}
For a nonzero block $(L,S_{Q \bar Q})J^{PC}$ with $L=0$,  the only nonzero row is $(0^{++},L_{Q \bar Q}=0)$ and the only nonzero entry in that row is $-G_{(0^{++},1^{--})\Sigma_g^+}(r)$ in the $(1^{--}, L_{Q\bar Q}^\prime=1)$ column.
For a nonzero block with $L>0$, the only nonzero row is $(0^{++},L_{Q \bar Q}=L)$ and it is the product of $-G_{(0^{++},1^{--})\Sigma_g^+}(r)$ and a unit vector whose two nonzero entries are the Clebsch-Gordan coefficients in Eqs.~\eqref{CG:L-1} and \eqref{CG:L+1}.
The transition $J^{PC}$ block of $\mathbf{V}_0(r)$ can be obtained by combining the $(L,S_{Q \bar Q})J$ blocks in Eq.~\eqref{V0JPConium1--} for $L$ and $S_{Q \bar Q}$ that satisfy $ \{ L, S_{Q \bar Q} , J\} = 1$.

\subsection{Radial Dimeson Basis}
\label{sec:OniumRadialDimeson}

The radial dimeson channels with total angular momentum $J$ can be labeled by the channels $\big( ( J_1^{P_1},J_2^{P_2})S,L_{Q \bar Q} \big)$. 
When the truncation is extended to include $0^{++}\Sigma_g^+$, the additional radial quarkonium channels can be labeled by $\big( ( \tfrac{1}{2}^+, \tfrac{1}{2}^-)S_{Q \bar Q} ,L_{Q \bar Q} \big)$, where  $S_{Q \bar Q}$ is the $Q \bar{Q}$ spin, or alternatively by $Q \bar Q(S_{Q \bar Q},L_{Q \bar Q})$.
We refer to the set of radial dimeson channels together with these radial quarkonium channels as the {\it radial dimeson basis}.
The angular-momentum states for the radial quarkonium channels in the radial dimeson basis differ from those in the radial tetraquark basis at most by a sign:
\begin{equation}
\vket{\bigr( ( \tfrac{1}{2}^+, \tfrac{1}{2}^-)S_{Q \bar Q} ,L_{Q \bar Q} \bigr)}{J}{M_J}{}
= (-1)^{J- L_{Q \bar Q}- S_{Q \bar Q}}
\vket{\bigl( (0^{++},L_{Q \bar Q}) L_{Q \bar Q}, S_{Q \bar Q} \bigr)}{J}{M_J}{} .
\label{QQbar:dimeson-tetra}
\end{equation}

The quarkonium $(L_{Q \bar Q},S_{Q \bar Q})J^{PC}$ block of the static potential $\mathbf{V}_0(r)$ in the radial dimeson basis can obtained by making the change of basis in Eq.~\eqref{QQbar:dimeson-tetra} to the quarkonium $(L_{Q \bar Q},S_{Q \bar Q})J^{PC}$ block in the radial tetraquark basis in Eq.~\eqref{V0JPConium}.
The quarkonium block is diagonal in $S_{Q \bar Q}$ and $L_{Q\bar Q}$.
The diagonal entries of the quarkonium $J^{PC}$ block  in the radial dimeson basis are
\begin{multline}	
\left( \mathbf{V}_0^{J^{PC}}(r)  \right)_{Q \bar Q(S_{Q \bar Q},L_{Q \bar Q}),Q \bar Q(S_{Q \bar Q},L_{Q \bar Q})} 
= \{ S_{Q \bar Q}, L_{Q \bar Q} , J\} \, \delta(P ,\ (-1)^{L_{Q \bar Q}+1})\,
\\
\times \delta\big(C P,(-1)^{S_{Q \bar Q}+1}\big)\, 
V_{0^{++}\Sigma_g^+}(r) .
\label{V0JPCtetra}
\end{multline}	
Each nonzero diagonal entry is the $0^{++}\Sigma_g^+$ quarkonium potential. 

The transition $J^{PC}$ block in the radial dimeson basis can be obtained most easily by starting from the transition $J$ block in the radial tetraquark basis. 
It can be obtained from Eq.~\eqref{V0JPConium1--} by omitting the Kronecker delta symbols that provide the constraints from $P$ and $CP$:
\begin{multline}
\left( \mathbf{V}_0^J(r)  \right)_{\big( (0^{++},L_{Q\bar Q})L_{Q\bar Q},S_{Q \bar Q} \big),\big( (1^{--}, L_{Q\bar Q}^\prime)L,S_{Q \bar Q}^\prime \big)} 
= \{ L_{Q\bar Q}, S_{Q \bar Q} , J \} \,  \delta(L_{Q \bar Q},L) \, \delta(S_{Q \bar Q}, S_{Q \bar Q}^\prime) 
\\
\times 
(-1)\, \cgr{1}{0}{L}{0}{L_{Q\bar Q}^\prime}{0}  \, G_{(0^{++},1^{--})\Sigma_g^+}(r).
\label{V0Jonium1--}
\end{multline}
The transition $J$ block in the radial dimeson basis can be obtained from the transition $J$ block in the radial tetraquark basis in Eq.~\eqref{V0Jonium1--} by using Eq.~\eqref{QQbar:dimeson-tetra} to change the quarkonium channel and using Eq.~\eqref{di-tetra12} with $j_1=j_2=\tfrac{1}{2}$ and $j=1$ to change the tetraquark channel to a dimeson channel:
\begin{multline}
\left( \mathbf{V}_0^J(r)  \right)_{Q\bar{Q}(S_{Q \bar Q},L_{Q\bar Q}),\big( (J_1^-,J_2^-)S, L_{Q\bar Q}^\prime \big)} 
=  (-1)^{J + S + L_{Q\bar Q}^\prime}\, \sqrt{3 \tilde{S}_{Q\bar Q} \tilde{L}_{Q\bar Q}} \sqrt{\tilde{J}_1 \tilde{J}_2 \tilde{S}}
\\
\times \cgr{1}{0}{L_{Q \bar Q}}{0}{L_{Q\bar Q}^\prime}{0}
\begin{Bmatrix}
J &  L_{Q \bar Q}^\prime & S  \\
1 &  S_{Q \bar Q}                              & L_{Q \bar Q}  \end{Bmatrix}
\begin{Bmatrix}
~\tfrac12~  & ~\tfrac12~ & J_1 \\
\tfrac12      & \tfrac12      & J_2 \\
1                & S_{Q \bar Q}                & S \end{Bmatrix}
G_{(0^{++},1^{--})\Sigma_g^+}(r).
\label{V0JPCQQbar1--}
\end{multline}
In the row for each quarkonium channel $Q\bar{Q}(S_{Q\bar Q},L_{Q\bar Q})$, the entries of Eq.~\eqref{V0JPCQQbar1--} are the product of the transition potential $G_{(0^{++},1^{--})\Sigma_g^+}(r)$ and the components of a unit vector whose entries are labeled by $( (J_1,J_2)S, L_{Q\bar Q}^\prime )$. 
That it is a unit vector follows from the orthogonality relation for Wigner $9j$ symbols and the orthogonality relation for Wigner $6j$ symbols.
The $J^P$ block is obtained by restricting the channels to those whose orbital angular momenta satisfy $P = (-1)^{L_{Q\bar Q}+1}$ and $P = (-1)^{L_{Q\bar Q}^\prime}$.
The $J^{PC}$ block is obtained by further requiring $S_{Q \bar Q}$ to satisfy $CP = (-1)^{S_{Q \bar Q}+1}$ and restricting the dimeson channels to those of the form in Eqs.~\eqref{dimesonCP+} for $CP = +1$ or in Eqs.~\eqref{dimesonCP-} for $CP = -1$.

\subsection{\texorpdfstring{$J^{PC} = 1^{++}$}{JPC = 1++} Block}
\label{sec:Onium1++}

When the diabatic potentials are truncated to $1^{--}\Sigma_g^+/\Pi_g$ and $0^{-+}\Sigma_u^-$, the three radial tetraquark channels $\bigl( (j^{\pi\gamma},L_{Q \bar Q}) L, S_{Q \bar Q} \bigr)$ with $J^{PC} = 1^{++}$ are enumerated in Section~\ref{sec:1++Channel}.
When the truncation is extended to include $0^{++}\Sigma_g^+$, the radial  diabatic channels with $J^{PC}$ include additional radial quarkonium channels of the form $\bigl( (0^{++},L_{Q \bar Q}) L, S_{Q \bar Q} \bigr)$ with $L_{Q \bar Q}= L$, $P=(-1)^{L+1}$, $CP=(-1)^{S_{Q \bar Q}+1}$, and $\{ L,S_{Q \bar Q},J \}=1$.
The only additional radial quarkonium channel with $J^{PC} = 1^{++}$ is $\big( (0^{++},1)1, 1 \big)$.
This is a $P$-wave channel.
The four radial channels for $J^{PC} = 1^{++}$ in the radial tetraquark basis are
\\
\indent \indent \indent 1: $\big( (1^{--},0) 1, 1 \big)$,\\
\indent \indent \indent 2: $\big( (1^{--},2) 1, 1 \big)$,\\
\indent \indent \indent 3: $\big( (1^{--},2) 2, 1 \big)$,\\
\indent \indent \indent 4. $\big( (0^{++},1)1, 1 \big)$. \\
The orbital-angular-momentum matrix  in Eq.~\eqref{L2:1++3} is extended to a $4 \times 4$ matrix whose 4th diagonal entry is 2:
\begin{equation}
\mathbf{L}_{Q \bar Q}^2 =  \mathrm{diag}\big( 0,6,6, 2 \big) .
\label{L2:1++4}
\end{equation}
The $3 \times 3$ matrix $\mathbf{V}^{1^{++}}_0(r)$ is extended to  a $4 \times4$ matrix that is the sum of the matrix in Eq.~\eqref{V0:1++} (with a 4th row and a 4th column of zeros) and two  additional matrices: 
\begin{equation}
 \Delta\mathbf{V}^{1^{++}}_0(r) =
V_{0^{++}\Sigma_g^+}(r)
\begin{pmatrix}
 \bm{0}_{3 \times 3} &  \bm{0}_{3 \times 1}     \\
\bm{0}_{1 \times 3}  & 1    \\
\end{pmatrix} 
+ G_{(0^{++},1^{--})\Sigma_g^+}(r)
\begin{pmatrix}
 \bm{0}_{3 \times 3} & \hat{\bm{N}}_4^T	\\
\hat{\bm{N}}_4                              & 0             \\
\end{pmatrix} ,
\label{V0:1++AC}
\end{equation}
where $\bm{0} _{m \times n}$ is the $m \times n$ zero matrix and $\hat{\bm{N}}_4$ is the 3-component unit row vector given by Eq.~\eqref{V0JPConium1--}:
\begin{equation}
 \hat{\bm{N}}_4 = \begin{pmatrix} \sqrt{\frac13} & -\sqrt{\frac23} & ~0~ \end{pmatrix}.
\label{Nhat1++:tetra}
\end{equation}

When the diabatic potentials are truncated to $1^{--}\Sigma_g^+/\Pi_g$  and $0^{-+}\Sigma_u^-$, the three radial dimeson channels $\big( ( J_1^-,J_2^-)S,L_{Q \bar Q} \big)$ with $J^{PC} = 1^{++}$ are enumerated in Section~\ref{sec:1++Channeldimeson} using the alternative notation $D^{(*)} \bar{D}^{(*)} (S,L_{Q \bar Q})$.
When the truncation is extended to include $0^{++}\Sigma_g^+$, the radial diabatic channels with $J^{PC}$ include additional radial quarkonium channels denoted by $Q \bar Q(S_{Q \bar Q},L_{Q \bar Q})$.
The additional radial quarkonium channel with $J^{PC} = 1^{++}$ is $Q \bar Q(1,1)$.
It differs by a factor $-1$ from the channel $\big( (0^{++},1)1, 1 \big)$ in the radial tetraquark basis.
The four radial channels for $J^{PC} = 1^{++}$ in the radial dimeson basis are\\
\indent \indent \indent 1:  $[D^* \bar D(1,0) + D \bar D^*(1,0)]/\sqrt2$, \\
\indent \indent \indent 2:  $[D^* \bar D(1,2) + D \bar D^*(1,2)]/\sqrt2$, \\
\indent \indent \indent 3:   $D^* \bar D^*(2,2)$,                                     \\
\indent \indent \indent 4. $Q \bar Q(1,1)$.  \\
The orbital-angular-momentum matrix $\mathbf{L}_{Q \bar Q}^2$ is given in Eq.~\eqref{L2:1++4}.
The $3 \times 3$ matrix $\mathbf{V}^{1^{++}}_0(r)$ is extended to a $4 \times4$ matrix that is the sum of the matrix in Eq.~\eqref{V0:1++3} (with a 4th row and a 4th column of zeros) and two  additional matrices with factors of $V_{0^{++}\Sigma_g^+}(r)$ and  $G_{(0^{++},1^{--})\Sigma_g^+}(r)$.
The additional matrices have the same form as in Eq.~\eqref{V0:1++AC} except that  $\hat{\bm{N}}_4$ is the 3-component unit row vector given by Eq.~\eqref{V0JPCQQbar1--}:
\begin{equation}
\hat{\bm{N}}_4 = \begin{pmatrix} \sqrt{\frac13} & ~\sqrt{\frac16}~ & \sqrt{\frac12} \end{pmatrix} .
\label{Nhat1++:dimeson}
\end{equation}
The complete $1^{++}$ block of the $4 \times 4$ radial diabatic static potential matrix in the radial dimeson basis is
\begin{align}
 \mathbf{V}^{1^{++}}_0(r) =&
V_{1^{--}\Pi_g}(r)  \begin{pmatrix}
 \bm{1}_{3 \times 3} &  \bm{0}_{3 \times 1}     \\
\bm{0}_{1 \times 3}  & 0    \\
\end{pmatrix} 
+ \frac{1}{6}\big[ V_{1^{--}\Sigma_g^+}(r) - V_{1^{--}\Pi_g}(r) \big]
\begin{pmatrix}
2            & \sqrt{2}  & \sqrt{6} & ~0~ \\
\sqrt{2}  & 1            & \sqrt{3} &   0    \\
\sqrt{6} & \sqrt{3} &   3          &   0    \\
0	     & 0            &   0          &   0   \\
\end{pmatrix}
\nonumber\\
& + V_{0^{++}\Sigma_g^+}(r)
\begin{pmatrix}
 \bm{0}_{3 \times 3} &  \bm{0}_{3 \times 1}     \\
\bm{0}_{1 \times 3}  & 1    \\
\end{pmatrix} 
+ G_{(0^{++},1^{--})\Sigma_g^+}(r)
\begin{pmatrix}
 \bm{0}_{3 \times 3} & \hat{\bm{N}}_4^T	\\
\hat{\bm{N}}_4                              & 0             \\
\end{pmatrix} .
\label{V0:1++4}
\end{align}
The $1^{++}$ block of the spin-splitting matrix  in Eq.~\eqref{VSS:1++3} is extended to a $4 \times 4$ matrix by adding a 4th diagonal entry 0:
\begin{equation}
\mathbf{V}^{1^{++}}_\text{SS}(r) = \mathrm{diag}\big( -1,-1, 1, 0 \big)
\left( \frac{1}{3}\Delta_{8,Q} \exp\big( -r^2/R_\text{SS}^2 \big)
+ \frac{1}{2}\Delta_Q \left[ 1 - \exp\big( -r^2/R_\text{SS}^2 \big) \right] \right) .
\label{VSS:1++4}
\end{equation}
The $1^{++}$ block of the kinetically improved mass matrix  is obtained by extending $\mathbf{M}_{1^{++}}$ in Eq.~\eqref{M:1++3} to a $4 \times 4$ matrix by adding a 4th diagonal entry $m_Q$:
\begin{equation}
\mathbf{M}_{1^{++}} =  \mathrm{diag}\big( 2\mu_{D^* D},2\mu_{D^* D},m_{D^*}, m_Q \big) .
\label{M:1++4}
\end{equation}


\section{Model B-O Potentials and Parameters}
\label{sec:Models}

In this section, we  present simple models for the diabatic potentials and transition potentials that appear in the radial diabatic Hamiltonians for $\chi_{c1}(3872)$ and its HQSS partners.
We also give the numerical values of all the parameters that appear in the radial diabatic Hamiltonians.

\subsection{Quarkonium potential and heavy-quark masses}
\label{sec:QuarkoniumPotential}

The $J^{PC}$ block of the radial diabatic Hamiltonian in the HQSS limit is given in Eq.~\eqref{H0-JPC}.
The kinetic term depends on the heavy-quark mass $m_Q$.
The heavy quark masses $m_c$ and $m_b$ can be determined by fitting spin-averaged charmonium and bottomonium energy levels in the $0^{++}\Sigma_g^+$ quarkonium potential.
A good parametrization of the quarkonium potential is the Cornell potential:
\begin{equation}
V_{0^{++}\Sigma_g^+}(r) = \frac{\kappa_1}{r} + E_0 + \sigma\, r.
\label{V-Cornell}
\end{equation}
The first term is the attractive color-Coulomb potential whose strength can be expressed as $\kappa_1=+ \tfrac16 \langle\alpha_s\rangle$, where $\langle\alpha_s\rangle$ can be interpreted by some averaged running coupling constant of QCD.
In the third term,  the coefficient $\sigma$ can be interpreted as the string tension of a gluon flux tube.
These parameters should be the same for hidden-charm mesons and hidden-bottom mesons.

In the 4-dimensional space defined by the parameters $(m_Q,\kappa_1,E_0,\sigma)$, the eigenvalues of the Schr\"odinger equation with the Cornell potential in Eq.~\eqref{V-Cornell} have a direction that is completely flat  \cite{Braa14}.
The energy levels for a given value of $m_Q$ can be reproduced for any other positive value of $m_Q$ by changing the values of $\kappa_1$, $E_0$, and $\sigma$.
Thus the heavy-quark  mass $m_Q$ cannot be determined by a fit to spin-averaged quarkonium energy levels.
However, the two heavy quark masses $m_c$ and $m_b$ can be determined by a simultaneous fit to spin-averaged charmonium and bottomonium energy levels if an additional constraint is imposed on the heavy-quark masses.
In Ref.~\cite{Braa14}, the heavy-quark mass difference $m_b-m_c$ was constrained to be equal to the mass difference $M_B - M_D$ between the  isospin-averaged spin-0 $S$-wave heavy mesons.
The heavy-quark masses and the parameters in the Cornell potential were determined in Ref.~\cite{Braa14} by fitting the spin-averaged $1S$, $2S$, and $1P$ energy levels of charmonium and the spin-averaged $1S$, $2S$, $1P$, and $2P$ energy levels of bottomonium with this additional constraint.
The charm and bottom quark masses in Ref.~\cite{Braa14} are
\begin{equation}
m_c = 1.48\,\mathrm{GeV} ,\qquad m_b = 4.89\,\mathrm{GeV} ,
\label{mc,mb}
\end{equation}
whose ratio is $m_b/m_c = 3.30$. 
The parameters in the Cornell potential in Ref.~\cite{Braa14} are 
\begin{equation}
\kappa_1 = -0.489, \qquad E_0= -0.242\,\mathrm{GeV}, \qquad \sigma=0.187\,\mathrm{GeV}^2.
\label{Cornellparams}
\end{equation}
The prediction for a spin-weighted quarkonium mass is the sum of $2 m_Q$ and the energy eigenvalue of a heavy quark-antiquark pair with quark mass $m_Q$ in the Cornell potential.

The difference between the mass of a heavy hadron in QCD and the heavy-quark mass $m_Q$ can be expanded in inverse powers of $m_Q$.
The constraint $m_b-m_c= M_B - M_D$ in Ref.~\cite{Braa14} ignored the spin splittings of order $1/m_Q$ between $D^*$ and $D$ and between $B^*$ and $B$.
It also ignored the spin-independent terms of order $1/m_Q$ in the masses of $D$ and $D^*$ and the masses of $B$ and $B^*$.
A quantitative estimate of the spin-independent contribution of order $1/m_Q$ to the mass of an $S$-wave heavy meson in the Minimal Renormalon Subtraction scheme for the heavy-quark masses was given in Ref.~\cite{Baza18}.
This contribution was denoted by $\mu_\pi^2/(2 m_Q)$, and the numerator was determined to be $\mu_\pi^2 = 0.05(22)\,\mathrm{GeV}^2$. 
The resulting contribution to the mass difference $M_B-M_D$ is $-13(56)$~MeV, which is consistent with zero. 
A more careful determination of $m_c$ and $m_b$ and the Cornell potential parameters from the spin-averaged charmonium and bottomonium energy levels should take into account all terms of order $1/m_Q$ in the masses of the $S$-wave heavy mesons.

\subsection{Tetraquark Potentials}
\label{sec:modelbopotentials}

In the absence of lattice QCD calculations of the B\nobreakdash-O potentials, we need to develop models for the B\nobreakdash-O potentials.
Our models must interpolate between the known behavior at small $r$ and the known behavior at large $r$.
The diabatic potentials most  relevant to $\chi_{c1}(3872)$ and its HQSS partners are the isospin-0 $1^{--}\Pi_g$, $1^{--}\Sigma_g^+$, and $0^{-+}\Sigma_u^-$ potentials.
At large $r$, all three potentials approach the triplet-meson-pair threshold. 
At small $r$, the $1^{--}\Pi_g$ and $1^{--}\Sigma_g^+$ potentials approach the repulsive color-Coulomb potential offset by the energy of the isospin-0 $1^{--}$ adjoint meson while the $0^{-+}\Sigma_u^-$ potential approaches  the repulsive color-Coulomb potential offset by the energy of the isospin-0 $0^{-+}$ adjoint meson.
We would like our models for the $1^{--}\Pi_g$ and $1^{--}\Sigma_g^+$ potentials to support a bound state whose energy can be tuned to the triplet-hadron-pair threshold.
We would like our model for the $0^{-+}\Sigma_u^-$ potential to not support any bound states or resonances.

At small $r$,  B\nobreakdash-O potentials can be constrained by BOEFT \cite{Ber24}.
The limiting behavior as $r\to0$ of a diabatic potential  with quantum numbers $j^{\pi\gamma}\Lambda^\epsilon_\eta$ is
\begin{equation}
V_{j^{\pi\gamma}\Lambda^\epsilon_\eta}(r) = \frac{\kappa_8}{r} + E_{8,j^{\pi\gamma}} + A_{j^{\pi\gamma}\Lambda^\epsilon_\eta}\, r^2 + \mathcal{O}(r^3).
\label{VjpiLambda}
\end{equation}
The strength $\kappa_8$ of the repulsive color-Coulomb potential is the same for all  tetraquark potentials $j^{\pi\gamma}\Lambda^\epsilon_\eta$.
The additive constant $E_{8,j^{\pi\gamma}}$ is the same for $\Lambda = 0, \dots,j$.
It can be interpreted as the energy of the $j^{\pi\gamma}$ adjoint meson.
BOEFT predicts that there is no term linear in $r$ in Eq.~\eqref{VjpiLambda}.
The $r^2$ term in Eq.~\eqref{VjpiLambda} is the first term in the expansion in powers of $r$ whose coefficient depends on $\Lambda$.
In pure $SU(3)$ gauge theory, the minimum of the $\Sigma_g^{+\prime}$ potential is lower than the minimum of the $\Pi_g$ potential \cite{Alas24}.
If this pattern also holds for the isospin-0 $1^{--}\Sigma_g^+$ and $1^{--}\Pi_g$ potentials in QCD, we would expect $A_{1^{--}\Sigma_g^+}$ to be smaller than $A_{1^{--}\Pi_g}$.

At large $r$, a tetraquark B\nobreakdash-O potential must approach a static-hadron-pair threshold given by the sum of the energies of a triplet hadron and an antitriplet hadron.
A simple way to ensure the correct behavior at large $r$ is to multiply an analytic expression for the deviation of $V_{(j^\pi)\Lambda}(r)$ from the threshold by a Gaussian factor $\exp( -r^2/R^2)$, where $R$ is a {\it relaxation length}.
Bicudo and Wagner and collaborators have used lattice QCD to calculate the adiabatic potentials for double-heavy hadrons  \cite{Bicu16,Hof24}, which are the discrete energies of QCD in the presence of two color-triplet ($\bm{3}$) sources separated by a distance $r$.
At small $r$, they reduce either to the attractive color-Coulomb potential associated with an antitriplet hadron or to the repulsive color-Coulomb potential associated with a sextet hadron.
The adiabatic potentials were parametrized by the product of a color-Coulomb potential $\kappa_\text{R}/r$ and a Gaussian relaxation factor $\exp( -r^2/R_\text{R}^2)$.
The parameters $\kappa_\text{R}$ and $R_\text{R}$ were determined by fitting lattice QCD results for the potentials and extrapolating them to the physical pion mass  \cite{Hof24}.
The Coulomb strengths were determined to be $\kappa_{3^\ast} = -0.34 \pm 0.03$ and $\kappa_{6} = +0.10 \pm 0.07$. 
Their ratio $\kappa_{3^\ast}/\kappa_{6} = -3.4 \pm 2.4$ is consistent with the leading-order perturbative QCD prediction $-2$ to within the large errors.
The relaxation lengths were determined to be $R_{3^\ast} = 0.45^{+0.12}_{-0.10}$~fm and $R_6 = 0.28 \pm 0.02$~fm.

A simple measure of the strength of the color source in a representation {\bf R} is the quadratic Casimir $C_\text{R}$.
The quadratic Casimirs for the $\bm{3^\ast}$, $\bm{6}$, and $\bm{8}$ representations of SU(3) are $C_{3^\ast}=4/3$, $C_6=10/3$, and $C_8=3$.
The ratio of the quadratic Casimirs for $\bm{6}$ and $\bm{3^\ast}$ is $C_6/C_{3^\ast} = (10/3)/(4/3) = 2.5$.
The square of the ratio of relaxation lengths determined in Ref.~\cite{Hof24} is $(R_{3^\ast}/R_6)^2 =2.6^{+1.4}_{-1.2}$.
Thus the relaxation lengths in Ref.~\cite{Hof24} are consistent with $R_\text{R}$ scaling as $C_\text{R}^{-1/2}$.
The quadratic Casimir $C_8$ for the adjoint representation is smaller than $C_6$ by only about 10\%.
If the approach of an adjoint-hadron potential to a static-hadron-pair threshold can be described by a Gaussian factor with relaxation length $R_8$ and if $R_\text{R}$ scales as $C_\text{R}^{-1/2}$, then we would expect $R_8$ to be approximately equal to $R_6$.

We choose to parametrize the $1^{--}\Pi_g$,  $1^{--}\Sigma_g^+$, and $0^{-+}\Sigma_u^-$ potentials by
\begin{equation}
V_{j^{\pi\gamma}\Lambda_\eta^\epsilon}(r) = \biggl( \frac{\kappa_8}{r} + E_{8,j^{\pi\gamma}} + \frac{\kappa_8}{R_{\Lambda_\eta^\epsilon}^2} r \biggr) \exp\bigl( -r^2/R_{\Lambda_\eta^\epsilon}^2 \bigr).
\label{bopots}
\end{equation}
We have chosen the zero of energy to be the triplet-meson-pair threshold.
The Gaussian factor in Eq.~\eqref{bopots} ensures that the potential approaches that threshold as $r\to\infty$.
The term proportional to $r$ in the factor multiplying the Gaussian in Eq.~\eqref{bopots} is a subtraction term that guarantees that the small-$r$ expansion has no term proportional to $r$ in accord with the constraint from BOEFT  in Eq.~\eqref{VjpiLambda}. 
The coefficient of the $r^2$ term in the small-$r$ expansion of Eq.~\eqref{bopots} is $A_{j^{\pi\gamma}\Lambda^\epsilon_\eta} = -  E_{8,j^{\pi\gamma}} /R_{\Lambda_\eta^\epsilon}^2$.

We choose the strength $\kappa_8$ of the repulsive color-Coulomb potential to be the value determined by fits to $\Pi_u$ and $\Sigma_u^-$ potentials in pure SU(3) gauge theory  \cite{Alas24}:
\begin{equation}
\kappa_8 = +0.037.
\label{kappa8}
\end{equation}
We treat the energy $E_{8,1^{--}}$ of the $1^{--}$ adjoint meson as an adjustable parameter that can be used to tune the energy of a bound state in the $1^{--}\Pi_g$ and  $1^{--}\Sigma_g^+$ potentials.
We choose the energy  of the $0^{-+}$ adjoint meson to be
\begin{equation}
E_{8,0^{-+}} = - 2 \kappa_8 / R_{\Sigma_u^-}.
\label{E80-+}
\end{equation}
With this choice, the $0^{-+}\Sigma_u^-$ potential in Eq.~\eqref{bopots} decreases to zero and then increases slightly before decreasing rapidly to zero.
The potential does not decrease below the triplet-meson-pair threshold, so it cannot support any bound states.
The local maximum is strongly suppressed by the Gaussian factor in Eq.~\eqref{bopots}, so the potential is unlikely to support any resonances.

The shapes of the potentials are controlled by the relaxation lengths in Eq.~\eqref{bopots}.
We wish to consider the fewest possible sets of relaxation lengths that can illustrate the range of behavior of solutions to the Schr\"odinger equations. 
We choose the relaxation lengths for each of the three potentials to be near one of the values $R_{3^\ast}$ and $R_6$ from the lattice QCD calculations in Ref.~\cite{Hof24}.
Since $V_{0^{-+}\Sigma_u^-}(r)$ is mostly repulsive, it should not have dramatic effects on the energies of bound states and resonances.
We therefore consider only a single value for $R_{\Sigma_u^-}$ near $R_6$.
If $V_{1^{--}\Sigma_g^+}(r)$ and $V_{1^{--}\Pi_g}(r)$ are equal, there are channels of the Schr\"odinger equations that decouple so we consider only distinct values for $R_{\Pi_g}$ and  $R_{\Sigma_g^+}$.
We would like to allow the difference $V_{1^{--}\Sigma_g^+}(r)-V_{1^{--}\Pi_g}(r)$ to be either positive or negative, so our parameter sets should include ones with both $R_{\Pi_g} > R_{\Sigma_g^+}$ and $R_{\Pi_g} < R_{\Sigma_g^+}$.
We choose $R_{\Pi_g}$ and $R_{\Sigma_g^+}$ to be either near $R_{3^\ast}$ and $R_6$ or near $R_6$ and $R_{3^\ast}$.
Our choices for the two sets of relaxation lengths are
\begin{equation}
(R_{\Pi_g},R_{\Sigma_g^+}) =  (0.25~\text{fm}, 0.50~\text{fm}) ~\text{or}~ ( 0.50~\text{fm}, 0.25~\text{fm}),
\qquad R_{\Sigma_u^-} =  0.25~\text{fm} .
\label{RSigma,RPi}
\end{equation}
For the first and second set of relaxation lengths, the deepest potential is $1^{--}\Sigma_g^+$ and $1^{--}\Pi_g$, respectively.
The $0^{-+}$ adjoint-meson energy in Eq.~\eqref{E80-+} is $-58.4$~MeV. 

\begin{figure}[t]
\centerline{ \includegraphics[width=12cm,clip=true]{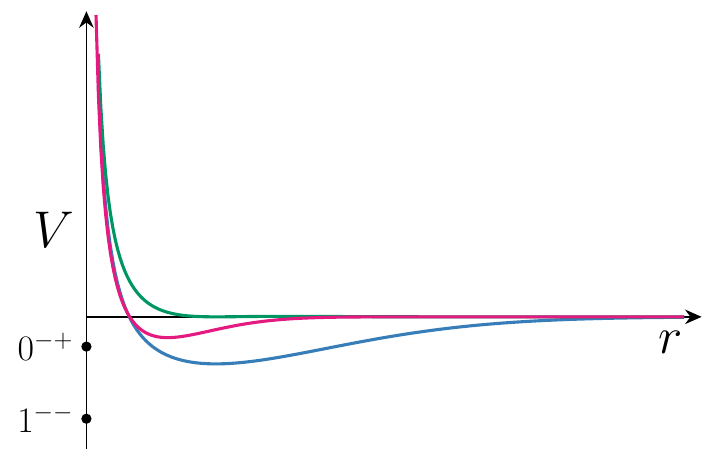} }
\caption{
Qualitative illustration of our models for the $1^{--}\Pi_g$,  $1^{--}\Sigma_g^+$, and $0^{-+}\Sigma_u^-$ potentials. 
The $0^{-+}\Sigma_u^-$ potential remains above the triplet-hadron-pair threshold. 
The two successively deeper potentials are either $1^{--}\Pi_g$ and  $1^{--}\Sigma_g^+$ or $1^{--}\Sigma_g^+$ and $1^{--}\Pi_g$.
The dots on the vertical axis are the energies of the $1^{--}$ and $0^{-+}$ adjoint mesons.
}
\label{fig:TetraquarkPotentials}
\end{figure}

Our two models for the $1^{--}\Pi_g$, $1^{--}\Sigma_g^+$, and $0^{-+}\Sigma_u^-$ potentials are Eq.~\eqref{bopots} with the two sets of relaxation lengths in Eq.~\eqref{RSigma,RPi}, $\kappa_8$ in Eq.~\eqref{kappa8}, $E_{8,0^{-+}}$ in Eq.~\eqref{E80-+}, and the adjustable adjoint-meson energy $E_{8,1^{--}}$.
A qualitative illustration of the three potentials is shown in Fig.~\ref{fig:TetraquarkPotentials}. 
The adjoint-meson energy $E_{8,j^{\pi\gamma}}$ can be obtained from the potential in Eq.~\eqref{bopots} by subtracting $\kappa_8/r$ and taking the limit as $r \to 0$.
The energies of the $1^{--}$ and $0^{-+}$ adjoint mesons are indicated in Fig.~\ref{fig:TetraquarkPotentials} by dots on the vertical axes.

\subsection{Spin-Dependent Potential}
\label{sec:SpinDependentPotentials}

The $J^{PC}$ block of the radial diabatic Hamiltonian with spin-splittings and kinetic improvement is given in Eq.~\eqref{H1-JPC}.
Our approximation for the spin-dependent terms in the $1/m_Q$ diabatic potential $\mathbf{V}_1$ is the spin-splitting potential $\mathbf{V}_{\text{SS}} (r, \bm{S}_1,\bm{S}_2)$ in Eq.~\eqref{VSS-r}.
It interpolates between the short-distance spin-splitting term $\mathbf{V}_{\text{SS},\text{short}} (\bm{S}_1+\bm{S}_2)$ at small $r$ and the long-distance spin-splitting term $ \mathbf{V}_{\text{SS},\text{long}} (\bm{S}_1,\bm{S}_2)$ at large $r$ using Gaussian relaxation factors with relaxation length $R_\text{SS}$.
We choose $R_\text{SS}$ to be near the Sommer scale $r_0$, which is roughly where the quarkonium potential $V_{0^{++}\Sigma_g^+}(r)$ crosses over from the attractive color-Coulomb potential to the repulsive linear potential. 
The Sommer scale is defined to be the radius $r_0$ where $r^2 dV_{0^{++}\Sigma_g^+}(r)/dr$ has the numerical value 1.65 \cite{Somm93}. 
For the Cornell potential in Eq.~\eqref{V-Cornell}, the Sommer scale is $r_0 = \sqrt{(1.65+\kappa_1)/\sigma}$. 
Inserting the values of the parameters for the Cornell potential in Eq.~\eqref{Cornellparams}, we obtain $r_0 =0.492$~fm. 
Our choice for the spin-splitting relaxation length is
\begin{equation}
R_\text{SS}= 0.50~\text{fm}.
\label{RSS}
\end{equation}

The $J$ block of the long-distance spin-splitting term $\mathbf{V}_{\text{SS},\text{long}}$ in the radial dimeson basis is given in Eq.~\eqref{VSSlong-J}.
It is determined by the spin splitting $\Delta_Q$ between the $S$-wave heavy mesons with $J^P=1^-$ and $0^-$.
We choose $\Delta_c$ and $\Delta_b$ to be the isospin-averaged $D^*$-$D$ and $B^*$-$B$ mass splittings: 
\begin{equation}
\Delta_c = 141.30\,\mathrm{MeV}, \qquad \Delta_b = 45.18\,\mathrm{MeV} .
\label{Deltac,Deltab}
\end{equation}
The ratio $\Delta_c/\Delta_b = 3.13$ is compatible with the ratio of the quark masses in Eq.~\eqref{mc,mb}.

The $J$ block of the short-distance spin-splitting term $\mathbf{V}_{\text{SS},\text{short}}$ in the radial dimeson basis  is given in Eq.~\eqref{VSSshort-J}.
It is determined by the spin-splitting $\Delta_{8,Q}$ for a heavy adjoint meson with quantum numbers $1^{--}$.
There are no lattice QCD calculations of this parameter.
The spin splitting $\Delta_{8,Q}$ is expressed in Eq.~\eqref{Delta8Q} in terms of the coefficient $C_{8,1^{--}}$ in the spin-splitting Hamiltonian for a heavy adjoint hadron with mass $2 m_Q$ in Eq.~\eqref{HSS-short}. 
To obtain estimates for $\Delta_{8,Q}$, we will assume that $C_{8,1^{--}}$ is comparable to analogous coefficients in the spin-splitting Hamiltonian for other heavy hadrons.
The spin splitting $\Delta_Q$ of $S$-wave heavy mesons is determined by the coefficient $C_{3,\nicefrac{1}{2}^-}$ in the spin-splitting Hamiltonian in Eq.~\eqref{HSS-meson} and is given in Eq.~\eqref{DeltaQ}. 
If we assume $C_{8,1^{--}} \approx C_{3,\nicefrac{1}{2}^-}$, we obtain $\Delta_{8,Q} \approx 3 \Delta_Q/2$. 
Given the spin splittings $\Delta_c$ and $\Delta_b$ in Eq.~\eqref{Deltac,Deltab}, our estimates for $\Delta_{8,Q}$ in the $c \bar{c}$ and $b \bar{b}$ sectors are 212~MeV and 68~MeV. 

We can also obtain an estimate for $\Delta_{8,Q}$ from the spin splittings of heavy baryons.
The light-QCD states in the lowest-energy heavy baryons can be approximated by adjoint baryons with $j^\pi$ quantum numbers  $0^+$ or $1^+$.
The heavy baryons with $j^\pi = 0^+$ are HQSS singlets. 
The heavy baryons with $j^\pi = 1^+$ form HQSS doublets with $J^P=\frac{1}{2}^+$ and $\frac{3}{2}^+$.
The spin splitting $\Delta_Q$ between the $\tfrac{3}{2}^+$ and $\tfrac{1}{2}^+$ heavy baryons is determined by the coefficient $C_{3,1^+}$ in the spin-splitting Hamiltonian in Eq.~\eqref{HSS-meson} and is given by Eq.~\eqref{ESS-JP}: $\Delta_Q = (3/2)C_{3,1^+}/m_Q$. 
The spin splitting $\Delta_{8,Q}$  for $1^{--}$ adjoint baryons is expressed in terms of $C_{8,1^{--}}$ in Eq.~\eqref{Delta8Q}.
If we assume $C_{8,1^{--}} \approx C_{3,1^+}$, we obtain $\Delta_{8,Q} \approx \Delta_Q$. 
The spin splittings between the $\tfrac{3}{2}^+$ and $\tfrac{1}{2}^+$ heavy baryons is $\Delta_c = 66.5$~MeV for charm baryons and $\Delta_b = 19.2$~MeV for bottom baryons. 
The ratio $\Delta_c /\Delta_b = 3.46$ is compatible with the ratio of the quark masses in Eq.~\eqref{mc,mb}.
We will use our estimates of $\Delta_{8,Q}$ from heavy-meson spin splittings and from heavy-baryon spin splittings as two alternative estimates for 
$\Delta_{8,c}$ and $\Delta_{8,b}$:
\begin{equation}
(\Delta_{8,c},\Delta_{8,b}) = (210\,\mathrm{MeV} ,  70\,\mathrm{MeV}) ~\mathrm{or}~ (70\,\mathrm{MeV}, 20\,\mathrm{MeV}) .
\label{Delta8Q-num}
\end{equation}

\begin{figure}[t]
\centerline{ \includegraphics[width=10cm,clip=true]{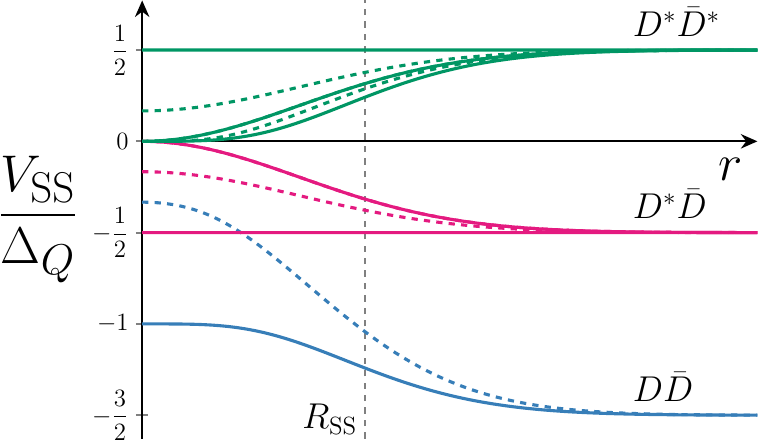}}
\caption{
The eigenvalues of the spin-splitting potential $\mathbf{V}_{\text{SS}}$ as functions of $r$ for $\Delta_{8,Q}/ \Delta_Q \approx 3/2$ (solid curves) or $\Delta_{8,Q}/ \Delta_Q \approx 1/2$ (dashed curves).
The horizontal axis is the spin-weighted $D^{(*)} \bar{D}^{(*)}$ threshold.
Two of the dashed curves that connect the $D^{(*)} \bar{D}^{(*)}$ threshold at small $r$ to the $D^* \bar{D}^*$ and $D^* \bar{D}$ thresholds at large $r$ are not visible because they are covered up by solid curves.
}
\label{fig:SpinDependentPotentials}
\end{figure}

The eigenvalues of the spin-splitting potential $\mathbf{V}_{\text{SS}}$ in Eq.~\eqref{VSS-r} are shown as functions of $r$ in Fig.~\ref{fig:SpinDependentPotentials} for $\Delta_c = 141.3$~MeV and for $\Delta_{8,c}  = 210$~MeV and 70~MeV
At small $r$, the spin splittings are those between the four distinct energy levels of the $1^{--}$ adjoint meson, which are described after Eq.~\eqref{VSSshort-J}.
At large $r$, the spin splittings  are those between the three distinct thresholds for pairs of $S$-wave heavy mesons. 
The six eigenvalues connect smoothly in the intermediate $r$ region.
Only four dashed curves for $\Delta_{8,c}  = 70$~MeV are visible in Fig.~\ref{fig:SpinDependentPotentials}, because the other two are covered up by solid curves for $\Delta_{8,c} = 210$~MeV.
The eigenvalues of $\mathbf{V}_{\text{SS}}$ for $\Delta_{8,c}  = 70$~MeV and 210~MeV approach each other at large $r$ but those for $\Delta_{8,c}  = 70$~MeV are smaller at small $r$ by a factor of 3.
For $\Delta_{8,c} = 210$~MeV, there are two potentials that appear to be horizontal because of the approximate equalities between $\pm\tfrac{1}{2}\Delta_c$ and $\pm\tfrac{1}{3}\Delta_{8,c}$.

\subsection{Kinetic Improvement}
\label{sec:MesonMasses}

The kinetically improved mass matrix $\mathbf{M}_{J^{PC}}$ in Eq.~\eqref{H1-JPC} is a diagonal matrix.
The diagonal entry of $\mathbf{M}_{J^{PC}}$ for a radial quarkonium channel is the heavy-quark mass $m_Q$.
Our values for the charm-quark and bottom-quark masses are given in Eq.~\eqref{mc,mb}.
The diagonal entry for a radial tetraquark channel that approaches the threshold for a pair of heavy mesons at large $r$ is twice the reduced mass of the heavy mesons.
The spin-weighted isospin-averaged masses of the $S$-wave charm mesons and the $S$-wave bottom mesons are
\begin{equation}
M_{D^{(*)}}=1973.23\,\mathrm{MeV}, \qquad M_{B^{(*)}}= 5313.45\,\mathrm{MeV}.
\label{MD(star),MB(star)}
\end{equation}
The masses of the $S$-wave charm mesons are 
\begin{equation}
M_D = M_{D^{(*)}} -\tfrac34 \Delta_c,  \qquad M_{D^*} = M_{D^{(*)}} + \tfrac14 \Delta_c, 
\label{MD,MDstar}
\end{equation}
where $\Delta_c$ is the $D^*-D$ splitting in Eq.~\eqref{Deltac,Deltab}.
The masses of the $S$-wave bottom mesons are determined in a similar way by $M_{B^{(*)}}$ in Eq.~\eqref{MD(star),MB(star)} and $\Delta_b$ in Eq.~\eqref{Deltac,Deltab}.

Since the mass of $\chi_{c1}(3872)$ is so close to the $D^{\ast0} \bar{D}^0$ threshold, it will be convenient to give the energies of $c \bar c$ states relative to the $D^*\bar{D}$ threshold at 3875.81~MeV. 
The $D \bar{D}$, $D^*\bar{D}$ and $D^*\bar{D}^*$ thresholds are then $-141.3$~MeV, 0, and $+141.3$~MeV.
We will give the energies of $b \bar b$ states relative to the $B^*\bar{B}$ threshold at 10604.31~MeV. 
The $B \bar{B}$, $B^*\bar{B}$ and $B^*\bar{B}^*$ thresholds are then $-45.2$~MeV, 0, and $+45.2$~MeV.

\subsection{Transition potential}
\label{sec:TransitionPotentials}

When our truncation of the diabatic potentials is extended to include the $0^{++}\Sigma_g^+$ quarkonium channel,  there is also a transition potential between $0^{++}\Sigma_g^+$ and $1^{--}\Sigma_g^+$. 
It is difficult to calculate diabatic potentials and transition potentials, such as $G_{(0^{++},1^{--})\Sigma_g^+}(r)$, directly using lattice QCD.
It is much easier to calculate adiabatic potentials.
The adiabatic potentials are the eigenvalues of the diabatic potential matrix.
Given a model for the diabatic potential matrix, the diabatic potentials and transition potentials can be determined by fitting the eigenvalues of the diabatic potential matrix to the calculated adiabatic potentials.

The adiabatic  isospin-0 $1\Sigma_g^+$, $2\Sigma_g^+$, and $3\Sigma_g^+$ potentials have been calculated most accurately in Ref.~\cite{Bul24} using lattice QCD with 2+1 light flavors of dynamical light quarks.
There is a narrow  avoided crossing between the $1\Sigma_g^+$ and $2\Sigma_g^+$ potentials at a radius near 1.2~fm and a narrower avoided crossing between the $2\Sigma_g^+$ and $3\Sigma_g^+$ potentials at a radius near 1.3~fm \cite{Bul24}.
The diabatic potentials and transition potentials were determined by fitting the three adiabatic potentials to the eigenvalues of a $3 \times 3$ diabatic potential matrix.  
The model for the diabatic potentials in the avoided-crossing region was the Cornell model for the $0^{++}\Sigma_g^+$ potential and constant energies for the $1^{--}\Sigma_g^+$ and $1^{-- \prime}\Sigma_g^+$ potentials.
The model for the transition potentials was a constant for the transition between $0^{++}\Sigma_g^+$ and $1^{--}\Sigma_g^+$, for the transition between $1^{--}\Sigma_g^+$ and $1^{-- \prime}\Sigma_g^+$, and 0 for the transition between $0^{++}\Sigma_g^+$ and $1^{-- \prime}\Sigma_g^+$. 

There are constraints  from BOEFT on the  transition potential  $G_{(0^{++},1^{--})\Sigma_g^+}(r)$ \cite{Ber24}.
They require it to go to zero as $r \to 0$ and as $r \to \infty$.
A simple model for the transition potential that satisfies these constraints and is consistent with the lattice QCD results in Ref.~\cite{Bul24} was introduced in Ref.~\cite{Bram24}:
\begin{equation}
G_{(0^{++},1^{--})\Sigma_g^+}(r) = \left\{ \begin{array}{lcr}
g\, r/r_1                  & r < r_1,                   \\
g                          & ~~~~~~r_1 < r < r_2, \\
g \,  e^{-(r-r_2)/r_0} & r > r_2.
\end{array} \right.
\label{gPi}
\end{equation}
This transition potential is continuous, it increases linearly below $r_1$, it is constant between $r_1$ and $r_2$, and it decreases exponentially beyond $r_2$.
The strength $g$ of the transition potential was chosen in Ref.~\cite{Bram24} to be the value in the range $r_1 < r < r_2$ calculated using lattice QCD in Ref.~\cite{Bul24}.
The relaxation length $r_0$ in the exponential in Eq.~\eqref{gPi} was chosen in Ref.~\cite{Bram24} to be the Sommer scale. 
The parameters in the transition potential in Ref.~\cite{Bram24} are
\begin{equation}
r_1 = 0.95~\text{fm} , \qquad  r_2 =1.51~\text{fm}, \qquad g= 50~\text{MeV}, \qquad r_0 = 0.5~\text{fm}.
\label{r1,r2,g,r0}
\end{equation}

\subsection{Resonance Energies and Widths using SPARSE}
\label{sec:EnergiesWidths}

The eigenvalue equation for the radial diabatic Hamiltonian described in the previous subsections is the Schr\"odinger equation for hidden-heavy tetraquarks.
Solutions with discrete real energies $E_n$ below the lowest heavy-meson-pair threshold correspond to bound states.
The  solutions with continuous real energies $E$ above the lowest heavy-meson-pair threshold correspond to scattering states of pairs of heavy mesons.
The $T$-matrix $\mathbf{T}(E)$ for the scattering states can be analytically continued to complex energies $E$.
If there are $n$ heavy-meson-pair thresholds, there are $2^n$ sheets of $E$ with $n$ square-root branch cuts.
The physical sheet of complex $E$ includes the physical real-$E$ axis.
The part of that axis above the lowest heavy-meson-pair threshold is the lower boundary of the physical sheet.
There are branch cuts that begin at each of the heavy-meson-pair thresholds. 
A bound state corresponds to a pole of $\mathbf{T}(E)$ at a real energy $E_n$ below the lowest heavy-meson-pair threshold on the physical sheet. 
Bound state poles are the only poles on the physical sheet.
A virtual state corresponds to a pole of $\mathbf{T}(E)$ at a real energy $E_n$ below the lowest heavy-meson-pair threshold but on an unphysical sheet. 
A resonance with energy $E_n$ and decay width $\Gamma_n$ corresponds to a pole of $\mathbf{T}(E)$ at a complex energy $E_n-i\Gamma_n/2$ on the unphysical sheet that extends below the branch cut that is the lower boundary of the physical sheet. 

It is relatively easy to calculate the energy $E_n$ of a bound state with high precision by solving the Schr\"odinger equation numerically.
It is more difficult to determine the complex pole $E_n-i\Gamma_n/2$ of a resonance with high precision by calculating the $T$-matrix $\mathbf{T}(E)$ as a function of the complex energy with $2^n$ sheets.
If the $T$-matrix $\mathbf{T}(E)$ is known only as a function of the real energy $E$ on the physical sheet, the complex poles could be determined by fitting the $T$-matrix elements to analytic functions of $E$ with the appropriate branch cuts and the appropriate poles on the various sheets of $E$.
However, the resonance parameters $E_n$ and $\Gamma_n$ are sensitive to assumptions about the poles.

The $T$-matrix can be expressed in terms of the $K$-matrix $\mathbf{K}(E)$ as described in Appendix~\ref{sec:SPARSE}.
If there is a narrow resonance corresponding to a complex pole $E_n-i\Gamma_n/2$ of $\mathbf{T}(E)$, then $\mathbf{K}(E)$  has a corresponding real pole $E_n^\prime$ near the resonance energy $E_n$.
The residue of the real pole can be used to obtain the simple estimate $\Gamma_n^\prime$ in Eq.~\eqref{Gamma_n} for the resonance width $\Gamma_n$. 
The $K$-matrix parameters $(E_n^\prime,\Gamma_n^\prime)$ provide the simple approximation $E_n^\prime-i\Gamma_n^\prime/2$ for the resonance pole $E_n-i\Gamma_n/2$.
It is relatively easy to calculate the $K$-matrix parameters $(E_n^\prime,\Gamma_n^\prime)$ with high precision by solving the Schr\"odinger equation with real energy $E$.
It requires no assumptions about the analytic continuation of $\mathbf{T}(E)$ such as the complex poles on the various sheets of $E$.

 We calculate the $K$-matrix at real energies $E$ by solving the Schr\"odinger equation numerically using the SPARSE algorithm \cite{Brus25}.
The SPARSE algorithm uses the finite-difference method with Dirichlet boundary conditions to reduce the system of coupled Schr\"odinger equations to a linear system with a sparse matrix of coefficients.
The crudeness of the finite difference method is offset by the numerical efficiency of this simple approach, which allows one to solve the Schr\"odinger equations with dozens of coupled channels numerically using a personal computer.
The $K$-matrix parameters $(E_n^\prime,\Gamma_n^\prime)$ provide a good approximation to the resonance parameters if the resonance is narrow, if it does not overlap with other resonances, and if the smooth background under the pole has mild energy dependence.
In this paper, we give results only for the $K$-matrix parameters $(E_n^\prime,\Gamma_n^\prime)$ of a resonance, which provide an approximation to the corresponding $T$-matrix pole $E_n-i\Gamma_n/2$. 
We will not attempt to determine the complex pole more accurately by developing a model for the analytic continuation of $\mathbf{T}(E)$.
We will also not attempt to estimate any of the poles for virtual states.

\subsection{Alternative B-O Schr\"odinger Equations}
\label{sec:SchrodingerEquations}

The Schr\"odinger equations in the B\nobreakdash-O approximation for  $X_c$ and its HQSS partners have also been studied recently in Ref.~\cite{Bram26}.
The equations were given in both the radial tetraquark basis and  the radial dimeson basis.
We describe below the differences between the Schr\"odinger equations in Ref.~\cite{Bram26} and those in this paper.

In Ref.~\cite{Bram26}, the heavy-quark mass was set equal to the spin-weighted heavy-meson mass.
This improves the accuracy in tetraquark channels at long distances at the expense of the accuracy in quarkonium channels at short distances, resulting in unnecesarily large errors in quarkonium binding energies.
We avoided this problem by incorporating kinetic improvement, which allows for different masses in  tetraquark channels and quarkonium channels.

In Ref.~\cite{Bram26}, a single model for the $1^{--}\Sigma_g^+$ , $1^{--}\Pi_g$ , and $0^{-+}\Sigma_u^-$ was considered.
These potentials interpolate between the repulsive color-octet potential at short distances and the triplet-meson-pair threshold at large distances.
They decrease below the threshold and then approach it from below.
Our models for the $1^{--}\Sigma_g^+$ and $1^{--}\Pi_g$ potentials are qualitatively similar.
Our model for the $0^{-+}\Sigma_u^-$ potential remains above the threshold.
The depths of the $1^{--}\Sigma_g^+$ and $1^{--}\Pi_g$ potentials were controlled by the $1^{--}$ adjoint-meson energy.
In Ref.~\cite{Bram26}, it was adjusted to tune the $X_c$ energy to 100~keV below the $D^*\bar{D}$ threshold.
We will tune the  $1^{--}$ adjoint-meson energy to the critical value for $X_c$ to be a bound state at the $D^*\bar{D}$ threshold.
Both choices for the $X_c$ energy are consistent with the measured energy of $\chi_{c1}(3872)$ to within errors.

In Ref.~\cite{Bram26}, the spin-dependent potential at all $r$ was approximated by the long-distance spin-splitting term, whose strength is determined by the heavy-meson spin splitting.
Our model for the spin-splitting potential is an interpolation between the short-distance spin-splitting term at small $r$ and the long-distance spin-splitting term at large $r$.
The strength of the short-distance spin-splitting term is determined by the unknown spin splitting of the $1^{--}$ adjoint meson.

The Schr\"odinger equations in Ref.~\cite{Bram26} were solved numerically to determine the energies of bound states but not resonances or virtual states.
We solve the Schr\"odinger equations numerically to determine the energies of bound states and resonances but not virtual states.
The energies of the HQSS partners of $X_c$ were not calculated in Ref.~\cite{Bram26} because they are resonances or virtual states.
The energies of the $0^{++}$, $1^{++}$, and $1^{+-}$ members of the HQSS multiplet of $X_b$ were calculated in Ref.~\cite{Bram26}, but the energy of the $2^{++}$ state was not calculated because it is a resonance.


\section{Tetraquark Spin Splittings}
\label{sec:DeltaQDependence}

In this Section, we study the dependence of the hidden-heavy tetraquark spectrum on the heavy-meson spin splitting $\Delta_Q$.
We present simple predictions from first-order perturbation theory in $\Delta_Q$
and compare them with the nonperturbative  results from solving the Schr\"odinger equation.

\subsection{Perturbative Heavy-Meson Spin Splitting}
\label{sec:SpinSplitting-PT}

If the energy of a pair of heavy hadrons is sufficiently close to their scattering threshold, their scattering is dominated by the $S$-wave channels, which have $L_{Q \bar Q}=0$.
Their scattering wavefunctions are dominated by the long-distance region beyond the range of the static potentials.
The short-distance spin-splitting term $\mathbf{V}_{\text{SS},\text{short}}$ in Eq.~\eqref{HSS-short} can be ignored.
If the spin splitting $\Delta_Q$ between the heavy mesons is sufficiently small, the long-distance spin-splitting term $\mathbf{V}_{\text{SS},\text{long}}$  in Eq.~\eqref{HSS-long} can be treated as a first-order perturbation to the static Hamiltonian $\mathbf{H}_0$.

The distinct energy eigenvalues of the static Hamiltonian $\mathbf{H}_0$ can be labeled by $L^P$, where $L$ is the B\nobreakdash-O angular momentum and $P$ is the parity.
The degenerate states with that energy can be labeled by radial tetraquark channels $\bigl( (j^{\pi \gamma},L_{Q \bar Q} )L,S_{Q \bar Q} \bigr) J^{PC}$.
We have assumed the only diabatic potentials with a bound state are $1^{--}\Pi_g$ and $1^{--}\Sigma_g^+$. 
The $j^{\pi \gamma}=1^{--}$ ground state has $L^P=1^+$ and we denote its energy by $\varepsilon_1$.
It is sensitive to the energy $E_{8,1^{--}}$ of the $1^{--}$ adjoint meson.

The first-order perturbation to the energy of the $\bigl( (j^{\pi \gamma},L_{Q \bar Q} )L,S_{Q \bar Q} \bigr) J^{PC}$ state is the expectation value of $\mathbf{V}_{\text{SS},\text{long}}$ in that state.
The expressions for the matrix elements of $\mathbf{V}_{\text{SS},\text{long}}$ are simpler in the radial dimeson basis and are given in Eq.~\eqref{VSSlong-J}.
The recoupling formula expressing the radial dimeson channels in terms of the radial tetraquark channels is given in Eq.~\eqref{di-tetra12}.
If the channels are restricted to $S$-wave channels with $L_{Q \bar Q}=0$, the radial tetraquark channels $\bigl( (j^{\pi \gamma},0)L,S_{Q \bar Q} \bigr) J^{PC}$ in the $J^{PC}$ block reduce to $(j^{\pi \gamma},S_{Q \bar Q})J^{PC}$. 
The $6j$ symbol in Eq.~\eqref{di-tetra12} reduces to
\begin{equation}
\begin{Bmatrix}
j & S_{Q \bar Q} & S \\
J & 0  & L \end{Bmatrix}
=   \{ j, S_{Q \bar Q}, S  \}\, \delta(L,j)\, \delta(S,J)\,  (-1)^{j + S_{Q \bar Q} +S}\, \frac{1}{\sqrt{ \tilde{j} \tilde{S}}}\,  .
\label{6j-Swave}
\end{equation}
The inverse of the recoupling formula in Eq.~\eqref{di-tetra12} with  $j_1=j_2 = \tfrac12$ and $L_{Q \bar Q} = 0$ reduces to
\begin{equation}
\vket{(j, S_{Q \bar Q})}{J}{M_J}{} =
(-1)\, \sqrt{ \tilde{j}\, \tilde{S}_{Q \bar Q} }
\sum_{J_1J_2} \sqrt{ \tilde{J}_1 \tilde{J}_2 }
\begin{Bmatrix}
\tfrac12          & \tfrac12                     & J_1 \\
\tfrac12          & \tfrac12                     & J_2 \\
j & S_{Q \bar Q} & J \end{Bmatrix}
\vket{(J_1,J_2)}{J}{M_J}{}.
\label{changebasis12-Swave}
\end{equation}
The matrix elements for the $J$ block of $\mathbf{V}_{\text{SS},\text{long}}$ between $S$-wave radial tetraquark states are
\begin{multline}
\Big( \mathbf{V}^J_{\text{SS},\text{long}} (\bm{S}_1,\bm{S}_2)  \Big)_{(j, S_{Q \bar Q}),(j^\prime, S_{Q \bar Q}^\prime)}
= \sqrt{ \tilde{j}\, \tilde{S}_{Q \bar Q}\,   \tilde{j}^\prime\, \tilde{S}_{Q \bar Q}^\prime }
\sum_{J_1J_2} \tilde{J}_1\, \tilde{J}_2 \,
\\
\times
\begin{Bmatrix}
\tfrac12          & \tfrac12                     & J_1 \\
\tfrac12          & \tfrac12                     & J_2 \\
j & S_{Q \bar Q} & J \end{Bmatrix}
\begin{Bmatrix}
\tfrac12          & \tfrac12                     & J_1 \\
\tfrac12          & \tfrac12                     & J_2 \\
j ^\prime & S_{Q \bar Q}^\prime & J \end{Bmatrix}
\frac{1}{2}   \big[ J_1(J_1+1) + J_2 (J_2+1) - 3 \big] \Delta_Q.
\label{<VSS>}
\end{multline}
The expectation value of $\mathbf{V}_{\text{SS},\text{long}}$ in the channel $(j, S_{Q \bar Q})$ is the diagonal entry with $(j^\prime, S_{Q \bar Q}^\prime) = (j, S_{Q \bar Q})$.

Our truncation of the diabatic potentials to $1^{--}\Pi_g$ and $1^{--}\Sigma_g^+$ implies that $j^{\pi \gamma} = 1^{--}$.
Given that $L_{Q \bar Q}=0$, the parity constraint in Eq.~\eqref{P-total} implies that the total parity is $P=+$.
We obtain the $J^{PC}$ block of $\mathbf{V}_{\text{SS},\text{long}}$ from the $J$ block in Eq.~\eqref{<VSS>} by setting $j = j^\prime =1$ and restricting $S_{Q \bar Q}$ to be either 0 or 1 in accord with the constraint $CP=(-1)^{S_{Q \bar Q}+1}$ from Eq.~\eqref{CP-total}.
The energy of the $j^{\pi \gamma}=1^{--}$ ground state with $L^P = 1^+$ is $\varepsilon_1$.
The spin splitting of the $J^{PC}$ state with $P=+$ and $C=(-1)^{S_{Q \bar Q}+1}$ from first-order perturbation theory in $\mathbf{V}_{\text{SS},\text{long}}$ is the diagonal entry of the $J$ block in Eq.~\eqref{<VSS>} with $j=1$.
The HQSS multiplet consists of spin-triplet states with $J^{PC}=0^{++}$, $1^{++}$, and $2^{++}$ and a spin-singlet $1^{+-}$ state.
The energies of the four $J^{PC}$ states from first-order perturbation theory in $\mathbf{V}_{\text{SS},\text{long}}$ are \cite{Braa24b}
\begin{equation}
E_{0^{++}} = \varepsilon_1 -  \Delta_Q, \qquad E_{1^{++}} = \varepsilon_1 - \tfrac12 \Delta_Q, 
\qquad E_{1^{+-}} = \varepsilon_1, \qquad E_{2^{++}} = \varepsilon_1 + \tfrac12 \Delta_Q.
\label{EJPC-triplet}
\end{equation}
These spin splittings can be understood from the amplitudes for $J_1$ and $J_2$ in Eq.~\eqref{changebasis12-Swave}, which give the particle content of each state.
The spin splittings of the $D \bar{D}$, $D^* \bar{D}$, and $D^* \bar{D}^*$ thresholds are $-\tfrac32 \Delta_Q$, $-\tfrac12 \Delta_Q$, and $+\tfrac12 \Delta_Q$, respectively.
The $S$-wave component of the $0^{++}$ state is the superposition of $D \bar{D}$ and $D^* \bar{D}^*$ with amplitudes $-\sqrt{3/4}$ and $1/2$, so its spin splitting is the corresponding weighted average of the spin splittings of the $D \bar{D}$ and $D^* \bar{D}^*$ thresholds, which is $-\Delta_Q$. 
The $S$-wave component of the $1^{++}$ state is $-(D^* \bar{D} + D \bar{D}^*)/\sqrt{2}$, so its spin splitting is that of the $D^* \bar{D}$ threshold.
The $S$-wave component of the $1^{+-}$ state is the sum of $-(D^* \bar{D} - D \bar{D}^*)/2$ and $-D^* \bar{D}^*/\sqrt{2}$, 
so its spin splitting is the average of the spin splittings of the $D^* \bar{D}$ and $D^* \bar{D}^*$ thresholds, which is 0.  
The $S$-wave component of the $2^{++}$ state is $-D^* \bar{D}^*$, so its spin splitting is that of the $D^* \bar{D}^*$ threshold.

We can use the results from first-order perturbation theory in $\mathbf{V}_{\text{SS},\text{long}}$ to make simple predictions for the spectrum of the $J^{PC}$ states as functions of $\Delta_Q$.
We use basic facts about low-energy scattering theory. 
A bound state corresponds to a pole in the $T$-matrix $T(E)$ on the physical real-$E$ axis below the threshold for its constituents.
When a bound state reaches a threshold to which it has an $S$-wave coupling, the universality of near-threshold $S$-wave resonances implies that it becomes a virtual state \cite{Braa04}. 
The pole in the $T$-matrix  then decreases below the threshold along the real-$E$ axis of an unphysical sheet. 
The square $|T(E)|^2$ of the $T$-matrix element is large in the threshold region with a local maximum at the threshold, but it has no peak that can be associated with a resonance.
There can also be other $S$-wave poles outside the universal region of the complex energy $E$.
In particular, there can be a pole in the lower half-plane of the unphysical sheet that becomes a resonance when the real part of its energy increases to above the threshold.
When a bound state crosses a threshold to which it has a $P$-wave or higher-wave coupling, it could become a narrow resonance.
If it becomes a narrow resonance, the pole in the $T$-matrix moves to below the branch cut on the unphysical sheet of the complex energy. 
There is a narrow peak in $|T(E)|^2$ at an energy near the real part of the pole.

To predict the behavior of a state when its energy reaches a heavy-meson-pair threshold, we need to know the lowest orbital angular momentum $L_{Q \bar{Q}}$ for its couplings to heavy-meson pairs.
The radial dimeson channels for $1^{++}$ are enumerated in Section~\ref{sec:1++Channeldimeson}.
The $1^{++}$ state has an $S$-wave coupling to $(D^* \bar D + D \bar D^*) / \sqrt2$ and a $D$-wave coupling to $D^* \bar D^*$.
The radial dimeson channels for other $J^{PC}$ are enumerated in Appendix~\ref{app:DiabaticH}.
The $0^{++}$ state has $S$-wave couplings to $D \bar D$ and $D^* \bar D^*$.
The $1^{+-}$ state has $S$-wave couplings to $(D^* \bar D - D \bar D^*)/\sqrt2$ and $D^* \bar D^*$.
The $2^{++}$ state has an $S$-wave coupling to $D^* \bar D^*$ and $D$-wave couplings to $D \bar D$ and $(D^* \bar D + D \bar D^*)/\sqrt2$.

To illustrate the perturbative predictions, it is convenient to use a small nonzero value of the energy $\varepsilon_1$ of the $j^{\pi\gamma} = 1^{--}$ ground state.
The energies of the four $J^{PC}$ states from first-order perturbation theory in the spin-splitting potential $\mathbf{V}_{\text{SS}}$  in Eq.~\eqref{DeltaE-SS} are
\begin{subequations}
\begin{eqnarray}
E_{0^{++}} &=& \varepsilon_1  - P_{1,\text{long}}\, \Delta_Q -  \tfrac23 (1-P_{1,\text{long}})\, \Delta_{8,Q}, 
\\
E_{1^{++}} &=& \varepsilon_1 - \tfrac12  P_{1,\text{long}}\,  \Delta_Q - \tfrac13 (1-P_{1,\text{long}})\, \Delta_{8,Q},  
\\
E_{1^{+-}} &=&  \varepsilon_1 ,
\\
E_{2^{++}} &=& \varepsilon_1 + \tfrac12 P_{1,\text{long}}\, \Delta_Q + \tfrac13  (1-P_{1,\text{long}}) \Delta_{8,Q},
\end{eqnarray}
\label{EJPC-longshort}%
\end{subequations}
where $P_{1,\text{long}}$ is the long-distance probability for the $1^{--}$ ground state defined in Eq.~\eqref{Plong}, which depends on $\varepsilon_1$.
Note that for the specific value $\Delta_{8,Q}= \tfrac{3}{2}\Delta_Q$ of the adjoint-meson energy, the dependence of the energies in Eq.~\eqref{EJPC-longshort} on $P_{1,\text{long}}$ cancels and the energies reduce to those in Eq.~\eqref{EJPC-triplet} for the limit $\varepsilon_1 \to 0$, which implies $P_{1,\text{long}} \to 1$.

\begin{figure}[t]
\centerline{ \includegraphics[width=12cm,clip=true]{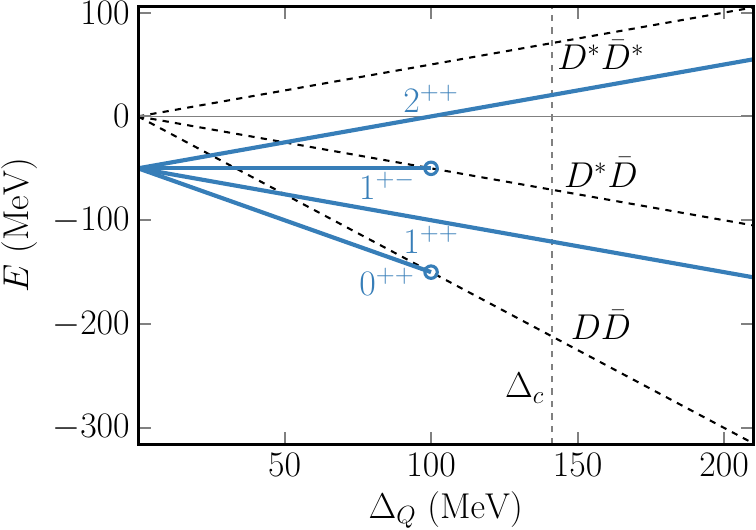}  }
\caption{
Energies of isospin-0 hidden-heavy tetraquarks as functions of the heavy-meson spin splitting $\Delta_Q$ from first-order perturbation theory in the spin-splitting potential $\mathbf{V}_{\text{SS}}$.
The energy of the $j^{\pi \gamma} = 1^{--}$ ground-state with $L^P=1^+$  relative to the $D^{(*)} \bar{D}^{(*)}$ threshold (thin horizontal solid line) is chosen to be $\varepsilon_1 = -50$~MeV.
The four solid lines are the energies of the $0^{++}$, $1^{++}$, $1^{+-}$, and $2^{++}$ states.  
The three dashed lines are the $D \bar{D}$, $D^* \bar{D}$, and $D^* \bar{D}^*$ thresholds.
The $0^{++}$ and $1^{+-}$ states are expected to become virtual states at the open dots.
The vertical dashed line is the charm-meson spin splitting $\Delta_c$.
}
\label{fig:SpectrumvsDelta:1OPT}
\end{figure}

We choose the $j^{\pi \gamma} = 1^{--}$ ground-state energy relative to the $D^{(*)} \bar{D}^{(*)}$ threshold to be $\varepsilon_1 = -50$~MeV.
The long-distance probability is $P_{1,\text{long}} = 0.501$.
We also choose $\Delta_{8,Q}= \tfrac{3}{2}\Delta_Q$, so the energies at first order in the spin splittings in Eqs.~\eqref{EJPC-longshort} reduce to those in Eqs.~\eqref{EJPC-triplet}.
The energies of the four $J^{PC}$ states as functions of $\Delta_Q$ are shown in Fig.~\ref{fig:SpectrumvsDelta:1OPT}.
At $\Delta_Q=0$, the $0^{++}$, $1^{++}$, $1^{+-}$, and $2^{++}$ states are all $S$-wave bound states with the same energy $\varepsilon_1$. 
The $0^{++}$ state reaches the $D \bar{D}$ threshold at $\Delta_Q = 2 |\varepsilon_1|$.  
It has an $S$-wave coupling to $D \bar{D}$, so it becomes a virtual state whose energy then decreases below the threshold.
The $1^{++}$ state crosses the $D \bar{D}$ threshold, but it has no couplings to $D \bar{D}$ so it remains a bound state.
The $1^{+-}$ state crosses the $D \bar{D}$ threshold, but it has no couplings to $D \bar{D}$ so it remains a bound state.
It reaches the $D^* \bar{D}$ threshold at $\Delta_Q = 2 |\varepsilon_1|$. 
It has $S$-wave couplings to $D^* \bar{D}$ and $D \bar{D}^*$, so it becomes a virtual state. 
The $2^{++}$ state crosses the $D \bar{D}$ threshold at $\Delta_Q = \tfrac{1}{2} |\varepsilon_1|$. 
It has $D$-wave couplings to $D \bar{D}$, so it becomes a narrow resonance.
It crosses the $D^* \bar{D}$ threshold at $\Delta_Q = |\varepsilon_1|$.  
It has $D$-wave couplings to $D^* \bar{D}$ and $D \bar{D}^*$, so it gains these additional decay channels. 
Note that for $\Delta_Q > 2 |\varepsilon_1|$, the spectrum in $\Delta_Q$ consists of only two states:  a $1^{++}$ bound state with the spin splitting of the $D^* \bar{D}$ threshold and a $2^{++}$ resonance with the spin splitting of the $D^* \bar{D}^*$ threshold.

We now describe the perturbative prediction for the spectrum as a function of $\Delta_Q$  in the limit $\varepsilon_1 \to 0$ from below.
At $\Delta_Q=0$, the $0^{++}$, $1^{++}$, $1^{+-}$, and $2^{++}$ states are all $S$-wave bound states at the threshold. 
As $\Delta_Q$ increases to just above 0,  the $0^{++}$ and $1^{+-}$ states become virtual states, but the $1^{++}$ state remains a bound state and the $2^{++}$ state becomes a narrow resonance.
At larger $\Delta_Q$, the perturbative approximation to the spectrum consists of  a $1^{++}$ bound state just below the $D^* \bar{D}$ threshold and a narrow $2^{++}$ resonance just below the $D^* \bar{D}^*$ threshold. 

\subsection{Nonperturbative Heavy-Meson Spin Splitting}
\label{sec:SpinSplitting-NPT}

The dependence of the tetraquark spectrum on the heavy-meson spin splitting can be calculated nonperturbatively by solving a Schr\"odinger equation that includes a spin-dependent potential that approaches the long-distance spin-splitting term $\mathbf{V}_{\text{SS},\text{long}}(\bm{S}_1,\bm{S}_2)$ in Eq.~\eqref{HSS-long} at large $r$.
We illustrate the nonperturbative dependence on the heavy-meson spin splitting $\Delta_Q$ by solving the diabatic Schr\"odinger equation for $c \bar c$ tetraquarks with the $1^{--} \Pi_g$, $1^{--}  \Sigma_g^+$, and $0^{-+} \Sigma_u^-$ potentials and the spin-splitting potential in Eq.~\eqref{VSS-r}.
We choose one of the four models for the diabatic potentials described in Section~\ref{sec:Models}.
Our model for the $1^{--} \Pi_g$, $1^{--}  \Sigma_g^+$, and $0^{-+} \Sigma_u^-$ potentials is determined by the first set of relaxation lengths in Eq.~\eqref{RSigma,RPi} with $(R_{\Pi_g},R_{\Sigma_g^+}) =  (0.50,0.25)$~fm.
We make simple changes in the model so that it coincides with the model when $\Delta_Q = \Delta_c$ but it reduces to the model without any spin splittings when $\Delta_Q=0$.
We take the adjoint-meson spin splitting to be $\Delta_{8,Q}= \tfrac{3}{2}\Delta_Q$. 
This is equal to  0 when $\Delta_Q=0$ and it is approximately equal to the first value of $\Delta_{8,c}$  in Eq.~\eqref{Delta8Q-num} if $\Delta_Q = 141.3$~MeV. 
The case of $c \bar c$ tetraquarks is considered by taking the charm-meson masses in the kinetically improved mass matrices to be
\begin{equation}
m_D = M_{D^{(*)}} - \tfrac{3}{4}\Delta_Q,  \qquad m_{D^*} =  M_{D^{(*)}} + \tfrac{1}{4}\Delta_Q,
\label{mD,mD*-Delta}
\end{equation}
where $M_{D^{(*)}}$ is the spin weighted charm-meson mass in Eq.~\eqref{MD(star),MB(star)} and $\Delta_Q$ is the variable heavy-meson spin splitting.
These both reduce to the spin-weighted charm-meson mass if $\Delta_Q = 0$ and they are equal to the physical charm-meson masses if $\Delta_Q$ is the physical charm-meson mass splitting $\Delta_c$ in Eq.~\eqref{Deltac,Deltab}.
The other parameters in the radial diabatic Hamiltonians are as described in Section~\ref{sec:Models}.

We calculate the dependence of the tetraquark energies on $\Delta_Q$ given a specified value $\varepsilon_1$ of the $j^{\pi\gamma} = 1^{--}$ ground-state energy at $\Delta_Q = 0$.
We first set $\Delta_Q = 0$ and tune the energy $E_{8,1^{--}}$ of the $1^{--}$ adjoint meson so $\varepsilon_1$ has the specified value. 
We then keep $E_{8,1^{--}}$ fixed and we calculate the energies of bound states and the $K$-matrix energies and widths of resonances as  functions of $\Delta_Q$ using the SPARSE algorithm \cite{Brus25}.
The dependence on $\Delta_Q$ comes primarily from the spin-splitting potential $\mathbf{V}_{\text{SS}}$.
There is also some dependence on $\Delta_Q$ from the kinetic improvement of the kinetic energy through the terms proportional to $\Delta_Q$ in the heavy-meson masses as in Eq.~\eqref{MD,MDstar}.
Its contribution to an energy at first order in $\Delta_Q$ is proportional to the expectation value $\langle p^2 \rangle$ in the $1^{--}$ ground state.

\begin{figure}[t]
\centerline{ \includegraphics[width=12cm,clip=true]{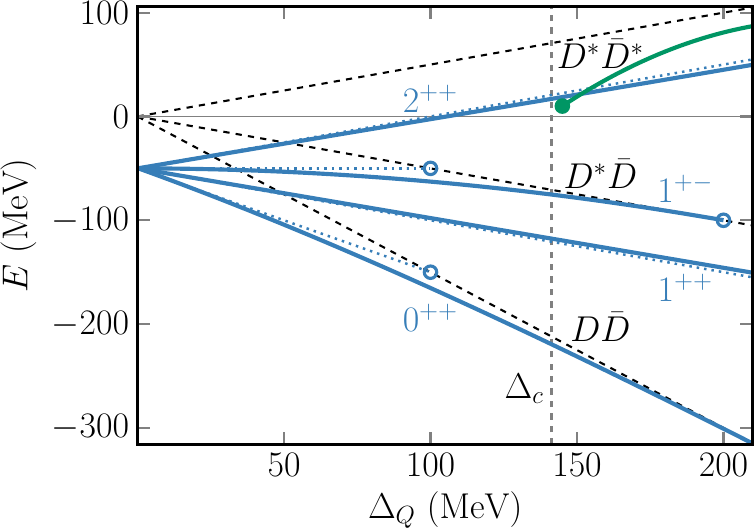} }
\caption{
Energies of isospin-0 $c \bar c$ tetraquarks as functions of the heavy-meson spin-splitting $\Delta_Q$ with $\Delta_{8,Q}= \tfrac{3}{2}\Delta_Q$ and $(R_{\Pi_g},R_{\Sigma_g^+}) =  (0.50,0.25)$~fm.  
The dependence on $\Delta_Q$ is calculated nonperturbatively by solving the Schr\"odinger equation.
The adjoint-meson energy $E_{8,1^{--}}$ is tuned so that the $j^{\pi \gamma} = 1^{--}$ ground-state energy at $\Delta_Q=0$ relative to the $D^{(*)} \bar{D}^{(*)}$ threshold (thin horizontal solid line) is $\varepsilon_1= -50$~MeV.
The solid curves are the energies of the $0^{++}$, $1^{++}$, $1^{+-}$, and $2^{++}$ states.
The dashed lines are the $D \bar{D}$, $D^* \bar{D}$, and $D^* \bar{D}^*$ thresholds.
The dotted lines are the predictions from first-order perturbation theory in $\Delta_Q$ in Fig.~\ref{fig:SpectrumvsDelta:1OPT}.
The vertical dashed line is the  charm-meson spin splitting $\Delta_c$.
An open dot indicates the disappearance of a bound state.
The solid dot indicates the appearance of a $1^{+-}$ resonance.
}
\label{fig:SpectrumvsDelta:SchrEq}
\end{figure}

To illustrate the nonperturbative predictions, we choose the $j^{\pi\gamma} = 1^{--}$ ground-state energy relative to the $D^{(*)} \bar{D}^{(*)}$ threshold  to be $\varepsilon_1 = -50$~MeV, which requires the $1^{--}$ adjoint-meson energy to be $-602$~MeV.
The long-distance probability for the $1^{--}$ ground state defined in Eq.~\eqref{Plong} is $P_{1,\text{long}} = 0.501$. 
The energies of $c \bar c$ tetraquarks with each of the four values of $J^{PC}$ are shown as functions of $\Delta_Q$  in Fig.~\ref{fig:SpectrumvsDelta:SchrEq}.
The straight dotted lines are the predictions from first-order perturbation theory in the spin splittings in Fig.~\ref{fig:SpectrumvsDelta:1OPT}.
The curves are the results from treating $\Delta_Q$ nonperturbatively by solving the Schr\"odinger equation.
The curves for the $1^{++}$ and $2^{++}$ states are close to the predictions from first-order perturbation theory, with the $2^{++}$ curve just below the prediction and the $1^{++}$ curve just above the prediction.
The energies of the $0^{++}$ and $1^{+-}$ states have obvious curvature.
However, their deviations from the extrapolations of the straight dotted lines are small enough that first-order perturbation theory remains a good approximation even for $\Delta_Q$ as large as $\Delta_c$.
The $0^{++}$ and $1^{+-}$ states remain bound states below the $D\bar{D}$ and $D^*\bar{D}$ thresholds out to much larger values of $\Delta_Q$ than predicted by first-order perturbation theory.
The $0^{++}$ bound state reaches the $D\bar{D}$ threshold and disappears at $\Delta_Q=223$~MeV. 
The $1^{+-}$ bound state reaches the $D^*\bar{D}$ threshold and disappears at $\Delta_Q=200$~MeV. 
The $2^{++}$ state becomes a narrow resonance after it crosses the $D \bar D$ threshold at $\Delta_Q=23$~MeV.
The width of the $2^{++}$ resonance is not shown in Fig.~\ref{fig:SpectrumvsDelta:SchrEq} but it is very small, compatible with $D$-wave couplings to $D \bar D$ and $(D^* \bar D + D \bar D^*)/\sqrt2$.
The $K$-matrix width increases from 0 at $\Delta_Q=23$~MeV to 0.04~MeV when it crosses the $D^*\bar{D}$ threshold at $\Delta_Q=51.2$~MeV and to 1.4~MeV at $\Delta_Q=\Delta_c$.

In addition to the $0^{++}$, $1^{++}$, $1^{+-}$, and $2^{++}$ states expected from perturbation theory in $\Delta_Q$, there is an additional $1^{+-}$ resonance in Fig.~\ref{fig:SpectrumvsDelta:SchrEq}.
Its $K$-matrix pole appears when $\Delta_Q$ is increased to above 144~MeV. 
When it appears, its $K$-matrix energy is below the $D^*\bar{D}^*$ threshold by about 65~MeV and its $K$-matrix width is extremely large.
When $\Delta_Q$ increases to 200~MeV, its $K$-matrix energy increases to 27~MeV below the $D^*\bar{D}^*$ threshold and its $K$-matrix width decreases to 178~MeV. 
The large width and the significant energy dependence of the background underneath the pole raise the question of whether there is a corresponding $T$-matrix pole. 

If the additional $1^{+-}$ $K$-matrix pole is a $D^*\bar{D}^*$ resonance, there must be attractive interactions in the $1^{+-}$ channel. 
One source of attraction is the coupling of the repulsive $0^{-+}\Sigma_u^-$ potential  to the attractive $1^{--}\Sigma_g^+/\Pi_g$  potentials through spin splittings, which would also produce an attraction in the $0^{++}$ channel.
For $\varepsilon_1=-50$ MeV, this source for the attraction is supported by the appearance of an additional $0^{++}$ $K$-matrix pole when $\Delta_Q$ is increased to above 240~MeV.
Another source of attraction for a $D^*\bar{D}^*$ resonance is the spin-splitting potential, whose eigenvalues are shown as functions of $r$ in Fig.~\ref{fig:SpinDependentPotentials}.

We now describe the nonperturbative spectrum in the limit $\varepsilon_1 \to 0$ from below.
At $\Delta_Q=0$, the $0^{++}$, $1^{++}$, $1^{+-}$, and $2^{++}$ states are all $S$-wave bound states at the threshold. 
When $\Delta_Q$ is increased to just above 0, the $2^{++}$ state becomes a narrow resonance near the $D^* \bar{D}^*$ threshold and the $0^{++}$, $1^{++}$, and $1^{+-}$ states become virtual states. 
At larger $\Delta_Q$, there may be additional resonances near the $D^* \bar{D}^*$ threshold that are not related by HQSS. Note that the relevant case for $X_c$ and its HQSS partners is obtained by tuning $\varepsilon_1$ such that there is a $1^{++}$ bound state just below the $D^* \bar{D}$ threshold at the physical value $\Delta_c$. This may correspond to a small but nonzero $|\varepsilon_1|$. 

\subsection{Molecular Dominance with Perturbative Spin Splittings}
\label{sec:MolecularDominance}

In molecular models for tetraquark mesons, they are assumed to be bound states or resonances of pairs of heavy mesons.
Qualitative aspects of these models can be reproduced in the B\nobreakdash-O framework by assuming that all the potentials that approach the heavy-meson-pair threshold support bound states.
We refer to this assumption as {\it molecular dominance}.
The diabatic potentials that approach the threshold for pairs of $S$-wave heavy mesons are the $1^{--}\Pi_g$, $1^{--}\Sigma_g^+$, and $0^{-+}\Sigma_u^-$ potentials.
We have denoted the energy of the ground state with $L^P = 1^+$ in the $1^{--}\Sigma_g^+/\Pi_g$ potentials by $\varepsilon_1$.
We denote the energy of the ground state with $L^P=0^+$ in the $0^{-+}\Sigma_u^-$ potential by $\varepsilon_0$.
It is sensitive to the energy $E_{8,0^{-+}}$ of the $0^{-+}$ adjoint meson.

The HQSS multiplet for the $j^{\pi\gamma}=0^{-+}$ ground state consists of a spin-singlet $0^{++}$ state and a spin-triplet $1^{+-}$ state.
We denote these states by $0^{++\prime}$ and $1^{+-\prime}$ to distinguish them from the $0^{++}$ and $1^{+-}$ states of the $j^{\pi\gamma}=1^{--}$ ground-state multiplet.
If the difference $\varepsilon_0 - \varepsilon_1$ between the ground-state energies is much larger than $\Delta_Q$, we can ignore the mixing between states with the same $J^{PC}$ in the $1^{--}$ and $0^{-+}$ ground-state multiplets.
The spin splitting of the $J^{PC}$ state with $P=+$ and $C=(-1)^{S_{Q \bar Q}+1}$ from first-order perturbation theory in $\mathbf{V}_{\text{SS},\text{long}}$ is the diagonal entry of the $J$ block of Eq.~\eqref{<VSS>} with $j=0$.
The energies of the $0^{++\prime}$ and $1^{+-\prime}$ states from first-order perturbation theory in $\mathbf{V}_{\text{SS},\text{long}}$ are
\begin{equation}
E_{0^{++\prime}} = \varepsilon_0, \qquad \qquad E_{1^{+-\prime}} = \varepsilon_0 .
\label{EJPC-singlet}
\end{equation}
The absence of spin splittings can be understood from the amplitudes for $J_1$ and $J_2$ in Eq.~\eqref{changebasis12-Swave}, which give the particle content of each state.
The $0^{++\prime}$ state is the superposition of $D \bar{D}$ and $D^* \bar{D}^*$ with amplitudes $-1/2 $ and $-\sqrt{3/4}$, 
so its spin splitting is the corresponding weighted average of the spin splittings of the $D \bar{D}$ and $D^* \bar{D}^*$ thresholds, which is 0.
The $1^{+-\prime}$ state is the sum of $(D^* \bar{D} - D \bar{D}^*)/2$ and $-D^* \bar{D}^*/\sqrt{2}$, so its spin splitting is the average of the spin splittings of the $D^* \bar{D}$ and $D^* \bar{D}^*$ thresholds, which is 0.

The energies of the two $0^{++}$ states and the two $1^{+-}$ states in Eqs.~\eqref{EJPC-triplet} and \eqref{EJPC-singlet} from first-order nondegenerate perturbation theory can be good approximations only if $\Delta_Q \ll |\varepsilon_1 - \varepsilon_0|$.
If  $\Delta_Q \gg |\varepsilon_1 - \varepsilon_0|$, degenerate perturbation theory gives better approximations for their energies.
The static Hamiltonian for the ground states in the $1^{--}\Sigma_g^+/\Pi_g$ and $0^{-+}\Sigma_u^-$ potentials is diagonal in the channels $(j^{\pi\gamma}, S_{Q \bar Q})$:
\begin{equation}
\big( \mathbf{H}_0  \big)_{(j^{\pi\gamma}, S_{Q \bar Q}),(j^{\pi\gamma}, S_{Q \bar Q})}
=  \varepsilon_1\,  \delta(j^{\pi\gamma},1^{--})  + \varepsilon_0\,  \delta(j^{\pi\gamma},0^{-+}).
\label{H0jpigamma}
\end{equation}
Its matrix elements do not depend on $S_{Q \bar Q}$.
The $J$ block of $\mathbf{H}_0$ can be obtained by multiplying by the triangle symbol $\{ j, S_{Q \bar Q},J \}$.
The $J$ block of the long-distance spin-splitting term $\mathbf{V}_{\text{SS},\text{long}}$ in the radial tetraquark basis is given by Eq.~\eqref{<VSS>}.
The parity for $S$-wave states is $P=+$.
The $J^{PC}$ block is obtained by restricting $S_{Q \bar Q}$ to be either 0 or 1 in accord with the $CP$ constraint in Eq.~\eqref{CP-total}.
The $1^{++}$ block and the $2^{++}$ block are the single energies $E_{1^{++}}$ and $E_{2^{++}}$ in Eq.~\eqref{EJPC-triplet}.
The $0^{++}$ and $1^{+-}$ blocks are $2 \times 2$ matrices that depend on $\varepsilon_0$, $\varepsilon_1$, and  $\Delta_Q$:
\begin{equation}
H_1^{0^{++}}  = \begin{pmatrix}
\varepsilon_1 - \Delta_Q & - \tfrac{\sqrt{3}}{2}\Delta_Q  \\
- \tfrac{\sqrt{3}}{2}\Delta_Q  & \varepsilon_0  \end{pmatrix}, \qquad
H_1^{1^{+-}}  = \begin{pmatrix}
\varepsilon_1 & \tfrac{1}{2}\Delta_Q  \\
\tfrac{1}{2}\Delta_Q  & \varepsilon_0  \end{pmatrix}.
\label{H1jpigamma}
\end{equation}
The off-diagonal terms mix states in which the heavy-quark pair is spin-triplet and spin-singlet.
The energies of the $0^{++}$ and $1^{+-}$ states are the eigenvalues of the matrices in Eq.~\eqref{H1jpigamma}. 
The energies of the six $J^{PC}$ states from first-order degenerate perturbation theory in $\mathbf{V}_{\text{SS},\text{long}}$ are
\begin{subequations}
\begin{align}
E_{0^{++}}^\pm &= \frac12 \left( \varepsilon_1 + \varepsilon_0  - \Delta_Q
 \pm \sqrt{( \varepsilon_1 - \varepsilon_0 - \Delta_Q)^2 + 3 \Delta_Q^2} \, \right),
\label{E-0++pm}
\\
E_{1^{++}} &= \varepsilon_1 - \tfrac12 \Delta_Q,
\\
E_{1^{+-}}^\pm &= \frac12 \left( \varepsilon_1 + \varepsilon_0  
 \pm \sqrt{( \varepsilon_1 - \varepsilon_0)^2 + \Delta_Q^2} \, \right),
\label{E-1+-pm}
\\
E_{2^{++}} &= \varepsilon_1 + \tfrac12 \Delta_Q.
\end{align}
\label{E0++,E1--}%
\end{subequations}
The states with energies $E_{0^{++}}^-$ and $E_{0^{++}}^+$ are $0^{++}$ and $0^{++\prime}$.
The states with energies $E_{1^{+-}}^-$ and $E_{1^{+-}}^+$ are $1^{+-}$ and $1^{+-\prime}$.
The energy eigenstates are superpositions of spin-triplet  and spin-singlet states. 
For the $0^{++}$ states, the mixing angle $\theta_0$ between the spin-triplet and spin-singlet states satisfies $\tan(2\theta_0) =  \sqrt{3}\, \Delta_Q/(\varepsilon_0 - \varepsilon_1 + \Delta_Q )$. 
For the $1^{+-}$ states, the mixing angle $\theta_1$ between the spin-triplet and spin-singlet states satisfies $\tan(2\theta_1) = \Delta_Q/(\varepsilon_1 - \varepsilon_0)$. 
At first order in $\Delta_Q$, the energies of the two $0^{++}$ states and the two $1^{+-}$ states reduce to those in Eqs.~\eqref{EJPC-triplet} and \eqref{EJPC-singlet}.

\begin{figure}[t]
\centerline{ \includegraphics[width=12cm,clip=true]{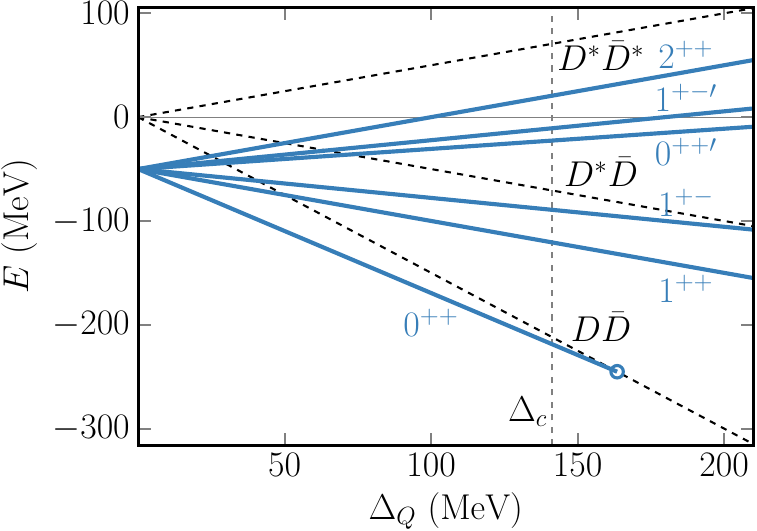} }
\caption{
Energies of isospin-0 hidden-heavy tetraquarks as functions of the heavy-meson spin splitting $\Delta_Q$ from first-order degenerate perturbation theory in the spin-splitting potential $\mathbf{V}_{\text{SS}}$ in the case of molecular dominance with ideal mixing.
The $j^{\pi \gamma} = 0^{-+}$  and $1^{--}$ ground-state energies relative to the $D^{(*)} \bar{D}^{(*)}$ threshold (thin horizontal solid line) are chosen to be $\varepsilon_1=\varepsilon_0 = -50$~MeV.
The six solid lines are the energies of the $0^{++}$, $1^{++}$, $1^{+-}$, $0^{++\prime}$, $1^{+-\prime}$, and $2^{++}$ states.  
The three dashed lines are the $D \bar{D}$, $D^* \bar{D}$, and $D^* \bar{D}^*$ thresholds.
The $0^{++}$ and $1^{+-}$ states are expected to become virtual states at the open dot and just beyond the frame, respectively.
The vertical dashed line is the charm-meson spin splitting $\Delta_c$.
}
\label{fig:SpectrumvsDelta:1OPT-ideal}
\end{figure}

The perturbative spectrum for molecular dominance is particularly simple in the limit  in which the $j^{\pi \gamma} = 0^{-+}$  and $1^{--}$ ground-state energies become degenerate.
We refer to the limit $\varepsilon_0 \to \varepsilon_1$ as {\it molecular dominance with ideal mixing}.
The energies of the $1^{++}$ and $2^{++}$ states in this limit are the same as those in Eq.~\eqref{EJPC-triplet}.  
The energies of the two $0^{++}$ and the two $1^{+-}$ states are obtained by setting $\varepsilon_0 = \varepsilon_1$ in Eqs.~\eqref{E0++,E1--}.
The energies of the six $J^{PC}$ states in Eq.~\eqref{E0++,E1--} reduce to  \cite{Braa24b}
\begin{subequations}
\begin{align}
&E_{0^{++}} = \varepsilon_1 -  \tfrac32 \Delta_Q, \qquad E_{1^{++}} = \varepsilon_1 - \tfrac12 \Delta_Q,  \qquad E_{1^{+-}} = \varepsilon_1 - \tfrac12 \Delta_Q , 
\\
&E_{0^{++\prime}} = \varepsilon_1 +  \tfrac12 \Delta_Q, \qquad  E_{1^{+-\prime}} = \varepsilon_1 + \tfrac12 \Delta_Q ,  \qquad E_{2^{++}} = \varepsilon_1 + \tfrac12 \Delta_Q.
\end{align}
\label{EJPC-ideal}%
\end{subequations}
The spin splitting of each state is equal to the spin splitting of one of the heavy-meson-pair thresholds.
The spin splitting of the $0^{++}$ state is equal to the spin splitting of the $D \bar{D}$ threshold.
The spin splittings of the $1^{++}$ and $1^{+-}$ states are equal to the spin splitting of the $D^* \bar{D}$ threshold.
The spin splittings of the $0^{++\prime}$, $1^{+-\prime}$, and $2^{++}$ states are all equal to the spin splitting of the $D^* \bar{D}^*$ threshold.
The spin splittings in Eq.~\eqref{EJPC-ideal} reveal the heavy-meson pair contents of the states.
The $0^{++}$ state is $D \bar{D}$. 
The $1^{++}$ and $1^{+-}$ states are $(D^* \bar{D} + D \bar{D}^*)/\sqrt{2}$ and $(D^* \bar{D} - D \bar{D}^*)/\sqrt{2}$.
The $0^{++\prime}$, $1^{+-\prime}$, and $2^{++}$ states are $D^* \bar{D}^*$.
The $0^{++}$ state is the superposition of spin-triplet and spin-singlet with amplitudes $\sqrt{3/4}$ and 1/2 and the $0^{++\prime}$ state is the complementary superposition.
The $1^{+-}$ state is the odd equal-amplitude superposition of spin-triplet and spin-singlet and the $1^{+-\prime}$ state is the even equal-amplitude superposition.

To illustrate the perturbative predictions, it is convenient to use a nonzero value of the equal energies $\varepsilon_1 = \varepsilon_0$ of the $j^{\pi \gamma} = 1^{--}$  and $0^{-+}$ ground states.
We also choose the specific value $\Delta_{8,Q}= \tfrac{3}{2}\Delta_Q$ of the adjoint-meson energy.
For this value of $\Delta_{8,Q}$, the energies of the six $J^{PC}$ states from first-order degenerate perturbation theory in the spin-splitting potential $\mathbf{V}_{\text{SS}}$ in Eq.~\eqref{DeltaE-SS} reduce to
\begin{subequations}
\begin{align}
E_{0^{++}}^\pm &= \varepsilon_1  +\tfrac{1}{2}  \left( -1 \pm \sqrt{ 1 +3 P_{01,\text{long}}^2}\,  \right) \Delta_Q ,
\\
E_{1^{++}} &= \varepsilon_1 - \tfrac12 \Delta_Q,
\\
E_{1^{+-}}^\pm &= \varepsilon_1  \pm \tfrac{1}{2} P_{01,\text{long}}\,  \Delta_Q ,
\\
E_{2^{++}} &= \varepsilon_1 + \tfrac12 \Delta_Q,
\end{align}
\label{EsmallDeltaQ}%
\end{subequations}
where $P_{01,\text{long}}$ is an off-diagonal analog of the long-distance probability defined in Eq.~\eqref{Plong}:
\begin{equation}
P_{01,\text{long}} = \int d^3r \, \psi_0(r)^*\,  \big[ 1 - \exp\big( -r^2/R_\text{SS}^2 \big) \big] \,\psi_{1S}(r).
\label{P01,long}
\end{equation}
The wavefunctions $\psi_0(r)$ and $\psi_{1S}(r)$ are those for the $0^{-+}$ ground state with energy $\varepsilon_0 = \varepsilon_1$ and the $S$-wave component of the $1^{--}$ ground state with energy $\varepsilon_1$.
For any value of $\Delta_{8,Q}$ other than $\tfrac{3}{2}\Delta_Q$, the expressions for $E_{0^{++}}^\pm$, $E_{1^{++}}$, and $E_{2^{++}}$ are more complicated, depending on $\Delta_{8,Q}$ and also on the long-distance probability $P_{1,\text{long}}$. 

We choose the $j^{\pi \gamma} = 0^{-+}$ and $1^{--}$ ground-state energies relative to the $D^{(*)} \bar{D}^{(*)}$ threshold to be $\varepsilon_0 = \varepsilon_1 = -50$~MeV.
The off-diagonal probability in Eq.~\eqref{EsmallDeltaQ} is then $P_{01,\text{long}}=0.556$. 
We also choose $\Delta_{8,Q}= \tfrac{3}{2}\Delta_Q$.
The energies of the six $J^{PC}$ states are shown as functions of $\Delta_Q$ in Fig.~\ref{fig:SpectrumvsDelta:1OPT-ideal}.
At $\Delta_Q=0$, the $0^{++}$, $1^{++}$, $1^{+-}$, $0^{++\prime}$, $1^{+-\prime}$, and $2^{++}$ states  are all $S$-wave bound states with the same energy. 
The $0^{++}$ state reaches the $D \bar{D}$ threshold at $\Delta_Q = 3.3\, |\varepsilon_1|$.  
It has an $S$-wave coupling to $D \bar{D}$, so it becomes a virtual state whose energy then decreases below the threshold.
The $1^{++}$ state crosses the $D \bar{D}$ threshold, but it has no couplings to $D \bar{D}$ so it remains a bound state.
The $1^{+-}$ state crosses the $D \bar{D}$ threshold, but it has no couplings to $D \bar{D}$ so it remains a bound state.
It reaches the $D^* \bar{D}$ threshold at $\Delta_Q = 4.5\,|\varepsilon_1|$. 
It has $S$-wave couplings to $D^* \bar{D}$ and $D \bar{D}^*$ so it becomes a virtual state.
The $0^{++\prime}$ state crosses the $D \bar{D}$ threshold then the $D^* \bar{D}$ threshold.
It has no couplings to $D \bar{D}$, $D^* \bar{D}$, or $D \bar{D}^*$ so it remains a bound state.
The $1^{+-\prime}$ state crosses the $D \bar{D}$ threshold and then the $D^* \bar{D}$ threshold. 
It has no couplings to $D \bar{D}$, $D^* \bar{D}$, or $D \bar{D}^*$ so it remains a bound state.
The $2^{++}$ state crosses the $D \bar{D}$ threshold at $\Delta_Q = \tfrac{1}{2} |\varepsilon_1|$.
It has $D$-wave couplings to $D \bar{D}$ so it becomes a narrow resonance. 
It crosses the $D^* \bar{D}$ threshold at $\Delta_Q = |\varepsilon_1|$.
It has $D$-wave couplings to $D^* \bar{D}$ and $D \bar{D}^*$ so it gains these additional decay channels. 
Note that for $\Delta_Q > 4.5|\varepsilon_1|$,  the spectrum consists of only four states:  a $1^{++}$ bound state with the spin splitting of the $D^* \bar{D}$ threshold, $0^{++\prime}$ and $1^{+-\prime}$ bound states between the $D^* \bar{D}$ and $D^* \bar{D}^*$ thresholds,
 and a narrow $2^{++}$ resonance with the spin splitting of the $D^* \bar{D}^*$ threshold.

We now describe the perturbative prediction for the spectrum with $\varepsilon_0 = \varepsilon_1$ as a function of $\Delta_Q$ in the limit $\varepsilon_1 \to 0$ from below.
As $\varepsilon_1 \to 0$, the long-distance probability $P_{1,\text{long}}$ for the $1^{--}$ ground state approaches 1.
At $\Delta_Q=0$, the $0^{++}$, $1^{++}$, $1^{+-}$, $0^{++\prime}$, $1^{+-\prime}$, and $2^{++}$ states are all $S$-wave bound states at the threshold. 
As $\Delta_Q$ increases to just above 0, the $2^{++}$ state becomes a narrow resonance and the $0^{++}$ and $1^{+-}$ states become virtual states.
At larger $\Delta_Q$, the spectrum consists of a $1^{++}$ bound state just below the $D^* \bar{D}$ threshold, $0^{++\prime}$ and $1^{+-\prime}$ resonances between the $D^* \bar{D}$ and $D^* \bar{D}^*$ thresholds, and a $2^{++}$ resonance just below the $D^* \bar{D}^*$ threshold.

\subsection{Molecular Dominance with Nonperturbative Spin Splittings}
\label{sec:SpinSplitting-SchrEq}

\begin{figure}[t]
\centerline{ \includegraphics[width=12cm,clip=true]{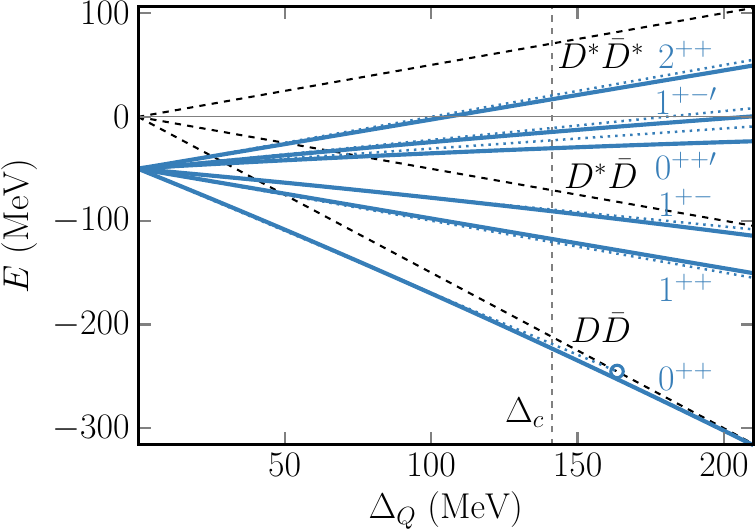} }
\caption{
Energies of isospin-0 $c \bar c$ tetraquarks as functions of the heavy-meson spin-splitting $\Delta_Q$ with $\Delta_{8,Q}= \tfrac{3}{2}\Delta_Q$ and $(R_{\Pi_g},R_{\Sigma_g^+}) =  (0.50,0.25)$~fm in the case of molecular dominance with ideal mixing.  
The dependence on $\Delta_Q$ is calculated nonperturbatively by solving the Schr\"odinger equation.
The adjoint-meson energies $E_{8,1^{--}}$ and $E_{8,0^{-+}}$ are tuned so that the $j^{\pi \gamma} = 1^{--}$ and $0^{-+}$ ground-state energies at $\Delta_Q=0$ relative to the $D^{(*)} \bar{D}^{(*)}$ threshold (thin horizontal solid line) are $\varepsilon_1=\varepsilon_0 = -50$~MeV.
The solid curves are the energies of the $0^{++}$, $1^{++}$, $1^{+-}$, $1^{+-\prime}$, $0^{++\prime}$, and $2^{++}$ states.
The dashed lines are the $D \bar{D}$, $D^* \bar{D}$, and $D^* \bar{D}^*$ thresholds.
The dotted lines are the predictions from first-order perturbation theory in $\Delta_Q$ in  Fig.~\ref{fig:SpectrumvsDelta:1OPT-ideal}.
The $0^{++}$ and $1^{+-}$ states become virtual states beyond the frame.
The vertical dashed line is the  charm-meson spin splitting $\Delta_c$.
}
\label{fig:SpectrumvsDelta:SchrEqMD}
\end{figure}

The dependence of the tetraquark spectrum on the heavy-meson spin splitting in the case of molecular dominance  can be calculated nonperturbatively by solving the Schr\"odinger equation with the $0^{-+}$ adjoint-meson energy $E_{8,0^{-+}}$ in Eq.~\eqref{E80-+} replaced by an adjustable parameter that determines the $j^{\pi \gamma} = 0^{-+}$ ground-state energy $\varepsilon_0$.
In the specific case of molecular dominance with ideal mixing, the adjoint-meson energies $E_{8,1^{--}}$ and $E_{8,0^{-+}}$ are adjusted so that the $1^{--}$  and $0^{-+}$ ground-state energies have equal values $\varepsilon_0 = \varepsilon_1$.
To illustrate the nonperturbative predictions in the case of molecular dominance with ideal mixing, we tune $E_{8,1^{--}}$ and $E_{8,0^{-+}}$ so  the $1^{--}$ and $0^{-+}$ ground-state energies at $\Delta_Q=0$ relative to the $D^{(*)} \bar{D}^{(*)}$ threshold are $\varepsilon_1 = \varepsilon_0 = -50$~MeV. 
This requires the adjoint-meson energies to be $E_{8,1^{--}} = -602$~MeV and $E_{8,0^{-+}}= -1344$~MeV. 
The long-distance probabilities defined in Eq.~\eqref{Plong} are $P_{1,\text{long}} = 0.501$ for the $1^{--}$ ground state with $\varepsilon_1 = -50$~MeV and $P_{0,\text{long}} = 0.642$ for the $0^{-+}$ ground state with $\varepsilon_0 = -50$~MeV.
The energies of $c \bar c$ tetraquarks for the six values of $J^{PC}$ are shown as functions of $\Delta_Q$  in Fig.~\ref{fig:SpectrumvsDelta:SchrEqMD}.
The straight dotted lines are the predictions from first-order perturbation theory in $\Delta_Q$ in Fig.~\ref{fig:SpectrumvsDelta:1OPT-ideal}.
The curves for $1^{++}$ and $2^{++}$ are very close to the predictions from first-order perturbation theory in $\Delta_Q$  in Fig.~\ref{fig:SpectrumvsDelta:1OPT-ideal}. 
The energies of the other four states have more obvious curvature as functions of $\Delta_Q$.
However, their deviations from the straight dotted lines are small enough that first-order perturbation theory remains a good approximation even for $\Delta_Q$ as large as $\Delta_c$.
The $0^{++}$ state reaches the $D \bar{D}$ threshold at $\Delta_Q= 245$~MeV and becomes a virtual state.  
The $1^{++}$ state remains a bound state near the $D^* \bar{D}$ threshold.
The $1^{+-}$ state reaches the $D^* \bar{D}$ threshold at $\Delta_Q= 323$~MeV and becomes a virtual state.
The $0^{++\prime}$ and $1^{+-\prime}$ states remain bound states with increasing energies and increasing binding energies relative to the $D^* \bar{D}^*$ threshold. 
The $2^{++}$ state becomes a narrow resonance when it crosses the $D \bar{D}$ threshold and it remains a narrow resonance near the $D^* \bar{D}^* $ threshold.

We now describe the nonperturbative spectrum with $\varepsilon_0 = \varepsilon_1$  in the limit $\varepsilon_1 \to 0$ from below.
At $\Delta_Q=0$, the $0^{++}$, $1^{++}$, $1^{+-}$, $0^{++\prime}$, $1^{+-\prime}$, and $2^{++}$ states are all $S$-wave bound states at the threshold. 
When $\Delta_Q$ is increased to just above 0, the $2^{++}$, $0^{++\prime}$,  and $1^{+-\prime}$ states become narrow resonances near the $D^* \bar{D}^*$ threshold and the $0^{++}$, $1^{++}$, and $1^{+-}$ states become virtual states. 


\section{Hidden-Charm Tetraquark Multiplet}
\label{sec:CharmoniumTetraquarks}

In this Section, we tune the energy of the $1^{--}$ adjoint meson to the critical value for the  $1^{++}$ $c \bar c$ tetraquark at the $D^*\bar{D}$ threshold.
We then predict the energies of its $0^{++}$, $1^{+-}$, and $2^{++}$ HQSS partners, first with the quarkonium potential ignored and then with the narrow avoided crossing taken into account.

\subsection{Static Potential Only}
\label{sec:noSpinSplittingsccbar}

If the diabatic potentials are truncated to $1^{--}\Sigma_g^+/\Pi_g$, the radial diabatic Hamiltonian $\mathbf{H}_0$ in the HQSS limit can be arranged into commuting blocks specified by the B\nobreakdash-O angular momentum $L$ and parity $P$. 
The $j^{\pi \gamma} = 1^{--}$ ground state has $L^P = 1^+$.
The $1^+$ block of $\mathbf{H}_0$ is given explicitly in Eq.~\eqref{H1+:diabatic}.
Our value for the charm-quark mass is $m_c=1.48$~GeV in Eq.~\eqref{mc,mb}.
We consider the two specific models for the $1^{--}\Pi_g$ and $1^{--}\Sigma_g^+$ potentials given by Eq.~\eqref{bopots} with the two pairs of relaxation lengths $(R_{\Pi_g},R_{\Sigma_g^+})$ in Eq.~\eqref{RSigma,RPi}.
We treat the adjoint-meson energy $E_{8,1^{--}}$ as an adjustable parameter. 
At large $r$, the potentials approach zero, which can be identified with the triplet-meson-pair threshold.
We tune $E_{8,1^{--}}$ to the critical value $E_8^\ast$ for the energy $\varepsilon_1$ of the $1^{--}$ ground state at the triplet-meson-pair threshold. 
For $(R_{\Pi_g},R_{\Sigma_g^+} ) = (0.25,0.50)$~fm, the critical adjoint-meson energy is $E_8^\ast = -595$~MeV. 
For $(R_{\Pi_g},R_{\Sigma_g^+} ) = (0.50,0.25)$~fm, the critical energy is $E_8^\ast = -410$~MeV. 

It was pointed out in Ref.~\cite{Braa24b} that the existence of  $\chi_{c1}(3872)$ with energy near the $D^{\ast0} \bar{D}^0$ threshold suggests that the isospin-0 $1^{--}$ adjoint meson is likely to be the lowest-energy adjoint hadron.
The critical $1^{--}$ adjoint-meson energy has been estimated previously using other models for the $1^{--}\Pi_g$ and $1^{--}\Sigma_g^+$  potentials.
These models also have a constant term at small $r$ that can be interpreted as the adjoint-meson energy and they approach a constant at large $r$ that can be identified with the  triplet-meson-pair threshold.
The estimate of the critical adjoint-meson energy relative to the triplet-meson-pair threshold in Ref.~\cite{Braa24b} is $E_8^\ast = -157$~MeV.
The estimate of the critical energy given in Ref.~\cite{Bram24} is +919~MeV. 
However, this is actually the critical energy in the pole-mass scheme \cite{Moha26}.
The critical energy relative to the triplet-meson-pair threshold obtained by solving the Schr\"odinger equation for the potential in Ref.~\cite{Bram24} is $-188$~MeV. 
Thus the estimates of the critical adjoint-meson energy $E_8^\ast$ from the various models for the potentials differ at most by hundreds of MeV.

With the $L^P=1^+$ state tuned to the triplet-meson-pair threshold, there are no excited states in  the $1^{--}\Pi_g$ potentials.
If the $1^+$ state was much more deeply bound, there would be excited states. 
The first few excited states may include $0^-$, $1^-$, and $2^-$, as listed in Table~\ref{tab:QQbarqqbar}, but their order in energy depends on details of the $1^{--}\Pi_g$ and $1^{--}\Sigma_g^+$ potentials.

\subsection{With Spin Splittings}
\label{sec:noNACccbar}

If the diabatic potentials are truncated to $1^{--}\Sigma_g^+/\Pi_g$ and $0^{-+}\Sigma_u^-$, the radial diabatic Hamiltonian $\mathbf{H}_1$ with spin splittings can be arranged into commuting blocks with quantum numbers $J^{PC}$. 
The $J^{PC}$ block has the general form in Eq.~\eqref{H1-JPC}.
For $J^{PC} =1^{++}$, the orbital-angular-momentum matrix $\mathbf{L}_{Q \bar Q}^2$, the radial diabatic static potential $\mathbf{V}_0^{J^{PC}}(r)$, the spin-splitting potential $\mathbf{V}^{J^{PC}}_\text{SS}(r)$, and the kinetically improved mass matrix $\mathbf{M}_{J^{PC}}$ are the $3 \times 3$ matrices in Eqs.~\eqref{L2:1++3}, \eqref{V0:1++3}, \eqref{VSS:1++3}, and \eqref{M:1++3}, respectively.
For $0^{++}$, $1^{+-}$, and $2^{++}$, the $3 \times 3$, $4 \times 4$, and $6 \times 6$ matrices in $\mathbf{H}_1^{J^{PC}}$ are given in Sections~\ref{sec:0++}, \ref{sec:1+-}, and \ref{sec:2++} of the Appendix~\ref{app:DiabaticH}, respectively.
Our models for the diabatic potentials and our choices for the parameters in the Schr\"odinger equation are described in Section~\ref{sec:Models}.
We treat the energy $E_{8,1^{--}}$ of the $1^{--}$ adjoint meson as an adjustable parameter.
The charm-meson mass splitting in the long-distance spin-splitting term is $\Delta_c = 141.3$~MeV in Eq.~\eqref{Deltac,Deltab}.
The charm-meson masses in the kinetically improved mass matrix are determined by $M_{D^{(*)}}$  in Eq.~\eqref{MD(star),MB(star)} and by Eq.~\eqref{MD,MDstar}.

\begin{table}
\centering
\begin{tabular}{|cccc|cccc|}
\hline
~$R_{\Pi_g}$ [fm]~ & ~$R_{\Sigma_g^+}$  [fm]~ & ~$\Delta_{8,c}$~ & ~$E_8^\ast$~ & ~$E_{0^{++}}$~ & ~$E_{1^{++}}$~ &  ~$E_{1^{+-}}$~ &  ~$E_{2^{++}}$~   \\
\hline
0.25 & 0.50 &  ~70 & $-554$ & $-$ & 0 & $(140.5, 12.3)$ & $(124.6, 1.6)$  \\
0.25 & 0.50 & 210  & $-471$ & $-$ & 0 & $-$ & $(140.8, 0.4)$ 	\\
0.50 & 0.25 &  ~70 & $-384$ & $-$  & 0 & $(141.2, 4.2)$ & $(128.1, 0.6)$ \\
0.50 & 0.25 & 210  & $-325$ & $-$ & 0 & $-$& ($141.1, 0.1$)  \\
\hline
\end{tabular}
\caption{
\label{tab:noNACccbar}
Energies and widths of $c \bar{c}$ bound states and resonances in the $1^{--}\Sigma_g^+/\Pi_g$ and $0^{-+}\Sigma_u^-$ tetraquark potentials in the $D^{(*)} \bar D^{(*)}$ threshold region.
The energy of the $j^{\pi\gamma} = 1^{--}$ adjoint meson is tuned to the critical energy $E_8^\ast$ for the $1^{++}$ $c \bar{c}$ tetraquark at the $D^* \bar{D}$ threshold. 
The parameters are relaxation lengths $R_{\Pi_g}$ and $R_{\Sigma_g^+}$ and the spin splitting $\Delta_{8,c}$ of the $1^{--}$ adjoint meson.
The critical energy $E_8^\ast$ is relative to the spin-weighted $D^{(*)} \bar{D}^{(*)}$ threshold at 3946.5~MeV.
A column $E_{J^{PC}}$ gives either the energy of a $J^{PC}$ bound state as a single number or the $K$-matrix energy and width of a $J^{PC}$ resonance as a pair of numbers $(E_{J^{PC}},\Gamma_{J^{PC}})$.
A dash $(-)$ indicates no bound state or resonance.
The energies $E_{J^{PC}}$ are relative to the $D^*\bar{D}$ threshold at 3875.8~MeV.
The energies $\Delta_{8,c}$, $E_8^\ast$, and $E_{J^{PC}}$ are in units of MeV. 
}
\end{table}

We consider the four models specified by the two pairs of relaxation lengths $(R_{\Pi_g},R_{\Sigma_g^+} )$ in Eq.~\eqref{RSigma,RPi} and the two adjoint-meson spin splittings $\Delta_{8,c}$ in Eq.~\eqref{Delta8Q-num}. 
For each model, we tune the energy $E_{8,1^{--}}$ of the $1^{--}$ adjoint meson to the critical value $E_8^\ast$ for the  $1^{++}$ $c \bar c$ tetraquark at the $D^*\bar{D}$ threshold.
The critical energies $E_8^\ast$ relative to the spin-weighted $D^{(*)} \bar{D}^{(*)}$ threshold are given in Table~\ref{tab:noNACccbar}.
They are higher than the critical energies without spin splittings by amounts ranging from 26 to 124~MeV. 
Given $E_8^\ast$, we use SPARSE to calculate the $K$-matrix energies and widths of the lowest-energy resonances with quantum numbers $0^{++}$, $1^{+-}$, and $2^{++}$. 
Their $K$-matrix energies relative to the $D^*\bar{D}$ threshold and their $K$-matrix widths are given in Table~\ref{tab:noNACccbar}.
For reference, the $D\bar{D}$, $D^*\bar{D}$, and $D^*\bar{D}^*$ thresholds are $-141.3$~MeV, 0, and +141.3~MeV.

The only bound state in Table~\ref{tab:noNACccbar} is the $1^{++}$ bound state whose energy has been tuned to the $D^* \bar{D}$ threshold.
None of the four models has a $0^{++}$ bound state or resonance.
None of the four models has a $1^{+-}$ bound state or resonance near the $D^* \bar{D}$ threshold.
The absence of $0^{++}$ and $1^{+-}$ states may be due to them becoming virtual states at values of $\Delta_Q$ between 0 and $\Delta_c$.
Each of the four models has a $2^{++}$ resonance near the $D^*\bar{D}^*$ threshold.
The spin splitting between the $2^{++}$ resonance and the $1^{++}$ bound state is close to the prediction 141.3~MeV from first-order perturbation theory in $\Delta_c$.
If $\Delta_{8,c} = 70$~MeV, its $K$-matrix energy is below the $D^*\bar{D}^*$ threshold by about 15~MeV and its $K$-matrix width is  less than 1.6~MeV.
If $\Delta_{8,c} = 210$~MeV, its $K$-matrix energy is below the $D^*\bar{D}^*$ threshold by 0.5 and 0.2~MeV in the 2nd and 4th model and its $K$-matrix width is 0.4 and 0.1~MeV.
Since the $K$-matrix width is comparable to the difference between the $D^*\bar{D}^*$ threshold and the $K$-matrix energy, the $T$-matrix energy of the $2^{++}$ resonance could be above or below the $D^*\bar{D}^*$ threshold.
The narrow width of the $2^{++}$ resonance can be explained by its $D$-wave coupling to $(D^*\bar{D}+D\bar{D}^*)/\sqrt{2}$ and to $D\bar{D}$.
These results are all compatible with the expectations from perturbation theory in $\Delta_c$.

There is a result in Table~\ref{tab:noNACccbar} that is not compatible with the expectations from perturbation theory in $\Delta_c$.
In the two models with $\Delta_{8,c} = 70$~MeV, there is a $1^{+-}$ resonance near the $D^*\bar{D}^*$ threshold. 
The $1^{+-}$ resonance disappears if $\Delta_{8,c}$ is increased to 210~MeV.
In the 1st and 3rd model, the $K$-matrix energy of the $1^{+-}$ resonance is below the $D^*\bar{D}^*$ threshold by 0.8 and 0.1~MeV and its $K$-matrix width is 12 and 4~MeV, respectively.
Its $T$-matrix energy could be either above or below the $D^*\bar{D}^*$ threshold.
The $K$-matrix width of the $1^{+-}$ resonance is about an order of magnitude larger than that of the $2^{++}$ resonance. 
The larger width can be explained by the $S$-wave coupling of a $1^{+-}$ resonance to $(D^*\bar{D}-D\bar{D}^*)/\sqrt{2}$ and $D^*\bar{D}^*$.
The attraction mechanism that produces this $1^{+-}$ resonance could be the same as that for the additional $1^{+-}$ resonance in Fig.~\ref{fig:SpectrumvsDelta:SchrEq} for $\varepsilon_1 = -50$~MeV.

The prominent features of Table~\ref{tab:noNACccbar} are consistent with the expectations from hadronic effective field theories with heavy-quark spin symmetry \cite{Niev12,Baru16}.
The solution of the Lippmann-Schwinger equations can produce hidden-heavy tetraquark states that are bound states, virtual states, or resonances with six $J^{PC}$ quantum numbers: $0^{++}$, $1^{++}$, $1^{+-}$, $0^{++\prime}$, $1^{+-\prime}$, and $2^{++}$.
A $1^{++}$ bound state at the $D^* \bar{D}$ threshold implies the existence of a $2^{++}$  bound state, virtual state, or resonance near the $D^* \bar{D}^*$ threshold.
The energies of the $0^{++}$, $1^{+-}$, $0^{++\prime}$, and $1^{+-\prime}$ states are sensitive to coupling constants other than the one to which the energies of the $1^{++}$ and $2^{++}$ states are most sensitive.

\subsection{With the Narrow Avoided Crossing}
\label{sec:NACccbar}

In the absence of the coupling between the $1^{--}\Sigma_g^+$ and $0^{++}\Sigma_g^+$ potentials, the spin-weighted charmonium masses can be calculated by solving the Schr\"odinger equation with charm-quark mass $m_c = 1.48$~GeV in the Cornell potential with the parameters in Eq.~\eqref{Cornellparams} and adding $2m_c$ to the eigenvalues.
The masses for $1S$, $1P$, and $2S$ differ from the measured spin-weighted masses by less than 20~MeV \cite{Braa14}.  
The first few $P$-wave and $F$-wave charmonium energy levels relative to the $D^*\bar{D}$ threshold are
\begin{subequations}
\begin{align}
E_{c \bar c,1P} &= -364.9\,\mathrm{MeV}, \qquad  E_{c \bar c,2P}  = +108.7\,\mathrm{MeV}, 
\label{EnP,c}
\\
E_{c \bar c,1F} &= +188.1\,\mathrm{MeV}.
\label{EiF,c}
\end{align}
\end{subequations}
The $2P$ energy level is between the $D^*\bar{D}$ threshold at 0 and the $D^*\bar{D}^*$ threshold at  $+141.3$~MeV while the $1F$ energy level is about 40~MeV above the $D^*\bar{D}^*$ threshold.

If the diabatic potentials are truncated to $1^{--}\Sigma_g^+/\Pi_g$, $0^{-+}\Sigma_u^-$, and $0^{++}\Sigma_g^+$, the quarkonium bound states with energies above the heavy-meson-pair thresholds become resonances and there are also tetraquark bound states and resonances.
Since the avoided crossing between the $1\Sigma_g^+$ and $2\Sigma_g^+$ potentials is narrow, the quarkonium energies are likely to be close to those in the absence of the coupling between $1^{--}\Sigma_g^+$ and $0^{++}\Sigma_g^+$.
The $J^{PC}$ block of the radial diabatic Hamiltonian $\mathbf{H}_1^{J^{PC}}$ with spin splittings has the general form in Eq.~\eqref{H1-JPC}.
For $J^{PC} =1^{++}$, $\mathbf{L}_{Q \bar Q}^2$, $\mathbf{V}_0^{J^{PC}}(r)$, $\mathbf{V}_\text{SS}^{J^{PC}}(r)$, and $\mathbf{M}_{J^{PC}}$ are the $4 \times 4$ matrices in Eqs.~\eqref{L2:1++4}, \eqref{V0:1++4}, \eqref{VSS:1++4}, and \eqref{M:1++4}, respectively.
For $J^{PC} = 0^{++}$, $1^{+-}$, and $2^{++}$, the $4 \times 4$, $5 \times 5$, and $8 \times 8$ matrices in $\mathbf{H}_1^{J^{PC}}$ are given in Sections~\ref{sec:0++}, \ref{sec:1+-}, and \ref{sec:2++} of the Appendix~\ref{app:DiabaticH}, respectively.

We consider four models specified by the two pairs of relaxation lengths $(R_{\Pi_g},R_{\Sigma_g^+})$ in Eq.~\eqref{RSigma,RPi} and the two adjoint-meson spin splittings $\Delta_{8,c}$ in Eq.~\eqref{Delta8Q-num}.
For each of our four models, we tune $E_{8,1^{--}}$ to the critical energy $E_8^\ast$  for the $1^{++}$ $c \bar c$ tetraquark at the $D^*\bar{D}$ threshold.
The critical energies $E_8^\ast$ relative to the spin-weighted $D^{(*)} \bar{D}^{(*)}$ threshold are given in Table~\ref{tab:ccbarNAC}.
They are higher than the values of $E_8^\ast$ in Table~\ref{tab:noNACccbar} with no coupling to the $0^{++}\Sigma_g^+$ potential by amounts ranging from 25 to 35~MeV. 
Given $E_8^\ast$, we use SPARSE to calculate the energies of $c \bar c$ bound states and resonances with quantum numbers $0^{++}$, $1^{++}$, $1^{+-}$, and $2^{++}$.
The $K$-matrix energies and widths of resonances in the $D^{(*)} \bar{D}^{(*)}$ threshold region are given in Table~\ref{tab:ccbarNAC}.
For reference, the $D\bar{D}$, $D^*\bar{D}$, and $D^*\bar{D}^*$ thresholds are $-141.3$~MeV, 0, and +141.3~MeV.

\begin{table}
\centering
\begingroup
\setlength{\tabcolsep}{4pt}
\begin{tabular}{|cccc|cccc|}
\hline
~$R_{\Pi_g}$ [fm]~ & ~$R_{\Sigma_g^+}$  [fm]~ & ~$\Delta_{8,c}$~ & ~$E_8^\ast$~ & ~$E_{0^{++}}$~ & ~$E_{1^{++}}$~ &  ~$E_{1^{+-}}$~ &  ~$E_{2^{++}}$~  \\
\hline
0.25 & 0.50 &  ~70 & $-521$ & $-$ & 0 & $-$ & $(134.6, 0.2)$  \\
        &         &         &             & $(101.3, 5.6)$ & $(97.9, 13.4)$ & $(100.1, 10.4)$ & $(95.9, 14.4)$ \\
0.25 & 0.50 & 210  & $-436$ & $-$ & 0 & $-$ & $-$ 	\\
       &         &         &             & $(101.2, 6.8)$ & $(98.1, 12.9)$ & $(100.7, 11.0)$ & $(96.8, 14.3)$ \\
0.50 & 0.25 &  ~70 & $-359$ & $-$                   & 0 & $-$ & $(136.1, 3.5)$ \\
        &         &         &             & $(101.5, 5.5)$ & $(97.9, 13.7)$ & $(99.6, 11.4)$ & $(97.2, 11.7)$ \\
0.50 & 0.25 & 210  & $-298$ & $-$ & 0 & $-$ & $-$  \\
       &         &         &             & $(101.4, 6.4)$ & $(98.1, 13.1)$ & $(100.3, 11.3)$ & $(97.6, 12.5)$ \\
\hline
\end{tabular}
\endgroup
\caption{
\label{tab:ccbarNAC} 
Energies and widths of $c \bar{c}$ bound states and resonances in the $1^{--}\Sigma_g^+/\Pi_g$ and $0^{-+}\Sigma_u^-$ tetraquark potentials and the $0^{++}\Sigma_g^+$ quarkonium potential in the $D^{(*)} \bar D^{(*)}$ threshold region.
The energy of the $j^{\pi\gamma} = 1^{--}$ adjoint meson is tuned to the critical energy $E_8^\ast$ for the $1^{++}$ $c \bar{c}$ tetraquark at the $D^* \bar{D}$ threshold. 
For each set of parameters, the energies $E_{J^{PC}}$ in the first and second rows are those for $c \bar{c}$  tetraquarks and $P$-wave charmonium.
The other aspects of the Table are described in the caption for Table~\ref{tab:noNACccbar}.
}
\end{table}

The effect of the coupling between the $1^{--}\Pi_g$ and $0^{++}\Sigma_g^+$ potentials on the $1P$ charmonium energy levels is very small.
In all four models, the spin-weighted energy of the $0^{++}$, $1^{++}$, $1^{+-}$, and $2^{++}$ states is smaller than the $1P$ energy in the Cornell model in Eq.~\eqref{EnP,c} by about 2~MeV. 
The spin splitting between the $2^{++}$ and $0^{++}$ states from the coupling to the $1^{--}\Pi_g$ potential is approximately $-0.3$~MeV.
This is much smaller than the physical spin splitting between $\chi_{c2}(1P)$ and $\chi_{c0}(1P)$ of +141.5~MeV. 
The physical spin splittings of $P$-wave quarkonium states are expected to be dominated by spin-dependent terms of order $1/m_Q^2$ that we have not taken into account.

The only bound state in Table~\ref{tab:ccbarNAC}  is the $1^{++}$ state whose energy has been tuned to the $D^* \bar{D}$ threshold.
All the remaining entries for $E_{0^{++}}$, $E_{1^{++}}$, $E_{1^{+-}}$, and $E_{2^{++}}$ in Table~\ref{tab:ccbarNAC} are $K$-matrix energies and widths of resonances.
Each of the four models has $0^{++}$, $1^{++}$, $1^{+-}$, and $2^{++}$ resonances with energies about 100~MeV above the $D^* \bar{D}$ threshold.
Their spin-weighted $K$-matrix energies are about 10~MeV below the spin-weighted $2P$ energy in the Cornell model in Eq.~\eqref{EnP,c}.
We identify these states as the $2P$ multiplet of charmonium. 
The spin splittings between the $2^{++}$ and $0^{++}$ states are all about $-4$~MeV. 
This is much smaller than the expected spin splittings of the $2P$ charmonium states, 
which should be dominated by spin-dependent terms of order $1/m_Q^2$ that we have not taken into account.
The $K$-matrix widths of the $0^{++}$, $1^{++}$, $1^{+-}$, and $2^{++}$ states range from 5.5 to 14.4~MeV.

The only other resonances in Table~\ref{tab:ccbarNAC} besides the $2P$ charmonium resonances are $2^{++}$ resonances with energies near the $D^* \bar{D}^*$ threshold in the two models with $\Delta_{8,c}=70$~MeV.
Their $K$-matrix energies are below the $D^* \bar{D}^*$ threshold by about 6~MeV and above the corresponding $2^{++}$ resonances in Table~\ref{tab:noNACccbar} by about 9~MeV. 
Their $K$-matrix widths differ significantly from those in Table~\ref{tab:noNACccbar}.
The width is smaller by a factor of 8 in the 1st model and larger by about a factor of 6 in the 3rd model.
The $K$-matrix poles of the $2^{++}$ tetraquark resonances disappear if $\Delta_{8,c}$ is increased to 210~MeV. 
The absence of a $2^{++}$ resonance may be due to it becoming a virtual state at some value of $\Delta_Q$ between 0 and $\Delta_c$.


\section{Hidden-bottom Tetraquark Multiplet}
\label{sec:BottomoniumTetraquarks}

In this section, we take the energy of the $1^{--}$ adjoint meson to be at the critical value for the $1^{++}$ $c \bar c$ tetraquark at the $D^*\bar{D}$ threshold.
We predict the energies of the $0^{++}$, $1^{++}$, $1^{+-}$, and $2^{++}$ $b \bar b$ bound states and resonances 
in the $B^{(*)} \bar{B}^{(*)}$ threshold region, first with the quarkonium potential ignored and then with the narrow avoided crossing taken into account.

\subsection{Static Potential Only}
\label{sec:noSpinSplittingsbbbar}

Since the bottom-quark mass is much larger than the charm-quark mass, a bound $b \bar b$ tetraquark state will have larger binding energy than the corresponding $c \bar c$ tetraquark state. 
In particular, if the energy of the $c \bar c$ ground state is tuned to the triplet-meson-pair threshold, the $b \bar b$ ground state will be more deeply bound and there may also be excited $b \bar b$ bound states.
We consider the two specific models for the $1^{--}\Pi_g$ and $1^{--}\Sigma_g^+$ potentials given by Eq.~\eqref{bopots} with the two pairs of relaxation lengths $(R_{\Pi_g},R_{\Sigma_g^+})$ in Eq.~\eqref{RSigma,RPi}.
Our value for the bottom-quark mass is $m_b=4.89$~GeV in Eq.~\eqref{mc,mb}.
The critical values $E_8^\ast$ of the adjoint-meson energy for the $j^{\pi \gamma}=1^{--}$ $c \bar c$  ground state at the triplet-meson-pair threshold are given in Section~\ref{sec:noSpinSplittingsccbar}.
For $(R_{\Pi_g},R_{\Sigma_g^+} ) = (0.25,0.50)$~fm, the critical energy is $E_8^\ast = -595$~MeV and the energy of the $1^{--}$ $b \bar b$ ground state is $-106.6$~MeV. 
For $(R_{\Pi_g},R_{\Sigma_g^+} ) = (0.50,0.25)$~fm, the critical energy is $E_8^\ast = -410$~MeV and the energy of the $1^{--}$ $b \bar b$ ground state is $-74.6$~MeV. 

If the $1^{--}$ $b \bar b$ ground state with $L^P=1^+$ is sufficiently deeply bound, there are also excited $b \bar b$ bound states with other $L^P$ quantum numbers.
For $(R_{\Pi_g},R_{\Sigma_g^+} ) = (0.25,0.50)$~fm, there is a single excited bound state with $L^P =0^-$ and energy
 $-52.1$~MeV. 
This energy is 54.5~MeV above the ground state.
For $(R_{\Pi_g},R_{\Sigma_g^+} ) = (0.50,0.25)$~fm, there are no excited bound states. 

\subsection{With Spin Splittings}
\label{sec:noNACbbbar}

The hidden-bottom tetraquarks with isospin 0 can be identified as $b \bar b$ energy levels in the  Schr\"odinger equation for the $1^{--}\Sigma_g^+/\Pi_g$ and $0^{-+}\Sigma_u^-$ potentials.
Our models for the potentials and our choices for the parameters in the Schr\"odinger equation are the same as those described in Section~\ref{sec:noNACccbar} with a few differences that take into account the change in the heavy quark from charm to bottom.
The bottom-meson mass splitting in the long-distance spin-splitting term is $\Delta_b= 45.2$~MeV in Eq.~\eqref{Deltac,Deltab}.
The bottom-meson masses in the kinetically improved mass matrix are determined by $M_{B^{(*)}}$ in Eq.~\eqref{MD(star),MB(star)} and by the analog of Eq.~\eqref{MD,MDstar}.

\begin{table}
\centering
\begin{tabular}{|cccc|cccc|}
\hline
~$R_{\Pi_g}$ [fm]~ & ~$R_{\Sigma_g^+}$  [fm]~ & ~$\Delta_{8,b}$ & ~$E_8^\ast$ & ~$E_{0^{++}}$  & ~$E_{1^{++}}$ & ~$E_{1^{+-}}$ &  ~$E_{2^{++}}$ \\
\hline
0.25 & 0.50 & 20 & $-554$ & ~$-102.7$~ & ~$-88.4$~ & ~$-77.5$~ & $-65.4$ \\
0.25 & 0.50 & 70 & $-471$ & ~$-88.0$ & $-63.6$ & $-43.9$ & $(-22.3, 0.15)$ \\
0.50 & 0.25 & 20  & $-384$ & ~$-79.5$ & $-61.4$ & $-49.0$ & $(-34.2, 0.01)$ \\
0.50 & 0.25 & 70 & $-325$ & ~$-71.9$ & $-44.9$ & $-25.2$ & $(-0.9, 0.25)$ \\
\hline
\end{tabular}
\caption{\label{tab:bbbarnoNAC}
Energies and widths of $b \bar{b}$ bound states and resonances in the $1^{--}\Sigma_g^+/\Pi_g$ and $0^{-+}\Sigma_u^-$ tetraquark potentials in the $B^{(*)} \bar B^{(*)}$ threshold region.
The energy of the $j^{\pi\gamma} = 1^{--}$ adjoint meson is tuned to the critical energy $E_8^\ast$ for the $1^{++}$ $c \bar{c}$ tetraquark at the $D^* \bar{D}$ threshold.
The parameters are relaxation lengths $R_{\Pi_g}$ and $R_{\Sigma_g^+}$ and the spin splitting $\Delta_{8,b}$ of the $1^{--}$ adjoint meson.
The critical energy $E_8^\ast$ is relative to the spin-weighted $B^{(*)} \bar{B}^{(*)}$ threshold at 10626.9~MeV and the energies $E_{J^{PC}}$ are relative to the $B^*\bar{B}$ threshold at 10604.3~MeV.
A column $E_{J^{PC}}$ gives either the energy of a $J^{PC}$ bound state as a single number or the $K$-matrix energy and width of a $J^{PC}$ resonance as a pair of numbers $(E_{J^{PC}},\Gamma_{J^{PC}})$.
The energies $\Delta_{8,b}$, $E_8^\ast$, and $E_{J^{PC}}$ are in units of MeV. 
}
\end{table}

We consider  four models specified by the two pairs of relaxation lengths $(R_{\Pi_g},R_{\Sigma_g^+} )$ in Eq.~\eqref{RSigma,RPi} and the two adjoint-meson spin splittings $\Delta_{8,b}$ in Eq.~\eqref{Delta8Q-num}.
In each of the four models, we set the adjoint-meson energy $E_{8,1^{--}}$ to the critical energy $E_8^\ast$  for the $1^{++}$ $c \bar{c}$ tetraquark at the $D^* \bar{D}$ threshold, which is given in Table~\ref{tab:noNACccbar}.
We then use SPARSE to calculate the energies of bound states and the $K$-matrix energies and widths of resonances with quantum numbers $0^{++}$, $1^{++}$, $1^{+-}$, and $2^{++}$. 
Their energies relative to the $B^*\bar{B}$ threshold and the $K$-matrix widths are given in Table~\ref{tab:bbbarnoNAC}.
For reference, the $B\bar{B}$, $B^*\bar{B}$, and $B^*\bar{B}^*$ thresholds are $-45.2$~MeV, 0, +45.2~MeV.

There is a $0^{++}$ bound state, a $1^{++}$ bound state, and a $1^{+-}$ bound state for each of the four models.
There is a $2^{++}$ bound state for the 1st model and a $2^{++}$ resonance for the other three models.
Its $K$-matrix energy is below the $B^*\bar{B}$ threshold, so its width comes from $D$-wave decays into $B \bar{B}$.
The spin-weighted energies range from $-24$ to $-79$~MeV.
The spin splittings between the $2^{++}$ and $0^{++}$ states are +37 and +45~MeV for the 1st and 3rd models with $\Delta_{8,b} = 20$~MeV
and they increase to +66 and +71~MeV for the 2nd and 4th models with $\Delta_{8,b} = 70$~MeV.
This suggests that much of the spin splitting comes from the short-distance spin-splitting term in the Hamiltonian.

\subsection{With the Narrow Avoided Crossing}
\label{sec:NACbbbar}

In the absence of the coupling between the $1^{--}\Sigma_g^+$ and $0^{++}\Sigma_g^+$ potentials, the spin-weighted bottomonium masses can be calculated by solving the Schr\"odinger equation with bottom-quark mass $m_b = 4.89$~GeV in the Cornell potential with the parameters in Eq.~\eqref{Cornellparams} and adding $2m_b$ to the eigenvalues.
The masses for $1S$, $1P$, $2S$, and $2P$ differ from the measured spin-weighted masses by less than 20~MeV \cite{Braa14}.  
The first few $P$-wave and $F$-wave bottomonium energy levels relative to the $B^*\bar{B}$ threshold are
\begin{subequations}
\begin{align}
E_{b \bar b,1P}  &= -691.4\,\mathrm{MeV}, \qquad E_{b \bar b,2P} = -334.3\,\mathrm{MeV}, \qquad E_{b \bar b,3P}= -46.2\,\mathrm{MeV},
\label{EnP,b}
\\
E_{b \bar b,1F}  &= -251.1\,\mathrm{MeV}, \qquad E_{b \bar b,2F} = +15.6\,\mathrm{MeV}.
\label{EnF,b}
\end{align}
\end{subequations}
Note that the $3P$ energy level is only about 1~MeV below the $B \bar{B}$ threshold at $-45.2$~MeV while the $2F$ energy level is about 16~MeV above the $B^* \bar B$ threshold at 0.

\begin{table}
\centering
\begin{tabular}{|cccc|cccc|}
\hline
~$R_{\Pi_g}$ [fm]~ & ~$R_{\Sigma_g^+}$  [fm]~ & ~$\Delta_{8,b}$~ & ~$E_8^\ast$ & ~$E_{0^{++}}$  & ~$E_{1^{++}}$ & ~$E_{1^{+-}}$ &  ~$E_{2^{++}}$ \\
\hline
0.25  & 0.50 & 20 & $-521$ & ~$-89.5$~ & ~$-74.3$~ & ~$-63.6$~ & $-49.3$  \\
        &         &       &             & $-57.6$ & $-55.6$ & $-54.9$ & $-56.6$ \\
        &         &       &             & $-$        & $-$        & $-$        & $(+2.6, 9.03)$ \\
0.25 & 0.50 & 70  & $-436$ & $-76.4$ & $-47.9$ & $-29.7$ & $(-7.6, 0.23)$ \\
        &         &       &             & $-56.3$ & $-57.7$ & $-56.1$ & $-55.9$  \\
        &         &       &             & $-$        & $-$        & $-$        & $(+2.8, 8.92)$ \\
0.50 & 0.25 & 20 & $-359$ & $-71.6$  & $-48.7$ & $-37.7$ & $(-22.7, 0.11)$ \\
        &         &      &             & $-54.8$  & $-57.1$ & $-55.6$ & $-55.3$  \\
        &         &      &             & $-$         & $-$        & $-$        & $(+3.5, 8.87)$ \\
0.50 & 0.25 & 70 & $-298$ & $-66.2$  & $-33.4$ & $-15.0$ & $(+9.8, 0.59)$ \\
        &         &      &             & $-52.4$  & $-56.1$ & $-55.4$ & $-55.3$ \\
        &         &      &             & $-$         & $-$        & $-$        & $(+3.5, 8.94)$ \\
\hline
\end{tabular}
\caption{\label{tab:bbbarNAC} 
Energies and widths of $b \bar{b}$ bound states and resonances in the $1^{--}\Sigma_g^+/\Pi_g$ and $0^{-+}\Sigma_u^-$ tetraquark potentials and  the $0^{++}\Sigma_g^+$ quarkonium potential in the $B^{(*)} \bar B^{(*)}$ threshold region.
The energy of the $j^{\pi\gamma} = 1^{--}$ adjoint meson is tuned to the critical energy $E_8^\ast$ for the $1^{++}$ $c \bar{c}$ tetraquark at the $D^* \bar{D}$ threshold.
For each set of parameters, the energies $E_{J^{PC}}$ in the first, second, and third rows are those for $b \bar{b}$  tetraquarks, $P$-wave bottomonium, and $F$-wave bottomonium.
The other aspects of the Table are as described in the caption for Table~\ref{tab:bbbarnoNAC}.
}
\end{table}

When we take into account the narrow avoided crossing between the $1\Sigma_g^+$ and $2\Sigma_g^+$ potentials, our models for the $1^{--}\Pi_g$, $1^{--}\Sigma_g^+$, $0^{-+}\Sigma_u^-$, and $0^{++}\Sigma_g^+$ potentials and our choices for the parameters in the Schr\"odinger equation for $b \bar b$ states are as described in Section~\ref{sec:NACccbar} with a few differences that take into account the change in the heavy quark from charm to bottom.
Our value for the bottom-quark mass $m_b$ is given in Eq.~\eqref{mc,mb} and the other differences  are described in Section~\ref{sec:noNACbbbar}.

We consider four models specified by the two pairs of relaxation lengths $(R_{\Pi_g},R_{\Sigma_g^+})$ in Eq.~\eqref{RSigma,RPi} and the two adjoint-meson spin splittings $\Delta_{8,b}$ in Eq.~\eqref{Delta8Q-num}. 
For each of the four models, we set the adjoint-meson energy $E_{8,1^{--}}$ to the critical value $E_8^\ast$ for the $1^{++}$ $c \bar c$ tetraquark at the $D^*\bar{D}$ threshold in Table~\ref{tab:ccbarNAC}.
We then use SPARSE to calculate the energies of $b \bar b$ bound states and resonances with quantum numbers $0^{++}$, $1^{++}$, $1^{+-}$, and $2^{++}$.
The energies of bound states and the $K$-matrix energies and widths of resonances in the $B^{(*)} \bar{B}^{(*)}$ threshold region are given in Table~\ref{tab:bbbarNAC}.
For reference, the $B\bar{B}$, $B^*\bar{B}$, and $B^*\bar{B}^*$ thresholds are $-45.2$~MeV, 0, +45.2~MeV.

For each of the four models, there are $0^{++}$, $1^{++}$, $1^{+-}$, and $2^{++}$ bound states whose energies are about 60~MeV below the $B^* \bar{B}$ threshold. 
Their spin-weighted $K$-matrix energy is about 10~MeV below the $3P$ bottomonium energy in Eq.~\eqref{EnP,b}.
We identify these four states as the $3P$ bottomonium multiplet.
Each of the four models also has a $2^{++}$ resonance whose energy is about 3~MeV above the $B^*\bar{B}$ threshold. 
Its $K$-matrix energy is 12 or 13~MeV below the $2F$ bottomonium energy in Eq.~\eqref{EnF,b}.
We identify this state as the $2^{++}$ member of the $2F$ bottomonium multiplet, whose other three members have quantum numbers $3^{++}$, $3^{+-}$, and $4^{++}$.

Having identified the bottomonium states in the region of the $B \bar{B}$, $B^* \bar{B}$, and $B^* \bar{B}^*$ thresholds, the remaining states must be $b \bar{b}$ tetraquark bound states and resonances.
There is a single tetraquark bound state for $0^{++}$, $1^{++}$, and $1^{+-}$. 
There is a $2^{++}$ tetraquark bound state for the 1st model and a $2^{++}$ tetraquark resonance for the last three models.
The spin-weighted energies for the four models range from $-62$ to $-14$~MeV and the spin splittings between the $2^{++}$ and $0^{++}$ states range from 40 to 76~MeV.  
The coupling between the $1^{--}\Sigma_g^+$ and $0^{++}\Sigma_g^+$ potentials increases the spin-weighted energies in Table~\ref{tab:bbbarnoNAC} by 10 to 15~MeV and it increases the spin splittings by 3 to 5~MeV.  


\section{Summary and Prospects}
\label{sec:SummaryProspects}

The Born-Oppenheimer (B\nobreakdash-O) approximation  for QCD is an approach to the exotic heavy-hadron problem that is based firmly on the fundamental theory QCD.
Many of the exotic heavy hadrons that have been discovered have mass near the threshold for a pair of heavy hadrons.
The spin splittings of heavy hadrons are small enough that first-order perturbation theory in the heavy-hadron spin splittings may provide a good qualitative approximation.
However, the spin splittings are large enough that a quantitative treatment of the exotic heavy hadrons requires that the spin splittings be treated nonperturbatively.
In this paper, we have presented the first quantitative application of the B\nobreakdash-O approximation for QCD in which heavy-meson and adjoint-meson spin splittings are treated nonperturbatively.

\subsection{Summary}
\label{sec:Summary}

The nonperturbative treatment of heavy-hadron spin splittings is most easily implemented using the diabatic representation of the B\nobreakdash-O approximation.
The diabatic B\nobreakdash-O approximation for QCD was introduced by Bruschini and Gonz\'alez \cite{Brus20}.
In Section~\ref{sec:BO-QCD}, we explained how the diabatic B\nobreakdash-O approximation allows the truncation of the diabatic potentials to be chosen so they have the full rotational and parity symmetries of light QCD.
The diabatic potentials can be labeled by $j^{\pi \gamma}\Lambda_\eta^\epsilon$, where $j^{\pi \gamma}$ specifies the quantum numbers of an adjoint meson and $\Lambda_\eta^\epsilon$ are B\nobreakdash-O  quantum numbers with $\eta =\gamma \pi$.
We also explained how the truncation can be chosen so the $J^{PC}$ quantum numbers are exactly conserved by the solutions to the Schr\"odinger equation.

The heavy-quark spin-symmetry (HQSS) multiplet for $\chi_{c1}(3872)$ ($X_c$) with $J^{PC} = 1^{++}$ also includes three other states with quantum numbers $0^{++}$, $1^{+-}$, and $2^{++}$.
In the HQSS limit, this multiplet corresponds to the ground state in diabatic potentials with quantum numbers $1^{--}\Pi_g$ and $1^{--}\Sigma_g^+$.
The problem of treating heavy-hadron spin splittings systematically was solved by Bruschini using the diabatic B\nobreakdash-O approximation \cite{Brus23a}.
In the case of $X_c$ and its partners, the solution requires extending the truncation of the diabatic potentials to include $0^{-+}\Sigma_u^-$ as well as $1^{--}\Sigma_g^+/\Pi_g$.
The numbers of coupled channels with quantum numbers $0^{++}$, $1^{++}$, $1^{+-}$, and $2^{++}$ required for exact conservation of $J^{PC}$  are 3, 3, 4, and 6, respectively.
The effects of the narrow avoided crossing between the $1\Sigma_g^+$ and $2\Sigma_g^+$ adiabatic potentials can be taken into account by further extending the truncation to include the $0^{++}\Sigma_g^+$ quarkonium potential.
The numbers of coupled channels with quantum numbers $0^{++}$, $1^{++}$, $1^{+-}$, and $2^{++}$ required for exact conservation of $J^{PC}$ are then 4, 4, 5, and 8, respectively.

The radial adiabatic Hamiltonian in the HQSS limit for the truncation to the $1\Pi_g$, $2\Sigma_g^+$, and $1\Sigma_u^-$ potentials is given in Section~\ref{sec:Adiabatic}.
The radial diabatic Hamiltonian in the HQSS limit for the truncation to the $1^{--}\Sigma_g^+/\Pi_g$ and $0^{-+}\Sigma_u^-$ potentials is given in Section~\ref{sec:Diabatic}.
Heavy-quark spins are introduced in Section~\ref{sec:RadTetraCh} and the diabatic Hamiltonian in the HQSS limit is given in the radial tetraquark basis.
Heavy-meson spin splittings are introduced in Section~\ref{sec:RadDiChannels} and the diabatic Hamiltonian in the HQSS limit is given in the radial dimeson basis.
The diabatic Hamiltonian with spin splittings and kinetic improvement is given in Section~\ref{sec:SchrEqSpinSplit} in the radial dimeson basis.
The truncation of the diabatic potentials is extended to include the $0^{++}\Sigma_g^+$ quarkonium potential in Section~\ref{sec:AvoidedCrossing} and the resulting diabatic Hamiltonian is given in both the radial tetraquark basis and the radial dimeson basis.
Explicit expressions for the $J^{PC} = 1^{++}$ blocks of the diabatic Hamiltonians in both the radial tetraquark basis and the radial dimeson basis are given explicitly in Sections~\ref{sec:RadTetraCh}, \ref{sec:RadDiChannels}, \ref{sec:SchrEqSpinSplit}, and \ref{sec:AvoidedCrossing}.
The corresponding $0^{++}$, $1^{+-}$, and $2^{++}$ blocks are given explicitly in Appendix~\ref{app:DiabaticH}.

In Section~\ref{sec:Models}, we introduced simple models for the diabatic potentials.
The $1^{--}\Pi_g$, $1^{--}\Sigma_g^+$, and $0^{-+}\Sigma_u^-$ diabatic potentials approach a repulsive color-Coulomb potential offset by the energy of the $1^{--}$ or $0^{-+}$ adjoint meson at small $r$ and they approach the triplet-meson-pair threshold at large $r$.
Our models for these potentials interpolate between those limits.
We introduced two models for the $1^{--}\Sigma_g^+/\Pi_g$ potentials that cross below the triplet-meson-pair threshold, reach a minimum, cross just barely above the threshold, and then rapidly approach the threshold.
The depths of the minima are controlled by the energy of the $1^{--}$ adjoint meson.
That energy can be adjusted so the ground-state energy is at the triplet-meson-pair threshold.
We introduced a model for the $0^{-+}\Sigma_u^-$ potential that never decreases below the triplet-meson-pair threshold and rapidly approaches the threshold.
It therefore cannot support a bound state.
We used a model introduced in Ref.~\cite{Bram24}  for the transition potential between $1^{--}\Sigma_g^+$ and $0^{++}\Sigma_g^+$.
We introduced a simple model for the spin-dependent $1/m_Q$ potential that interpolates between constant spin-splitting terms at small $r$ and at large $r$.
The spin-splitting term at large $r$ is determined by the known spin splitting $\Delta_Q$ of $S$-wave  heavy mesons.
The spin-splitting term at small $r$ is determined  by the unknown spin splitting $\Delta_{8,Q}$ of the $1^{--}$ adjoint meson.
We considered two possible values of $\Delta_{8,c}$ for the $c \bar c$ case and two possible values of $\Delta_{8,b}$ for the $b \bar b$ case.
We used the SPARSE algorithm to calculate the $K$-matrix for scattering states as a function of the real energy $E$ \cite{Brus25}.
The $T$-matrix energy and width of a resonance were approximated by the $K$-matrix energy and width.

In Section~\ref{sec:DeltaQDependence}, we compared the predictions of first-order perturbation theory in the spin splittings with nonperturbative results 
from solving the Schr\"odinger equation.
We chose one of our two models for the  $1^{--}\Sigma_g^+/\Pi_g$ potentials and we set $\Delta_{8,Q} = (3/2)\Delta_Q$.
We tuned the energy of the $1^{--}$ adjoint meson so there is a $c \bar c$ bound state 50~MeV below the $D^{(*)}\bar{D}^{(*)}$ threshold.
We then calculated the energies of $0^{++}$, $1^{++}$, $1^{+-}$, and $2^{++}$ $c \bar c$ bound states and resonances as functions of $\Delta_Q$ nonperturbatively by solving the Schr\"odinger equation.
The results in Fig.~\ref{fig:SpectrumvsDelta:SchrEq} indicate that first-order perturbation theory remains a good approximation even for $\Delta_Q$ as large as the physical value $\Delta_c$ of the charm-meson spin splitting.
We also considered the case of molecular dominance with ideal mixing by replacing the $0^{-+}\Sigma_u^-$ potential by an attractive potential and tuning the energy of the $0^{-+}$ adjoint meson so there is a $c \bar c$ bound state 50~MeV below the $D^{(*)}\bar{D}^{(*)}$ threshold.
The results in Fig.~\ref{fig:SpectrumvsDelta:SchrEqMD} again indicate that first-order perturbation theory remains a good approximation even for $\Delta_Q$ as large as the physical value $\Delta_c$.

In Section~\ref{sec:CharmoniumTetraquarks}, we calculated the spin splittings of hidden-charm tetraquarks nonperturbatively. 
We considered four specific models defined by our two choices for the $1^{--}\Sigma_g^+/\Pi_g$ potentials and our two choices for $\Delta_{8,c}$.
For each model, we tuned the energy of the $1^{--}$ adjoint meson to the critical value for a $1^{++}$ $c \bar c$ bound state at the $D^*\bar{D}$ threshold.
We then calculated the energies of $0^{++}$, $1^{+-}$, and $2^{++}$ $c \bar c$ bound states and resonances in the $D^{(*)}\bar{D}^{(*)}$ threshold region.
The results are given in Table~\ref{tab:noNACccbar}.
There is a $2^{++}$ tetraquark resonance near the $D^*\bar{D^*}$ threshold in all four models, which is compatible with expectations from first-order perturbation theory in the spin splittings, but there is also an additional  $1^{+-}$ resonance near the $D^*\bar D^*$ threshold in two of the models.
We also considered the effects of transitions between the $1^{--}\Sigma_g^+$ and $0^{++}\Sigma_g^+$ potentials.  
The results are given in Table~\ref{tab:ccbarNAC}.
The states in the $D^{(*)}\bar{D}^{(*)}$ threshold region are a complete $2P$ charmonium multiplet in all four models and a $2^{++}$ tetraquark resonance near the $D^*\bar{D^*}$ threshold in two of the models. 

In Section~\ref{sec:BottomoniumTetraquarks}, we calculated the spin splittings of hidden-bottom tetraquarks nonperturbatively. 
We used the same four models as above except that the spin-splitting  potential is determined by the known spin splitting $\Delta_b$ of $S$-wave bottom mesons and the unknown spin splitting $\Delta_{8,b}$ of the $1^{--}$ adjoint meson.
For each model, we kept the energy of the $1^{--}$ adjoint meson at the critical value for a $1^{++}$ $c \bar c$ bound state at the $D^*\bar{D}$ threshold.
We then calculated the energies of $0^{++}$, $1^{++}$, $1^{+-}$, and $2^{++}$ $b \bar b$ bound states and resonances in the $B^{(*)}\bar{B}^{(*)}$ threshold region.
The results are given in Table~\ref{tab:bbbarnoNAC}.
In all four models, the states form a complete HQSS multiplet of $b \bar{b}$ tetraquark bound states or resonances.
We also considered the effects of  transitions between the $1^{--}\Sigma_g^+$ and $0^{++}\Sigma_g^+$ potentials. 
The results are given in Table~\ref{tab:bbbarNAC}.
In all four models, the states in the $B^{(*)}\bar{B}^{(*)}$ threshold region consist of bound states that form the complete $3P$ bottomonium multiplet, a $2^{++}$ resonance that belongs to the $2F$ bottomonium multiplet,  and a complete HQSS multiplet of $b \bar{b}$ tetraquark bound states or resonances.  

\subsection{Improvements}
\label{sec:Improvements}

We have presented the simplest possible treatment of $X_c$ and its HQSS partners using the diabatic B\nobreakdash-O approximation for QCD in which charm-meson spin splittings are taken into account nonperturbatively.
There are many improvements that would be required before the predictions for the HQSS partners could be regarded as quantitative.
Many of them would be straightforward to implement within the framework we have presented.

We have used charm and bottom quark masses determined by a fit to spin-weighted charmonium and bottomonium energy levels in a Cornell potential \cite{Braa14}.
It would be better to use heavy-quark masses with a more direct connection to the parameters of QCD, such as those in the Minimal Renormalon Subtraction scheme \cite{Baza18}.

We have introduced extremely simple models for the $1^{--}\Sigma_g^+/\Pi_g$ and $0^{-+}\Sigma_u^-$ diabatic  potentials.
Their behavior at small $r$ could be improved by lattice QCD calculations of the energies of the $1^{--}$ and $0^{-+}$ adjoint mesons.
Their behavior at large $r$ could be improved by taking into account the effects of pion exchange.
Their behavior at intermediate $r$ could be improved by lattice QCD calculations of the $1\Pi_g$, $2\Sigma_g^+$, and $1\Sigma_u^-$  adiabatic potentials.
Our models for the $1^{--}\Pi_g$ and $1^{--}\Sigma_g^+$ potentials decrease to a minimum and then approach the triplet-meson-pair threshold rapidly at larger $r$. 
A possible improvement to the models would be to allow the potentials to increase well above the triplet-meson-pair threshold before approaching it rapidly at larger $r$.
This would allow the possibility of $S$-wave resonances in the HQSS limit with energies above the threshold that are trapped in the short-distance region of the potentials and decay into heavy-meson pairs by tunneling through the classically forbidden region.

We have taken into account transitions between the $1^{--}\Sigma_g^+$ and $0^{++}\Sigma_g^+$ diabatic potentials using the model in Ref.~\cite{Bram24}, which was informed by the lattice QCD calculation of the narrow avoided crossing between the $1\Sigma_g^+$ and $2\Sigma_g^+$ adiabatic potentials in Ref.~\cite{Bul24}.  
This avoided crossing involves the creation or annihilation of a light-quark pair $u \bar{u} + d \bar{d}$.
However, the narrow avoided crossing between the $2\Sigma_g^+$ and $3\Sigma_g^+$ adiabatic potentials was also calculated in Ref.~\cite{Bul24} and it is nearby in energy. 
This avoided crossing involves the creation or annihilation of a strange-quark pair $s \bar{s}$.
To take into account this second avoided crossing, our truncation of the diabatic potentials would have to be extended to include the $1^{--}\Sigma_g^+$ potential associated with a $1^{--}$ adjoint meson that contains a strange-quark pair.

We used spin-averaged  energy levels for charmonium and bottomonium multiplets below the heavy-meson-pair threshold as inputs to determine the Cornell potential.
The spin splittings of $P$-wave quarkonium states are dominated by $1/m_Q^2$ spin-dependent terms in the Hamiltonian.
Taking these spin-dependent terms into account would be important  for accurate predictions of the spin splittings of the HQSS multiplet for $X_c$ and the spin splittings of  the nearby $2P$ charmonium multiplet.

We have taken into account spin splittings between charm mesons, but we have ignored isospin splittings.
The mass of $X_c$ is extremely close to the $D^{\ast0} \bar{D}^0$ threshold, which is 8.2~MeV below the $D^{\ast+} D^-$ threshold.
To predict the masses of the spin-symmetry partners of $X_c$ with a precision much smaller than 10~MeV, it would be necessary to take into account the isospin splittings of charm mesons.
Taking into account charm-meson-pair channels with $u \bar{u}$ and $d \bar{d}$ would double the number of scattering channels.
If we also take into account $s \bar{s}$ channels, the number of scattering channels is tripled.
This proliferation of the number of channels for each $J^{PC}$ can easily be handled by the SPARSE algorithm for calculating $K$-matrix elements with real energy $E$.

The SPARSE algorithm calculates the $K$-matrix energy and width of a narrow isolated resonance efficiently even if there are dozens of scattering channels.
Determining the complex $T$-matrix energy of the resonance accurately would require developing a model for the analytic continuation of the $K$-matrix into the multi-sheeted complex $E$ plane.
This analytic continuation would also be essential to determine the pole parameters of virtual states.

\subsection{Other Applications}
\label{sec:Extensions}

The Schr\"odinger equations given explicitly in this paper can be directly applied to other hidden-heavy mesons whose diabatic potentials approach the threshold for a pair of $S$-wave heavy mesons at large $r$.
They include the isospin-1 tetraquark mesons $T_{b \bar b1}(10610)$ and $T_{b \bar b1}(10650)$ with quantum numbers $1^{+-}$.
Their properties require a bound state near threshold in the $0^{-+}\Sigma_u^-$ potential as well as in the $1^{--}\Sigma_g^+/\Pi_g$ potentials \cite{Braa24b}.
A quantitative analysis of these $b \bar{b}$ tetraquark mesons, their spin-symmetry partners, and the corresponding $c \bar{c}$ tetraquark mesons using the diabatic B\nobreakdash-O approximation for QCD is in progress \cite{Alas26x}. 
The Schr\"odinger equations for hidden-heavy hadrons whose diabatic potentials approach other heavy-hadron-pair thresholds can be determined using results given in this paper.
One interesting application is hidden-charm pentaquarks for which the B\nobreakdash-O approximation gives predictions for their $J^{PC}$ quantum numbers that differ from most previous predictions \cite{Alas25}.
Another interesting application is the first isospin-1 tetraquark to be discovered, $T_{c \bar c1}(4430)$, whose energy is near the threshold for an $S$-wave charm meson and a $P$-wave charm meson.

A  qualitative pattern for the exotic-hidden-heavy hadrons was proposed in Ref.~\cite{Braa24b}, where it was applied to hidden-heavy tetraquarks.
We applied that qualitative pattern to hidden-charm pentaquarks in Ref.~\cite{Alas25}.
In this paper, we developed a strategy for the quantitative analysis of hidden-heavy hadrons in which the effects of heavy-hadron spin splittings are taken into account nonperturbatively.
We illustrated that strategy by applying it to the $X_c$ multiplet.
The accuracy of the approach can be improved by incorporating results from lattice QCD as they become available and by imposing additional constraints from experiment as they become available.
The successful quantitative descriptions of several multiplets of hidden-heavy hadrons could confirm the pattern proposed in Ref.~\cite{Braa24b}.
It would provide compelling evidence that an understanding of the hidden-heavy hadrons based firmly on QCD has indeed been achieved.

\begin{acknowledgments}
This research was supported in part by the Department of Energy under grant DE-FG02-05ER15715.
This work contributes to the goals of the US DOE ExoHad Topical Collaboration, Contract DE-SC0023598.
We thank Abhishek Mohapatra for valuable discussions.
\end{acknowledgments}


\appendix

\section{Blocks of Radial  Diabatic Hamiltonian for other \texorpdfstring{$J^{PC}$}{JPC}}
\label{app:DiabaticH}

In this Appendix, we give the radial diabatic Hamiltonian matrices for the other three quantum numbers $J^{PC}$ in the HQSS multiplet that includes the $1^{++}$ state we identify with $X_c$.  
The other members of the multiplet have $J^{PC}= 0^{++}$, $2^{++}$, and $1^{+-}$.

\subsection{\texorpdfstring{$J^{PC}$}{JPC} Blocks of Hamiltonian}
\label{sec:JPC}

The number of radial diabatic channels in the $J^{PC}$ block of the radial diabatic Hamiltonian depends on $J^{PC}$.
The radial tetraquark channels $\big( (j^{\pi\gamma},L_{Q \bar Q}) L, S_{Q \bar Q} \big)$ in the $J^{PC}$ block are constrained by the triangle relations $\{ j,L_{Q \bar Q}, L \}=1$ and $\{ L, S_{Q \bar Q},J \}=1$ and by the eigenvalues $P = \pi (-1)^{L_{Q \bar Q}+1}$ and $CP = \gamma\pi (-1)^{S_{Q \bar Q}+1}$.
The  radial dimeson channels $\big((J_1^-,J_2^-)S,L_{Q\bar Q} \big)$ in the $J^P$ block are constrained by the triangle relations $\{ J_1, J_2, S \}=1$ and $\{ S, L_{Q \bar Q},J \}=1$ and the parity eigenvalue $P = (-1)^{L_{Q\bar Q}}$.
The $J^{PC}$ block consists of the channels in Eqs.~\eqref{dimesonCP+} for $CP = +1$ and the channels in Eqs.~\eqref{dimesonCP-} for $CP = -1$. 
The transformation from the basis of radial tetraquark channels to the basis of radial dimeson channels is given in  Eq.~\eqref{di-tetra12}. 
If the truncation of the diabatic potentials includes the quarkonium potential, there are also radial quarkonium channels $Q \bar{Q}(S_{Q\bar Q},L_{Q\bar Q})$ that differ from $\big( (0^{++},L_{Q \bar Q}) L_{Q \bar Q}, S_{Q \bar Q} \big)$ at most  by the sign in Eq.~\eqref{QQbar:dimeson-tetra}.

The general form of the $J^{PC}$ block of the radial diabatic Hamiltonian $\mathbf{H}_0$ in the HQSS limit is given in Eq.~\eqref{H0-JPC}.
The general form of the $J^{PC}$ block of the radial diabatic Hamiltonian $\mathbf{H}_1$ with spin splittings and kinetic improvement is given in Eq.~\eqref{H1-JPC}.
The orbital-angular-momentum matrix $\mathbf{L}_{Q \bar Q}^2$ is diagonal with diagonal entries $L_{Q \bar Q}(L_{Q \bar Q}+1)$.
In the radial tetraquark basis, the radial diabatic static potential $\mathbf{V}_0(r)$ is block diagonal in the quantum numbers $(L,S_{Q\bar Q})$.
The entries of the $(L,S_{Q\bar Q})J^{PC}$ block are given by Eq.~\eqref{V0JPC} in terms of the entries of the $L$ block of $\mathbf{V}^L(r)$, which are given in Eq.~\eqref{V0-L}.
In the radial dimeson basis, the entries of the $J$ block of $\mathbf{V}_0(r)$ are given in Eq.~\eqref{Vstatic-J} in terms of the entries of the $L$ block of $\mathbf{V}^L(r)$.
The spin-splitting potential $\mathbf{V}_\text{SS}$ is the interpolation in Eq.~\eqref{VSS-r} between the short-distance spin-splitting term $\mathbf{V}_{\text{SS},\text{short}}$ at small $r$ and the long-distance spin-splitting term $\mathbf{V}_{\text{SS},\text{long}}$ at large $r$.
The $J$ blocks of $\mathbf{V}_{\text{SS},\text{short}}$ and $\mathbf{V}_{\text{SS},\text{long}}$ in the radial dimeson basis are given by Eqs.~\eqref{VSSshort-J} and \eqref{VSSlong-J}.

If the diabatic potentials are truncated to $1^{--}\Sigma_g^+$, $1^{--}\Pi_g$, and $0^{-+}\Sigma_u^-$, the entries of the $L$ block of $\mathbf{V}_0(r)$ are given in Eq.~\eqref{VLQQbar}.
They are linear combinations of the diabatic potentials $V_{1^{--}\Sigma_g^+}(r)$, $V_{1^{--}\Pi_g}(r)$, and $V_{0^{-+}\Sigma_u^-}(r)$.
In this Appendix, we denote these three tetraquark potentials more concisely as $V_{\Sigma_g^+}(r)$, $V_{\Pi_g}(r)$, and $V_{\Sigma_u^-}(r)$.
If the truncation is extended to include $0^{++}\Sigma_g^+$, the additional entries of the $L$ block of $\mathbf{V}_0(r)$ are given in Eqs.~\eqref{VLQQbarAC}.
They have a factor of either the diabatic quarkonium potential $V_{0^{++}\Sigma_g^+}(r)$, which we denote more concisely as $V_{Q \bar Q}(r)$, or the transition potential $G_{(0^{++},1^{--})\Sigma_g^+}(r)$, which we denote more concisely as $G_{\Sigma_g^+}(r)$.

\subsection{\texorpdfstring{$J^{PC} = 0^{++}$}{JPC = 0++} Block}
\label{sec:0++}

We first give the $0^{++}$ block of $\mathbf{H}_0$ in the radial tetraquark basis. 
We then give the $0^{++}$ block of $\mathbf{H}_1$ in the radial dimeson basis.

\subsubsection{HQSS Limit}

When the diabatic potentials are truncated to $0^{-+}\Sigma_u^-$,  $1^{--}\Sigma_g^+$, and  $1^{--}\Pi_g$,
there are 3 radial tetraquark channels $\big( (j^{\pi\gamma},L_{Q \bar Q}) L, S_{Q \bar Q} \big)$ for $J^{PC}=0^{++}$. 
We enumerate them in order of increasing $L_{Q \bar Q}$, increasing $L$, and then increasing $S_{Q \bar Q}$:\\
\indent \indent \indent 1: $\big( (0^{-+},0) 0, 0 \big)$,\\
\indent \indent \indent 2: $\big( (1^{--},0) 1, 1 \big)$,\\
\indent \indent \indent 3:  $\big( (1^{--},2) 1, 1 \big)$.\\
Channels 1 and 2 are $S$-wave and channel 3 is $D$-wave.

We now give the $3 \times 3$ matrices in the $0^{++}$ block of $\mathbf{H}_0$ in the radial tetraquark basis.
The orbital-angular-momentum matrix is
\begin{equation}
\mathbf{L}_{Q \bar Q}^2 =  \mathrm{diag}\big( 0,0,6 \big) .
\label{L2:0++3}
\end{equation}
The radial diabatic static potential for $0^{++}$  in the radial tetraquark basis is
\begin{eqnarray}
 \mathbf{V}^{0^{++}}_0(r) &=& V_{\Pi_g}(r) \bm{1}_{3 \times 3} 
+ \big[ V_{\Sigma_u^-}(r)  - V_{\Pi_g}(r) \big] \,
 \mathrm{diag}\big(1,0,0 \big)
\nonumber\\
&&  \hspace{2cm}+ \frac{1}{3}\big[ V_{\Sigma_g^+}(r) - V_{\Pi_g}(r) \big]
\begin{pmatrix}
~0~	& 0	        & 0             \\
0	& 1		& -\sqrt{2}  \\
0	& -\sqrt{2}	& 2		 \end{pmatrix}.
\label{V0++tetra}
\end{eqnarray}
The only nonzero off-diagonal entries are those that couple channels 2 and 3.
If $V_{\Sigma_g^+}(r) = V_{\Pi_g} (r)$, each of the three channels is decoupled.

When the truncation of the diabatic potentials is extended to include $0^{++}\Sigma_g^+$, there are additional radial quarkonium channels of the form $\big((0^{++}, L_{Q \bar Q}) L, S_{Q \bar Q} \big)$, with $L_{Q \bar Q} = L$, $P=(-1)^{L+1}$, and $CP=(-1)^{S_{Q \bar Q}+1}$.
The additional radial quarkonium channel with $J^{PC} = 0^{++}$ is \\
\indent \indent \indent 4.  $\big( (0^{++},1) 1, 1 \big)$.  \\
This is a $P$-wave channel.

We now give the $4 \times 4$ matrices in the $0^{++}$ block of $\mathbf{H}_0$ in the radial tetraquark basis.
The orbital-angular-momentum matrix in Eq.~\eqref{L2:0++3} is extended to a $4 \times 4$ diagonal matrix:
\begin{equation}
\mathbf{L}_{Q \bar Q}^2 =  \mathrm{diag}\big( 0,0,6,2 \big) .
\label{L2:0++4}
\end{equation}
The static potential for $0^{++}$ is the sum of the matrix in Eq.~\eqref{V0++tetra} (with a 4th row and a 4th column of zeros) and two additional $4 \times 4$ matrices:
\begin{equation}
 \Delta\mathbf{V}^{0^{++}}_0(r) =
V_{Q \bar Q}(r)
\begin{pmatrix}
\bm{0}_{3 \times 3} & \bm{0}_{3 \times 1}\\
\bm{0}_{1 \times 3} & 1                                                \\
\end{pmatrix} 
+ G_{\Sigma_g^+}(r)
\begin{pmatrix}
\bm{0}_{3 \times 3} & \hat{\bm{N}}_4^T\\
\hat{\bm{N}}_4 & 0                                                \\
\end{pmatrix} ,
\label{V0++:4}
\end{equation}
where $\bm{0}_{m \times n}$ is the $m \times n$ zero matrix and $\hat{\bm{N}}_4$ is the 3-component unit  row vector 
\begin{equation}
\hat{\bm{N}}_4 = \begin{pmatrix} 0 & \sqrt{\frac13} & -\sqrt{\frac23}  \end{pmatrix} .
\end{equation}

\subsubsection{With Spin Splittings}

When the diabatic potentials are truncated to $0^{-+}\Sigma_u^-$,  $1^{--}\Sigma_g^+$, and $1^{--}\Pi_g$, there are 3 radial dimeson channels $\big((J_1^-, J_2^-) S,L_{Q\bar Q} \big)$ for  $J^{PC}=0^{++}$.
We enumerate them in order of increasing  $L_{Q \bar Q}$, increasing  $J_1+J_2$, and then increasing $S$.
We take channels 1, 2 and 3 to be $\big( ( 0^-,0^-)0,0 \big)$, $\big( ( 1^-,1^-)0,0 \big)$, and $\big( ( 1^-,1^-)2,2 \big)$.
Our 3 radial dimeson channels labeled by charm-meson pairs and by $(S,L_{Q \bar Q})$ are\\
\indent \indent \indent 1: $D \bar D(0,0)$,\\
\indent \indent \indent 2: $D^* \bar D^*(0,0)$,\\
\indent \indent \indent 3:  $D^* \bar D^*(2,2)$.

We now give the $3 \times 3$ matrices in the $0^{++}$ block of $\mathbf{H}_1$ in the radial dimeson  basis.
The orbital-angular-momentum matrix $\mathbf{L}_{Q \bar Q}^2$ is given in Eq.~\eqref{L2:0++3}.
The static potential for $0^{++}$ in the radial dimeson basis is
\begin{eqnarray}
 \mathbf{V}^{0^{++}}_0(r) &=& V_{\Pi_g}(r) \bm{1}_{3 \times 3} 
+ \frac14 \big[ V_{\Sigma_u^-}(r) - V_{\Pi_g}(r) \big] 
\begin{pmatrix}
~1~	    & \sqrt{3} & ~0~ \\
\sqrt{3} & 3	  & 0      \\
0	    & 0     & 0          \end{pmatrix}
\nonumber\\
&& \hspace{2cm} + \frac{1}{12}\big[ V_{\Sigma_g^+}(r) - V_{\Pi_g}(r) \big]
\begin{pmatrix}
3             & -\sqrt{3}	& -2\sqrt{6} \\
-\sqrt{3}   & 1		& 2\sqrt{2}  \\
-2\sqrt{6} & 2\sqrt{2}	& 8		  \end{pmatrix}.
\label{V0++dimeson}
\end{eqnarray}
The spin-splitting potential is
\begin{multline}
\mathbf{V}^{0^{++}}_\text{SS}(r) = \frac{1}{6}\, \Delta_{8,Q}
\begin{pmatrix}
-3         & \sqrt{3} & ~0~ \\
\sqrt{3} & -1          & 0  \\
~0~       & 0          & 2 \end{pmatrix}
\exp\big( - r^2/R_\text{SS}^2 \big)
\\ 
+ \frac{1}{2}\, \Delta_Q\,\mathrm{diag}\big( -3, 1, 1 \big) \left[ 1 - \exp\big( -r^2/R_\text{SS}^2 \big) \right] ,
\label{VSS:0++3}
\end{multline}
where  $\Delta_{8,Q}$ is the adjoint-meson spin splitting and $\Delta_Q$ is the heavy-meson spin splitting.
If $V_{\Sigma_g^+}(r) = V_{\Pi_g}(r)$, channels 1 and 2 are decoupled from channel 3. 
The kinetically improved mass matrix is
\begin{equation}
\mathbf{M}_{0^{++}} =  \mathrm{diag}\big( m_D, m_{D^*}, m_{D^*} \big) .
\label{M:0++2}
\end{equation}

When the truncation of the diabatic potentials is extended to include $0^{++}\Sigma_g^+$, there are additional radial quarkonium channels of the form $Q \bar{Q}(S_{Q \bar Q},L_{Q \bar Q})$, with $P=(-1)^{L_{Q \bar Q}+1}$ and $CP=(-1)^{S_{Q \bar Q}+1}$.
The radial quarkonium channel $Q \bar{Q}(S_{Q \bar Q},L_{Q \bar Q})$ with quantum numbers $J^{PC}$ is equal to $\big((0^{++}, L_{Q \bar Q}) L_{Q \bar Q}, S_{Q \bar Q} \big)$ multiplied by $(-1)^{J - L_{Q \bar Q} - S_{Q \bar Q}}$.
The additional radial quarkonium channel with $J^{PC} = 0^{++}$ is\\
\indent \indent \indent 4. $Q \bar Q(1,1)$. \\
This is equal to the radial quarkonium channel $\big( (0^{++},1) 1, 1 \big)$ in the radial tetraquark basis.

We now give the $4 \times 4$ matrices in the $0^{++}$ block of $\mathbf{H}_1$ in the radial dimeson basis.
The orbital-angular-momentum matrix $\mathbf{L}_{Q \bar Q}^2$ is given in Eq.~\eqref{L2:0++4}.
The static potential $\mathbf{V}^{0^{++}}_0(r)$ is the sum of the matrix in Eq.~\eqref{V0++dimeson} (with a 4th row and a 4th column of zeros) and two additional $4 \times 4$ matrices with the form in Eq.~\eqref{V0++:4} except that $\hat{\bm{N}}_4$ is the 3-component unit  row vector 
\begin{equation}
\hat{\bm{N}}_4 = \begin{pmatrix} -\frac12 & \sqrt{\frac{1}{12}} & \sqrt{\frac23} \end{pmatrix} .
\end{equation}
The spin-splitting potential $\mathbf{V}^{0^{++}}_\text{SS}(r)$ in Eq.~\eqref{VSS:0++3} is extended to a $4\times 4$ matrix with a 4th row and a 4th column of zeros.
The mass matrix in Eq.~\eqref{M:0++2} is extended to a $4 \times 4$ matrix whose 4th diagonal entry is $m_Q$:
\begin{equation}
\mathbf{M}_{0^{++}} =  \mathrm{diag}\big( m_D, m_{D^*}, m_{D^*},m_Q \big) .
\label{M:0++4}
\end{equation}

\subsection{\texorpdfstring{$J^{PC} = 1^{+-}$}{JPC = 1+-} Block}
\label{sec:1+-}

We first give the $1^{+-}$ block of $\mathbf{H}_0$ in the radial tetraquark basis.
We then give the $1^{+-}$  block of $\mathbf{H}_1$ in the radial  dimeson basis.

\subsubsection{HQSS Limit}

When the diabatic potentials are truncated to $0^{-+}\Sigma_u^-$,  $1^{--}\Sigma_g^+$, and $1^{--}\Pi_g$,
there are 4 radial tetraquark channels $\big( (j^{\pi\gamma},L_{Q \bar Q}) L, S_{Q \bar Q} \big)$ for $J^{PC}=1^{+-}$.
We enumerate them in order of increasing $L_{Q \bar Q}$, increasing $L$, and then increasing $S_{Q \bar Q}$:\\
\indent \indent \indent  1: $\big( (0^{-+},0) 0, 1 \big)$,\\
\indent \indent \indent  2: $\big( (1^{--},0) 1, 0 \big)$,\\
\indent \indent \indent  3:  $\big( (1^{--},2) 1, 0 \big)$,\\
\indent \indent \indent  4: $\big( (0^{-+},2) 2, 1 \big)$.\\
Channels 1 and 2 are $S$-wave and channels 3 and 4 are $D$-wave.

We now give the $4 \times 4$ matrices in the $1^{+-}$ block of $\mathbf{H}_0$ in the radial tetraquark basis.
The orbital-angular-momentum matrix is
\begin{equation}
\mathbf{L}_{Q \bar Q}^2 =  \mathrm{diag}\big( 0,0,6,6 \big) .
\label{L2:1+-4}
\end{equation}
The radial diabatic static potential for $1^{+-}$  in the radial tetraquark basis is
\begin{multline}
 \mathbf{V}^{1^{+-}}_0(r) = V_{\Pi_g}(r) \bm{1}_{4 \times 4} 
+ \big[  V_{\Sigma_u^-}(r) - V_{\Pi_g}(r) \big]\,
 \mathrm{diag}\big( 1,0,0,1 \big)\\
+ \frac{1}{3}\big[ V_{\Sigma_g^+}(r)- V_{\Pi_g}(r) \big]
\begin{pmatrix}
~0~ & 0            & 0            & ~0~ \\
0     & 1            & -\sqrt{2}	& 0    \\
0     & -\sqrt{2} & 2             & 0     \\
0     & 0           & 0             & 0    \end{pmatrix}.
\label{V0:1+-}
\end{multline}
The only nonzero off-diagonal entries are those that couple channels 2 and 3.
If $V_{\Sigma_g^+}(r) = V_{\Pi_g} (r)$, each of the four channels is decoupled.

When the truncation of the diabatic potentials is extended to include $0^{++}\Sigma_g^+$, the only additional radial quarkonium channel with  $J^{PC} = 1^{+-}$ is \\
\indent \indent \indent 5.  $\big( (0^{++},1) 1, 0 \big)$.  \\
This is a $P$-wave channel.

We now give the $5 \times 5$ matrices in the $1^{+-}$ block of $\mathbf{H}_0$ in the radial tetraquark basis.
The orbital-angular-momentum matrix in Eq.~\eqref{L2:1+-4} is extended to a $5 \times 5$ diagonal matrix:
\begin{equation}
\mathbf{L}_{Q \bar Q}^2 =  \mathrm{diag}\big( 0,0,6,6,2 \big) .
\label{L2:1+-5}
\end{equation}
The static potential for $1^{+-}$ is the sum of the matrix in Eq.~\eqref{V0:1+-} (with a 5th row and a 5th column of zeros) and two additional matrices:
\begin{equation}
 \Delta\mathbf{V}^{1^{+-}}_0(r) =
V_{Q \bar Q}(r)
\begin{pmatrix}
\bm{0}_{4 \times 4} & \bm{0}_{4 \times 1}\\
\bm{0}_{1 \times 4} & 1                                                \\
\end{pmatrix} 
+ G_{\Sigma_g^+}(r)
\begin{pmatrix}
\bm{0}_{4 \times 4} & \hat{\bm{N}}_5^T \\
\hat{\bm{N}}_5 & 0                                                \\
\end{pmatrix} ,
\label{V1+-:4}
\end{equation}
where $\hat{\bm{N}}_5$ is the 4-component unit  row vector
\begin{equation}
\hat{\bm{N}}_5 = \begin{pmatrix} 0 & \sqrt{\frac13} & -\sqrt{\frac23} & 0 \end{pmatrix} .
\end{equation}

\subsubsection{With Spin Splittings}

When the diabatic potentials are truncated to $0^{-+}\Sigma_u^-$,  $1^{--}\Sigma_g^+$, and $1^{--}\Pi_g$, there are 4 radial dimeson channels $\big( (J_1^-, J_2^-) S,L_{Q\bar Q} \big)$ for $J^{PC}=1^{+-}$.
We enumerate them in order of increasing  $L_{Q \bar Q}$, increasing  $J_1+J_2$, and then increasing $S$.
We take channel 1 to be the odd superposition of $\big( ( 1^-,0^-)1,0 \big)$ and $\big( ( 0^-,1^-)1,0 \big)$.
We take channel 2 to be $\big( ( 1^-,1^-)1,0 \big)$.
We take channel 3 to be the odd superposition of $\big( ( 1^-,0^-)1,2 \big)$ and $\big( ( 0^-,1^-)1,2 \big)$.
We take channel 4 to be $\big( ( 1^-,1^-)1,2 \big)$.
Our 4 radial dimeson channels for  $J^{PC}=1^{+-}$  labeled by charm-meson pairs and by $(S,L_{Q \bar Q})$ are\\
\indent \indent \indent  1: $[D^* \bar D(1,0) - D \bar D^*(1,0)]/\sqrt2$, \\
\indent \indent \indent  2:  $D^* \bar D^*(1,0)$,\\
\indent \indent \indent  3: $[D^* \bar D(1,2) - D \bar D^*(1,2)]/\sqrt2$,\\
\indent \indent \indent  4: $D^* \bar D^*(1,2)$.

We now give the $4 \times 4$ matrices in the $1^{+-}$ block of $\mathbf{H}_1$ in the radial dimeson  basis.
The orbital-angular-momentum matrix $\mathbf{L}_{Q \bar Q}^2$ is given in Eq.~\eqref{L2:1+-4}.
The static potential for $1^{+-}$ in the radial dimeson basis  is
\begin{multline}
 \mathbf{V}^{1^{+-}}_0(r) =
V_{\Pi_g}(r) \bm{1}_{4 \times 4} 
+ \frac{1}{2} \big[ V_{\Sigma_u^-}(r) - V_{\Pi_g}(r) \big]
\begin{pmatrix}
 ~1~ & ~-1~ & ~0~ & ~0~ \\
-1      & 1	  & 0     & 0     \\
0      & 0	  & 1	    & -1     \\
0      & 0     & -1      & 1    \end{pmatrix}
\\
+ \frac{1}{6}\big[ V_{\Sigma_g^+}(r) - V_{\Pi_g}(r) \big]
\begin{pmatrix}
1            & 1           & -\sqrt{2} & -\sqrt{2}  \\
1           & 1           & -\sqrt{2}   & -\sqrt{2}          \\
-\sqrt{2} & -\sqrt{2}  & 2            & 2 \\
-\sqrt{2}  & -\sqrt{2} & 2           & 2           \end{pmatrix}.
\label{V0:1+-4}
\end{multline}
The spin-splitting  potential is
\begin{equation}
\mathbf{V}^{1^{+-}}_\text{SS}(r) = 
\frac{\Delta_Q}{2}\, \mathrm{diag}\big( -1, 1, -1, 1 \big) \left[ 1 - \exp\big( -r^2/R_\text{SS}^2 \big) \right] .
\label{VSS:1+-4}
\end{equation}
Note that there is no contribution from the short-distance spin-splitting potential.
If $V_{\Sigma_g^+}(r) = V_{\Pi_g}(r)$, the only coupled channels are channels 1 and 2 and channels 3 and 4.
The kinetically improved mass matrix is
\begin{equation}
\mathbf{M}_{1^{+-}} =  \mathrm{diag}\big( 2 \mu_{D^* D}, m_{D^*}, 2 \mu_{D^* D}, m_{D^*} \big) .
\label{M:1+-4}
\end{equation}

When the truncation of the diabatic potentials is extended to include $0^{++}\Sigma_g^+$, the only additional radial quarkonium channel with $J^{PC} = 1^{+-}$ is \\
\indent \indent \indent 5. $Q \bar Q(0,1)$.  \\
This is equal to the quarkonium channel $\big( (0^{++},1) 1, 0 \big)$ in the radial tetraquark basis.

We now give the $5 \times 5$ matrices in the $1^{+-}$ block of $\mathbf{H}_1$ in the radial dimeson basis.
The orbital-angular-momentum matrix $\mathbf{L}_{Q \bar Q}^2$ is given in Eq.~\eqref{L2:1+-5}.
The static potential $\mathbf{V}^{1^{+-}}_0(r)$ is the sum of the matrix in Eq.~\eqref{V0:1+-4} (with a 5th row and a 5th column of zeros) and two additional matrices of the form in Eq.~\eqref{V1+-:4} except that $\hat{\bm{N}}_5$ is the 4-component unit  row vector
\begin{equation}
\hat{\bm{N}}_5 = \begin{pmatrix}  -\sqrt{\frac{1}{6}} ~& -\sqrt{\frac{1}{6}} ~& \sqrt{\frac{1}{3}} ~&~  \sqrt{\frac{1}{3}} \end{pmatrix} .
\end{equation}
The spin-splitting potential $\mathbf{V}^{1^{+-}}_\text{SS}(r)$ in Eq.~\eqref{VSS:1+-4} is extended to a $5 \times 5$ matrix whose 5th diagonal entry is 0.
The mass matrix in Eq.~\eqref{M:1+-4} is extended to a $5 \times 5$ matrix whose 5th diagonal entry is $m_Q$:
\begin{equation}
\mathbf{M}_{1^{+-}} =  \mathrm{diag}\big( 2 \mu_{D^* D}, m_{D^*}, 2 \mu_{D^* D}, m_{D^*} ,m_Q \big) .
\label{M:1+-5}
\end{equation}

\subsection{\texorpdfstring{$J^{PC} = 2^{++}$}{JPC = 2++} Block}
\label{sec:2++}

We first give the $2^{++}$ block of $\mathbf{H}_0$ in the radial tetraquark basis.
We then give the $2^{++}$ block of $\mathbf{H}_1$ in the radial  dimeson basis.

\subsubsection{HQSS Limit}

When the diabatic potentials are truncated to $0^{-+}\Sigma_u^-$,  $1^{--}\Sigma_g^+$, and $1^{--}\Pi_g$, there are 6 radial  tetraquark channels $\big( (j^{\pi\gamma},L_{Q \bar Q}) L, S_{Q \bar Q} \big)$ for $J^{PC}=2^{++}$. 
We enumerate them in order of increasing $L_{Q \bar Q}$, increasing $L$, and then increasing $S_{Q \bar Q}$:\\
\indent \indent \indent 1: $\big( (1^{--},0) 1, 1 \big)$,\\
\indent \indent \indent 2: $\big( (1^{--},2) 1, 1 \big)$,\\
\indent \indent \indent 3:  $\big( (0^{-+},2) 2, 0 \big)$,\\
\indent \indent \indent 4: $\big( (1^{--},2) 2, 1 \big)$,\\
\indent \indent \indent 5: $\big( (1^{--},2) 3, 1 \big)$,\\
\indent \indent \indent 6: $\big( (1^{--},4) 3, 1 \big)$.\\
Channel 1 is $S$-wave, channels 2$-$5 are $D$-wave, and channel 6 is $G$-wave.

We now give the $6\times 6$ matrices in the $2^{++}$ block of $\mathbf{H}_0$ in the radial tetraquark basis.
The orbital-angular-momentum matrix is
\begin{equation}
\mathbf{L}_{Q \bar Q}^2 = \mathrm{diag} \big( 0,6,6,6,6,20\big) .
\label{L2:2++6}
\end{equation}
The radial diabatic static potential for $2^{++}$ in the radial tetraquark basis is
\begin{multline}
\mathbf{V}^{2^{++}}_0 = V_{\Pi_g}(r) \bm{1}_{6 \times 6} 
+\big[ V_{\Sigma_u^-}(r) - V_{\Pi_g}(r) \big]\, \mathrm{diag}\big( 0,0,1,0,0,0 \big)
\\
+ \frac{1}{21}\big[ V_{\Sigma_g^+}(r) -  V_{\Pi_g}(r) \big]
\begin{pmatrix}
7              & -7\sqrt{2} &~ 0~ & ~0~ & 0              & 0              \\
-7\sqrt{2} & 14            & 0     & 0     & 0              & 0              \\
0              & 0              & 0     & 0     & 0              & 0             \\
0              & 0              & 0     & 0     & 0              & 0             \\
0              & 0              & 0     & 0     & 9              & -6\sqrt{3} \\
0              & 0              & 0     & 0     & -6\sqrt{3} & 12            \end{pmatrix}.
\label{V2++:tetra}
\end{multline}
Only 2 of the 15 entries above the diagonal are nonzero.
They couple channels 1 and 2 and channels 5 and 6.
If $V_{1^{--}\Sigma_g^+}(r) = V_{1^{--}\Pi_g}(r)$, each of the six channels is decoupled.

When the truncation of the diabatic potentials is extended to include $0^{++}\Sigma_g^+$, the additional radial quarkonium channels with $J^{PC} = 2^{++}$ are\\
\indent \indent \indent 7.  $\big( (0^{++},1) 1, 1 \big)$ \\
\indent \indent \indent 8.  $\big( (0^{++},3) 3, 1 \big)$.  \\
Channel 7 is $P$-wave and channel 8 is $F$-wave.

We now give the $8\times 8$ matrices in the $2^{++}$ block of $\mathbf{H}_0$ in the radial tetraquark basis.
The orbital-angular-momentum matrix in Eq.~\eqref{L2:2++6} is extended to a diagonal $8 \times 8$ matrix:
\begin{equation}
\mathbf{L}_{Q \bar Q}^2 = \mathrm{diag} \big( 0,6,6,6,6,20, 2,12 \big) .
\label{L2:2++8}
\end{equation}
The static potential for $2^{++}$ is the sum of the matrix in Eq.~\eqref{V2++:tetra} (with a 7th and 8th row of zeros and a 7th and 8th column of zeros) and two additional  $8 \times 8$ matrices:
\begin{equation}
 \Delta\mathbf{V}^{2^{++}}_0(r) =
V_{Q \bar Q}(r)
\begin{pmatrix}
\bm{0}_{6 \times 6} & \bm{0}_{6 \times 2}\\
\bm{0}_{2 \times 6} & \bm{1}_{2 \times 2}  \\
\end{pmatrix} 
+ G_{\Sigma_g^+}(r)
\begin{pmatrix}
\bm{0}_{6 \times 6} & \hat{\bm{N}}_7^T & \hat{\bm{N}}_8^T\\
\hat{\bm{N}}_7 & 0 & 0                                                \\
\hat{\bm{N}}_8 & 0  & 0                                              \\
\end{pmatrix} ,
\label{V2++:4}
\end{equation}
where $\hat{\bm{N}}_7$ and $\hat{\bm{N}}_8$ are 6-component unit  row vectors: 
\begin{subequations}
\begin{align}
\hat{\bm{N}}_7 &= 
\begin{pmatrix} \sqrt{\frac{1}{3}} ~& -\sqrt{\frac{2}{3}} ~&~  0  ~&~ 0  ~&~ 0  ~&~ 0 \end{pmatrix} , 
\\
\hat{\bm{N}}_8 &= 
\begin{pmatrix}  0  ~&~ 0  ~&~ 0  ~&~ 0  ~&~ \sqrt{\frac{3}{7}} ~&~ - \sqrt{\frac{4}{7}}\end{pmatrix} . 
\end{align}
\end{subequations}

\subsubsection{With Spin Splittings}

When the diabatic potentials are truncated to $0^{-+}\Sigma_u^-$,  $1^{--}\Sigma_g^+$, and $1^{--}\Pi_g$, there are 6 radial dimeson channels $\big( (J_1^-, J_2^-) S,L_{Q\bar Q} \big)$ for $J^{PC}=2^{++}$. 
We enumerate them in order of increasing  $L_{Q \bar Q}$, increasing  $J_1+J_2$, and then increasing $S$.
We take channels 1 and 2 to be $\big( ( 1^-,1^-)2,0 \big)$ and $\big( ( 0^-,0^-)0,2 \big)$.
We take channel 3 to be the even superposition of $\big( ( 1^-,0^-)1,2 \big)$ and $\big( ( 0^-,1^-)1,2 \big)$.
We take channels 4, 5, and 6 to be $\big( ( 1^-,1^-)0,2 \big)$, $\big( ( 1^-,1^-)2,2 \big)$, and $\big( ( 1^-,1^-)2,4 \big)$.
Our 6 radial dimeson channels for $J^{PC}=2^{++}$  labeled by charm-meson pairs and by $(S,L_{Q \bar Q})$ are\\
\indent \indent \indent 1: $D^* \bar D^*(2,0)$,\\
\indent \indent \indent 2: $D \bar D(0,2)$,\\
\indent \indent \indent 3: $[D^* \bar D(1,2) + D \bar D^*(1,2)]/\sqrt2$,\\
\indent \indent \indent 4:  $D^* \bar D^*(0,2)$,\\
\indent \indent \indent 5: $D^* \bar D^*(2,2)$,\\
\indent \indent \indent 6: $D^* \bar D^*(2,4)$.

We now give the $ 6\times 6$ matrices in the $2^{++}$ block of $\mathbf{H}_1$ in the radial dimeson basis.
The orbital-angular-momentum matrix $\mathbf{L}_{Q \bar Q}^2$ is given in Eq.~\eqref{L2:2++6}.
The static potential for $2^{++}$ is
\begin{multline}
\mathbf{V}^{2^{++}}_0 = V_{\Pi_g}(r) \bm{1}_{6 \times 6} 
+ \frac14 \big[ V_{\Sigma_u^-}(r) - V_{\Pi_g}(r) \big]
\begin{pmatrix}
~0~  & ~0~      & ~0~ & ~0~       & ~0~ & ~0~ \\
0      & 1          & 0      & \sqrt{3} & 0      & 0 \\
0      & 0          & 0      & 0          & 0       & 0 \\
0      & \sqrt{3} & 0      & 3          & 0       & 0 \\
0      & 0          & 0      & 0          & 0       & 0 \\
0      & 0          & 0      & 0          & 0       & 0 \end{pmatrix}
\\
+ \frac{1}{420}\big[ V_{\Sigma_g^+}(r) - V_{\Pi_g}(r) \big]
\begin{pmatrix}
140              & -14\sqrt{30}   & -42\sqrt{10} & 14\sqrt{10} & -14\sqrt{70} & 0                     \\
-14\sqrt{30} & 105                & 0                 & -35\sqrt{3}  & 20\sqrt{21}   & -12\sqrt{105}   \\
-42\sqrt{10} & 0                    & 210             & 0                 & 30\sqrt{7}     & 24\sqrt{35}      \\
14\sqrt{10}  & -35\sqrt{3}      & 0                 & 35               & -20\sqrt{7}   & 12\sqrt{35}      \\
-14\sqrt{70} & 20\sqrt{21}     & 30\sqrt{7}    & -20\sqrt{7}  & 110              & -24\sqrt{5}       \\
0                  & -12\sqrt{105} & 24\sqrt{35}  & 12\sqrt{35}  & -24\sqrt{5}   & 240                 \end{pmatrix}.
\label{V2++:dimeson}
\end{multline}
All but 3 of the 15 entries above the diagonal are nonzero. 
The spin-splitting potential is
\begin{multline}
\mathbf{V}^{2^{++}}_\text{SS}(r) = \frac{1}{6}\, \Delta_{8,Q}
\begin{pmatrix}
~2~ & ~0~       & ~0~ & ~0~      & ~0~  & ~0~ \\
  0   &  -3         &   0   & \sqrt{3} &   0     & 0     \\
  0   &   0         &  -2   &   0         &   0    & 0     \\
  0   & \sqrt{3} &   0    &  -1        &    0    & 0     \\
  0   &   0        &    0   &   0         &   2     & 0   \\
  0  &   0        &    0   &   0         &   0     & 2    \end{pmatrix}
 \exp\big( -r^2/R_\text{SS}^2 \big) 
\\
+  \frac{1}{2}\, \Delta_Q\,  \mathrm{diag}\big( 1, -3, -1, 1, 1, 1 \big)  \left[ 1 - \exp\big( -r^2/R_\text{SS}^2 \big) \right] .
\label{VSS:2++6}
\end{multline}
If $V_{\Sigma_g^+}(r) = V_{\Pi_g}(r)$, channels 2 and 4 are coupled by both $\mathbf{V}^{2^{++}}_0$ and $\mathbf{V}^{2^{++}}_\text{SS}$, but each of the other 4 channels is decoupled. 
The kinetically improved mass matrix is
\begin{equation}
\mathbf{M}_{2^{++}} =  \mathrm{diag}\big( m_{D^*},m_D,2 \mu_{D^* D}, m_{D^*},  m_{D^*}, m_{D^*} \big) .
\label{M:2++6}
\end{equation}

When the truncation of the diabatic potentials is extended to include $0^{++}\Sigma_g^+$, there are additional radial quarkonium channels of the form $Q \bar{Q}(S_{Q \bar Q},L_{Q \bar Q})$.
The additional radial quarkonium channels with $J^{PC} = 2^{++}$ are\\
\indent \indent \indent 7.  $Q \bar Q(1,1)$, \\
\indent \indent \indent 8. $Q \bar Q(1,3)$. \\
They are equal to the 7th and 8th channels in the radial tetraquark basis.

We now give the $8 \times 8$ matrices in the $2^{++}$ block of $\mathbf{H}_1$ in the radial dimeson basis.
The orbital-angular-momentum matrix $\mathbf{L}_{Q \bar Q}^2$ is given in Eq.~\eqref{L2:2++8}.
The static potential for $2^{++}$ is the sum of the matrix in Eq.~\eqref{V2++:dimeson} (with a 7th and 8th row of zeros and a 7th and 8th column of zeros) and two additional $8 \times 8$ matrices of the form in Eq.~\eqref{V2++:4} except that $\hat{\bm{N}}_7$ and $\hat{\bm{N}}_8$ are the 6-component unit  row vectors: 
\begin{subequations}
\begin{align}
\hat{\bm{N}}_7 =& 
\begin{pmatrix} -\sqrt{\frac{1}{3}} ~&~ \sqrt{\frac{1}{10}} ~&~ \sqrt{\frac{3}{10}} ~& -\sqrt{\frac{1}{30}} ~&~ \sqrt{\frac{7}{30}} ~& ~0~ \end{pmatrix} ,\\
\hat{\bm{N}}_8 =& 
\begin{pmatrix}  0  ~& -\sqrt{\frac{3}{20}} ~&~ \sqrt{\frac{1}{5}} ~&~ \sqrt{\frac{1}{20}} ~& -\sqrt{\frac{1}{35}} ~&~ \sqrt{\frac{4}{7}} \end{pmatrix} . 
\end{align}
\end{subequations}
The spin-splitting potential in Eq.~\eqref{VSS:2++6} is extended to an $8 \times 8$ matrix with a 7th and 8th row of zeros and a 7th and 8th column of zeros.
The mass matrix in Eq.~\eqref{M:2++6} is extended to a $8 \times 8$ matrix whose 7th and 8th diagonal entries are $m_Q$:
\begin{equation}
\mathbf{M}_{2^{++}} =  \mathrm{diag}\big( m_{D^*},m_D,2 \mu_{D^* D}, m_{D^*},  m_{D^*}, m_{D^*}, m_Q, m_Q \big) .
\label{M:2++8}
\end{equation}


\section{Resonance Energies and Widths using SPARSE}
\label{sec:SPARSE}

The $S$-matrix $\mathbf{S}(E)$ is the complex-valued matrix of scattering amplitudes between asymptotic states at large $r$ as functions of the energy $E$.
The dimension of the $S$-matrix is the number of scattering channels.
The $S$-matrix is a unitary matrix for real energy $E$.
Time reversal invariance implies that it is also a symmetric matrix.
The $T$-matrix $\mathbf{T}(E)$ is related to the $S$-matrix by $\mathbf{S}(E) = \mathbf{I} + 2 i\, \mathbf{T}(E)$.
The $T$-matrix $\mathbf{T}(E)$ is an analytic function of the complex energy $E$.

An introduction to the $K$-matrix formalism is given in a 1972 paper by Aitchison  \cite{Aitc72}.\\
The $K$-matrix $\mathbf{K}(E)$ can be defined by expressing the $S$-matrix and $T$-matrix as
\begin{equation}
\mathbf{S}(E) = \frac{\mathbf{I} +i\, \mathbf{K}(E)}{\mathbf{I} - i\, \mathbf{K}(E)}, 
\qquad \mathbf{T}(E) = \frac{\mathbf{K}(E)}{\mathbf{I} - i\, \mathbf{K}(E)}.
\label{S-T}
\end{equation}
The $K$-matrix is a hermitian matrix for real energy $E$.
Time reversal invariance implies that it is a real symmetric matrix. 
The $K$-matrix has poles at real energies $E_n$. 
Some of the poles are related to the poles of the $T$-matrix, but there can also be additional spurious poles.
The limiting behavior of the $K$-matrix for real energy $E$ approaching $E_n$ must have the form
\begin{equation}
\mathbf{K}(E) \longrightarrow  \mp\,  \mathbf{g}_n\, \frac{1}{E - E_n}\, \mathbf{g}_n^T,
\label{K-E}
\end{equation}
where $\mathbf{g}_n$ is a column vector of real amplitudes and the prefactor can be $-$ or +.
Note that the matrix $\mathbf{g}_n\, \mathbf{g}_n^T$ is a projector onto vectors proportional to $\mathbf{g}_n$.
Inserting the limiting expression in Eq.~\eqref{K-E} into the $T$-matrix in Eq.~\eqref{S-T}, we get an approximation to $\mathbf{T}(E)$ for real energy sufficiently close to $E_n$:
\begin{equation}
\mathbf{T}(E) \approx \mp\, \mathbf{g}_n\, \frac{1}{E - (E_n  \mp i\, \mathbf{g}_n \cdot  \mathbf{g}_n ) }\, \mathbf{g}_n^T .
\label{T-En}
\end{equation}
This equation suggests that the analytic continuation of $\mathbf{T}(E)$ into the complex energy plane has a pole near $E_n  \mp i\, \mathbf{g}_n \cdot  \mathbf{g}_n$.
However, the imaginary part of a pole of $\mathbf{T}(E)$ must be negative.
Hence $\mathbf{T}(E)$ cannot have a pole near $E_n  + i\,\mathbf{g}_n \cdot  \mathbf{g}_n$.
A pole with the prefactor + in Eq.~\eqref{K-E} must therefore be a spurious pole that does not correspond to a pole of the $T$-matrix.
If $\mathbf{g}_n  \cdot \mathbf{g}_n$ is small, the background under the spurious pole in Eq.~\eqref{K-E} must have strong energy dependence to ensure that there is no complex $T$-matrix pole corresponding to the real $K$-matrix pole.
A pole with the prefactor $-$ in Eq.~\eqref{K-E} may correspond to a resonance pole of the $T$-matrix whose width can be approximated by
\begin{equation}
\Gamma_n = 2\, \mathbf{g}_n  \cdot \mathbf{g}_n.
\label{Gamma_n}
\end{equation}
The complex energy $E_n - i\, \Gamma_n/2$ may be a good approximation to the resonance pole if the pole is isolated, if it is close enough to the real axis, and if the smooth background near the pole in Eq.~\eqref{K-E} varies sufficiently slowly with $E$ \cite{Svar12}.
We will refer to the real pole $E_n$ of the $K$-matrix associated with a resonance as its {\it $K$-matrix energy}.
We will refer to $\Gamma_n$ in Eq.~\eqref{Gamma_n} as the {\it $K$-matrix width} of the resonance.
To avoid the inference that $E_n- i \Gamma_n/2$ is a pole in the analytic continuation of $\mathbf{T}(E)$ into the complex $E$ plane, we will represent these $K$-matrix parameters by $(E_n,\Gamma_n)$.

The approximation for the $T$-matrix in  Eq.~\eqref{T-En} with prefactor $-$ can be used to define the {\it $K$-matrix branching fractions} $B_{n,i}$ for the resonance $n$ into each of the possible scattering channels $i$:
\begin{equation}
B_{n,i} = (\mathbf{g}_{n,i})^2/( \mathbf{g}_n  \cdot \mathbf{g}_n ).
\label{B-ni}
\end{equation}
The diagonal entries of the $T$-matrix in Eq.~\eqref{T-En} at complex energies $E$ near the real $K$-matrix pole $E_n$ can be approximated by
\begin{equation}
\bigl( \mathbf{T}(E) \bigr)_{ii} \approx i \frac{B_{n,i}(i \, \Gamma_n/2)}{E - ( E_n - i\,  \Gamma_n/2)}  .
\label{eq:kmatparam}
\end{equation}

The $K$-matrix $\mathbf{K}(E)$ for pairs of particles interacting through potentials can be calculated as a function of the real energy $E$ using the SPARSE algorithm \cite{Brus25}.
The SPARSE algorithm solves the Schr\"odinger equation numerically as a linear system with a sparse matrix of coefficients.
For energies $E$ above the lowest particle-pair threshold, the linear system admits a set of linearly independent solutions that represent scattering states of pairs of particles.
SPARSE calculates the $K$-matrix for the scattering process by matching the numerical solutions to the known asymptotic behavior of the analytic solutions with real boundary conditions as $r \to \infty$.
SPARSE can also calculate the energies of bound states below the lowest particle-pair threshold from the eigenvalues of the sparse matrix of coefficients \cite{Brus25}. 

The $K$-matrix is calculated by SPARSE as a function of the real energy $E$.
It can be used to calculate scattering amplitudes and to give estimates for resonance poles and branching fractions.
The $K$-matrix energy $E_n$, width $\Gamma_n$, and branching fractions $B_{n,i}$ provide good approximations to the resonance energy, width, and branching fractions if $E_n$ is an isolated $K$-matrix pole, $\Gamma_n$ is positive and sufficiently small, and if the smooth background under the pole varies sufficiently slowly with $E$.
The approximations may not be justified for resonances that are broad, are overlapping, or whose background has strong energy dependence.
The $K$-matrix parameters can alternatively be extracted from experimental data for reaction rates $|\mathbf{T}_{ij}(E)|^2$ with real energy $E$ by fitting the $K$-matrix to an expression that includes poles on the real $E$ axis like that in Eq.~\eqref{eq:kmatparam}.

We will use SPARSE to calculate the $K$-matrix only on the physical axis with real energy $E$ above the lowest threshold.
This gives us information about the resonance poles of $\mathbf{T}(E)$ only on the unphysical sheet of $E$ just below the physical real axis.
Our calculations will provide no information about the poles of $\mathbf{T}(E)$ far from the physical real axis or on other sheets of $E$.
We will be able to follow a resonance pole as a function of the parameters in the Schr\"odinger equation only as long as the pole remains on the unphysical sheet near the physical real axis.
A pole that moves to another sheet is no longer evident from the $K$-matrix at real energies.
In particular, a bound state that becomes a virtual state when it reaches a threshold disappears from the real $K$-matrix spectrum.
The virtual state corresponds to a pole below the threshold on the real axis of an unphysical sheet of the energy. 
There is no corresponding real pole of the $K$-matrix, so SPARSE cannot determine its dependence on the parameters in the Schr\"odinger equation.

\bibliography{BO-X3872-arXiv}
\bibliographystyle{JHEP}

\end{document}